\documentclass[%
 reprint,
 superscriptaddress,
 amsmath,amssymb,
 prl,
]{revtex4-2}

\usepackage{graphicx}
\usepackage{dcolumn}
\usepackage{bm}
\usepackage{gensymb}
\usepackage{svg}
\usepackage{afterpage}
\usepackage{physics}
\usepackage{tabularx}
\usepackage{xr}
\usepackage{natbib}
\usepackage{amsmath}
\usepackage{amsfonts}
\usepackage[T1]{fontenc}
\usepackage[utf8]{inputenc}

\usepackage{xr}
\begin{document}

\preprint{APS/123-QED}

\title{Emergent universal long-range structure in random-organizing systems}

\author{Satyam Anand}
\thanks{Equal contribution}
\email{sa7483@nyu.edu}
\affiliation{Courant Institute of Mathematical Sciences, New York University, New York, NY 10003, USA}
\affiliation{\mbox{Center for Soft Matter Research, Department of Physics, New York University, New York, NY 10003, USA}}

\author{Guanming Zhang}
\thanks{Equal contribution}
\email{gz2241@nyu.edu}
\affiliation{\mbox{Center for Soft Matter Research, Department of Physics, New York University, New York, NY 10003, USA}}
\affiliation{\mbox{Simons Center for Computational Physical Chemistry, Department of Chemistry, New York University, New York, NY 10003, USA}}

\author{Stefano Martiniani}
\email{sm7683@nyu.edu}
\affiliation{Courant Institute of Mathematical Sciences, New York University, New York, NY 10003, USA}
\affiliation{\mbox{Center for Soft Matter Research, Department of Physics, New York University, New York, NY 10003, USA}}
\affiliation{\mbox{Simons Center for Computational Physical Chemistry, Department of Chemistry, New York University, New York, NY 10003, USA}}
\affiliation{\mbox{Center for Neural Science, New York University, New York, NY 10003, USA}}

\begin{abstract}

Self-organization through noisy interactions is ubiquitous across physics, mathematics, and machine learning, yet how long-range structure emerges from local noisy dynamics remains poorly understood. Here, we investigate three paradigmatic random-organizing particle systems drawn from distinct domains: models from soft matter physics (random organization, biased random organization) and machine learning (stochastic gradient descent), each characterized by distinct sources of noise. We discover universal long-range behavior across all systems, namely the suppression of long-range density fluctuations, governed solely by the noise correlation between particles. Furthermore, we establish a connection between the emergence of long-range order and the tendency of stochastic gradient descent to favor flat minima---a phenomenon widely observed in machine learning. To rationalize these findings, we develop a fluctuating hydrodynamic theory that quantitatively captures all observations. Our study resolves long-standing questions about the microscopic origin of noise-induced hyperuniformity, uncovers striking parallels between stochastic gradient descent dynamics on particle system energy landscapes and neural network loss landscapes, and should have wide-ranging applications---from the self-assembly of hyperuniform materials to ecological population dynamics and the design of generalizable learning algorithms.

\end{abstract}

\maketitle

While typically associated with disorder, noise can paradoxically drive the emergence of diverse forms of order, such as pattern formation \cite{sagues2007spatiotemporal}, self-organization \cite{corte2008random, hexner2017enhanced}, suppression of chaos \cite{matsumoto1983noise}, selection of ordered states \cite{villain1980order}, and swarming \cite{jhawar2020noise}. 
Physical systems exhibit a broad spectrum of order: perfect crystals and ideal gases mark the extremes, while intermediate regimes display correlated disorder, such as hyperuniformity---where local disorder coexists with the anomalous suppression of long-range density fluctuations \cite{torquato2018hyperuniform}. Hyperuniformity can emerge either at criticality \cite{donev2005unexpected, hexner2015hyperuniformity, wilken2021random}, or away from it \cite{lei2019hydrodynamics, hexner2017noise, huang2021circular, galliano2023two}. Away from criticality, in equilibrium, hyperuniformity requires long-range interactions \cite{torquato2018hyperuniform}, whereas out of equilibrium, it can emerge from long- or short-range, and even noisy interactions \cite{lei2019hydrodynamics, hexner2017noise, huang2021circular, galliano2023two}. The process by which long-range spatial structure develops away from criticality---particularly in systems interacting solely via short-range, noisy dynamics---is a long-standing question that remains poorly understood. 

Non-equilibrium particle systems with short-range noisy interactions---such as random organization (RO) \cite{corte2008random, hexner2015hyperuniformity, tjhung2015hyperuniform, tjhung2016criticality, milz2013connecting, menon2009universality}, biased random organization (BRO) \cite{wilken2021random, wilken2023dynamical, hexner2017noise, lei2023random, galliano2023two, milz2013connecting}, and stochastic gradient descent (SGD) \cite{zhang2024absorbing}---provide an ideal framework to investigate the noise-driven emergence of long-range spatial structure. These systems have been studied in a wide variety of contexts, such as sheared colloidal suspensions \cite{pine2005chaos, corte2008random, hexner2015hyperuniformity, wilken2020hyperuniform}, random close packing \cite{wilken2021random, wilken2023dynamical}, two-dimensional crystallization \cite{galliano2023two}, and self-supervised learning \cite{zhang2024absorbing}. All systems undergo a phase transition as the particle volume fraction increases; from a low-density state where all motion ceases (absorbing state), to a high-density state where motion persists forever (active state) \cite{hinrichsen2000non, henkel2008non, martiniani2019quantifying}. 
Irrespective of microscopic details, RO, BRO, and SGD belong to the same universality class, i.e., display the same critical behavior \cite{menon2009universality, wilken2021random, zhang2024absorbing}. 
Away from criticality in the active phase, however, variations in microscopic interactions significantly influence the emergent long-range structure \cite{tjhung2015hyperuniform, tjhung2016criticality, hexner2017noise, galliano2023two}.
Despite extensive experimental, numerical, and theoretical research over two decades, a quantitative microscopic understanding of dynamics and structure far from criticality remains elusive \cite{pine2005chaos, corte2008random, hexner2015hyperuniformity, tjhung2015hyperuniform, tjhung2016criticality, milz2013connecting, wilken2021random, wilken2023dynamical, hexner2017noise, lei2023random, galliano2023two, milz2013connecting, zhang2024absorbing}. 
Fundamental questions remain unanswered: How does macroscopic structure emerge from noisy interactions?
Moreover, what universal principles govern the variability in emergent structures within and across different random-organizing systems?
Finally, is the emergent long-range structure in SGD related to its ability to discover flat minima---a feature linked to robust generalization in machine learning \cite{jastrzkebski2017three}?

\begin{figure*}[htpb!]
    \centering
    \includegraphics[width=\linewidth]{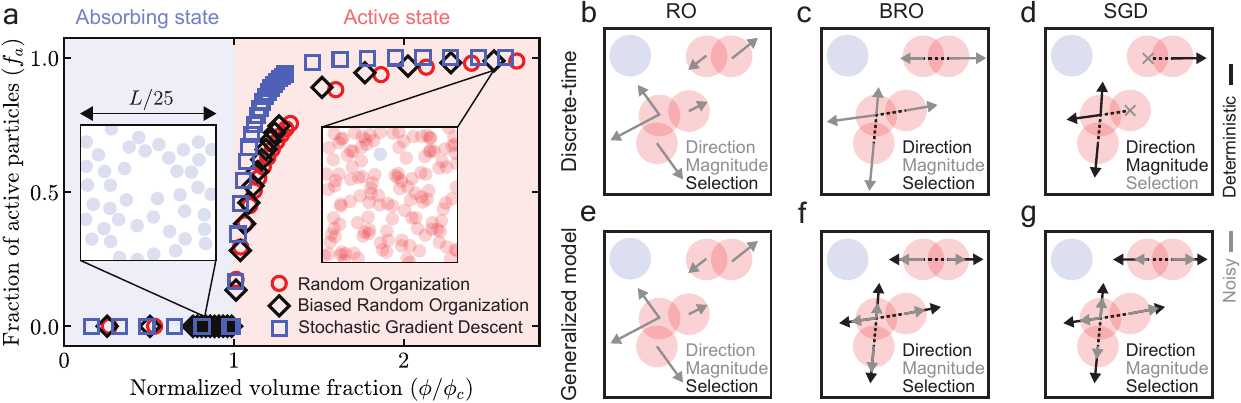}
    \caption{\small \textbf{Random-organizing systems.} (a) Absorbing-to-active phase transition. The steady-state fraction of active (overlapping) particles ($f_a$) plotted as a function of normalized volume fraction $\phi/\phi_c$, where $\phi_c$ is the critical volume fraction. Starting from an initial random configuration, at time $t \to \infty$, in the absorbing state ($\phi/\phi_c < 1$), $f_a = 0$, while in the active state ($\phi/\phi_c > 1$), $f_a > 0$. Insets show zoomed in exemplar configurations of discrete-time BRO, and $L$ is the side length of the simulation box. Blue particles have no overlapping neighbors whereas red particles have at least one overlapping neighbor. Schematics of discrete-time RO (b) (Eq.~\ref{eq:discrete_ro}), BRO (c) (Eq.~\ref{eq:discrete_bro}), SGD (d) (Eq.~\ref{eq:discrete_sgd}), and the corresponding continuous-time approximation (generalized model, Eq.~\ref{eq:generic_sde}) of RO (e), BRO (f), and SGD (g). Solid gray colors denote noise, while solid black colors denote deterministic interactions. Dashed black lines connect the centers of a pair of overlapping particles. Crosses in panel (d) denote unselected particles. For RO, the direction and the magnitude of the kicks are both noisy; for BRO, only the magnitude of the kicks are noisy; while for SGD, only the selection of active particles is noisy. Notice that in the generalized model (Eq.~\ref{eq:generic_sde}), the selection noise of SGD is approximated by a noise in the magnitude of the kicks. 
}
    \label{fig:figure1}
\end{figure*}

Here, combining particle and continuum simulations with hydrodynamic theory, we provide a quantitative, microscopic understanding of the active phases of RO, BRO, and SGD---random-organizing systems with distinct sources of noise. 
We discover universal long-range behavior across all three systems governed by a single parameter: the noise correlation coefficient between particles.
All systems self-organize to suppress density fluctuations below a crossover length scale, which diverges as the noise becomes anti-correlated (reciprocal interactions), resulting in strong (class I \cite{torquato2018hyperuniform}) hyperuniformity.
Further, by directly coarse-graining the microscopic dynamics, we develop a fluctuating hydrodynamic theory for random-organizing systems that quantitatively predicts both the emergence of long-range structure and the crossover length scale across all systems. 
Finally, we demonstrate how noise correlation, batch size, and learning rate bias SGD towards flatter minima---a finding that aligns with empirical observations in machine learning \cite{jastrzkebski2017three, xie2020diffusion, keskar2016large, huang2020understanding}---and reveal a novel connection between the emergence of long-range structure and the ability of SGD to generalize effectively. 
Our study underscores the critical role of noise correlations in facilitating long-range structure, and has wide-ranging applications---from providing a robust method for self-assembling hyperuniform structures, to offering insights into other systems having correlated noise, such as neural population activity in the brain \cite{averbeck2006neural}, ecological population dynamics \cite{pal2022tipping}, and gene expression dynamics in cells \cite{tsimring2014noise}.


\subsection{Setup}

Random-organizing systems---RO, BRO, and SGD---are discrete-time systems consisting of $N$ interacting spherical particles of radius $R$ in $d$-dimensional space. 
At any given time-step, the positions of isolated particles (blue in Figs.~\ref{fig:figure1}b-d) do not evolve, and those of overlapping (active) particles (red in Figs.~\ref{fig:figure1}b-d) evolve according to system specific rules designed to resolve particle overlaps.
All systems undergo an absorbing-to-active phase transition as the particle volume fraction $\phi$ increases (Fig.~\ref{fig:figure1}a). Starting from a random initial configuration, for $\phi < \phi_c$, the system finds an absorbing configuration ($f_a = 0$), whereas for $\phi > \phi_c$, the system never finds such a configuration and reaches a non-equilibrium steady state ($f_a > 0$) (Fig.~\ref{fig:figure1}a). Here, $f_a$ is the fraction of active particles in the system. We report all our results in the active phase when the system has reached steady state (Methods). 

\subsection{Random Organization}

RO was originally introduced to model experiments on sheared colloidal suspensions at high P\'eclet number \cite{pine2005chaos, corte2008random}. Consider a system undergoing periodic shear cycles. After every shear cycle, overlapping particles are given a ``kick'' in a random direction, and non-overlapping particles return to their original position \cite{corte2008random, hexner2015hyperuniformity}. RO was subsequently simplified to a model without external shearing, which retains all the essential properties of the original version \cite{milz2013connecting, tjhung2015hyperuniform, tjhung2016criticality, martiniani2019quantifying}. Here, we study this simpler isotropic version of RO \cite{milz2013connecting, tjhung2015hyperuniform, tjhung2016criticality, martiniani2019quantifying}.

In RO, the dynamics of the position of particle $i$ at time-step $m+1$ ($\mathbf{x}_{i}^{m+1}$) is given by,
\begin{equation}
    \mathbf{x}_{i}^{m+1} = \mathbf{x}_{i}^{m} + \epsilon \sum_{j \in \Gamma_i^m} u_{ji}^m \boldsymbol{\zeta}^m_{ji}, 
    \label{eq:discrete_ro}
\end{equation}
where $\epsilon$ controls the magnitude of the pairwise kick given by particle $j$ to $i$, $u_{ji}^m$ is a random number sampled from a standard uniform distribution ($U[0,1]$) at time-step $m$, $\boldsymbol{\zeta}^m_{ji}$ is a random unit vector sampled uniformly on the surface of a $d$-dimensional unit hypersphere at time-step $m$, and $\Gamma_i^m = \{j \mid |\mathbf{x}_j^m - \mathbf{x}_i^m| < 2R,\ j \neq i \}$ is the set containing all particles that overlap with particle $i$ at time-step $m$ (Fig.~\ref{fig:figure1}b). 
We set $\boldsymbol{\zeta}^m_{ji} = - \boldsymbol{\zeta}^m_{ij}$ so that particles effectively move away from (``repel'') each other. 
Finally, $c \in [-1,0]$ is the Pearson correlation coefficient between corresponding components of the complete noise vectors $\omega_{ij, \alpha}^m = \epsilon u_{ij}^m {\zeta}^m_{ij, \alpha}$ and $\omega_{ji, \alpha}^m$, defined by $ \langle \omega_{ij, \alpha}^m \, \omega_{ji, \beta}^m \rangle / \langle \omega_{ij, \alpha}^m \rangle \langle \omega_{ji, \beta}^m \rangle = c \, \delta_{\alpha \beta}$. $c = 0$ denotes pairwise kicks which are uncorrelated, and $c=-1$ denotes pairwise kicks which are anti-correlated (equal in magnitude and opposite in direction, i.e., conserve pairwise center of mass). 

\subsection{Biased Random Organization}

BRO was originally introduced to study random close packing of spheres \cite{wilken2021random, wilken2023dynamical}. Similar to RO, overlapping particles are given a kick, however, the kick is ``biased'' along the direction of the line joining the centers of overlapping particles (Fig.~\ref{fig:figure1}c). An interesting feature of BRO is that for $\epsilon \to 0$, its critical point was proposed as an alternative definition of random close packing \cite{wilken2021random, wilken2023dynamical}, which, despite decades of research, still lacks a clear definition \cite{torquato2000random, kamien2007random, parisi2005ideal, martiniani2017numerical, anzivino2023estimating}.

In BRO, the dynamics of the position of particle $i$ at time-step $m+1$ ($\mathbf{x}_{i}^{m+1}$) is given by,
\begin{equation}
    \mathbf{x}_{i}^{m+1} = \mathbf{x}_{i}^{m} + \epsilon \sum_{j \in \Gamma_i^m} u_{ji}^m \mathbf{\hat{x}}^m_{ji}, 
    \label{eq:discrete_bro}
\end{equation}
where $\epsilon$ and $u_{ji}^m$ are defined as in RO, and $\mathbf{\hat{x}}^m_{ji} = -(\mathbf{x}^m_{j} - \mathbf{x}^m_{i})/|\mathbf{x}^m_{j} - \mathbf{x}^m_{i}|$ is the deterministic unit vector pointing from the center of particle $j$ to $i$ at time-step $m$ (Fig.~\ref{fig:figure1}c). 
$c \in [-1,0]$ is the Pearson correlation coefficient between corresponding components of the complete noise vectors $\omega_{ij, \alpha}^m = \epsilon u_{ij}^m {\hat{x}}^m_{ij, \alpha}$ and $\omega_{ji, \alpha}^m$ (notice that $\mathbf{\hat{x}}^m_{ji} = -\mathbf{\hat{x}}^m_{ij}$).

\subsection{Stochastic Gradient Descent}

SGD is a widely used optimization algorithm, e.g., in artificial neural networks, to minimize a loss function composed of a sum of many terms \cite{bishop2023deep}. Stochasticity in SGD comes from the random selection of a subset of terms in the sum at every step. In the context of interacting particle systems, we take the loss to be the total energy $E = \sum_i \sum_{j>i} V (\mathbf{x}_{i}, \mathbf{x}_{j})$, where $V (\mathbf{x}_{i}, \mathbf{x}_{j})$ is any pairwise potential. SGD then corresponds to randomly selecting a subset of terms in $E$ and updating the corresponding particle positions—either one or both at once—to minimize the partial energy. This is in contrast to simultaneously moving all active particles, which corresponds to (noiseless) gradient descent \cite{zhang2024absorbing}.

In SGD, the dynamics of the position of particle $i$ at time-step $m+1$ ($\mathbf{x}_{i}^{m+1}$) is given by,
\begin{equation}
    \begin{aligned}
        \mathbf{x}_{i}^{m+1} = \mathbf{x}_{i}^{m} - \alpha \sum_{j \in \Gamma_i^m} \theta_{ji}^m \nabla_{i} V_{ji}^{m}, 
    \end{aligned}
\label{eq:discrete_sgd}
\end{equation}
where $\nabla_{i} = \nabla_{\mathbf{x}_{i}}$, $\alpha$ is the learning rate having units of $\text{length}/\text{force}$, $V_{ji}^m = V(|\mathbf{x}^m_{j} - \mathbf{x}^m_{i}|)$ is the pairwise interaction potential between particle $i$ and $j$, and $\theta_{ji}^m$ is a random number sampled from a Bernoulli distribution having parameter $b_f$ (batch fraction) at time-step $m$ (Fig.~\ref{fig:figure1}d). $b_f$ represents the average fraction of active particle pairs $(i,j)$ that move at any given time-step. 
$c \in [-1,0]$ is the Pearson correlation coefficient between corresponding components of the complete noise vectors $\omega_{ij, \alpha}^m = - \alpha \theta_{ij}^m \partial_{i, \alpha} V_{ij}^{m}$ and $\omega_{ji, \alpha}^m$ (notice that $\nabla_{i} V_{ji}^{m} = -\nabla_{j} V_{ji}^{m}$), originating from the pairwise correlated selection noise $\theta_{ij}^m$. While $V_{ji}$ can be any short- or long-range potential in SGD, here, we consider a class of short-range, repulsive potentials given by
\begin{equation}
    V_{ij}(r) =     
    \begin{cases}
      \frac{E}{p}\left(1 - \frac{r_{ij}}{2R}\right)^p, & \text{if}\ 0 < r_{ij} < 2R, \\
      0, & \text{otherwise},
    \end{cases}  
    \label{eq:v_ji_family}
\end{equation}
where $r_{ij} = |\mathbf{x}_j - \mathbf{x}_i|$, $E$ is the characteristic energy scale, and $p$ controls the stiffness of the potential. 

\begin{figure*}[htpb!]
    \centering
    \includegraphics[width=\linewidth]{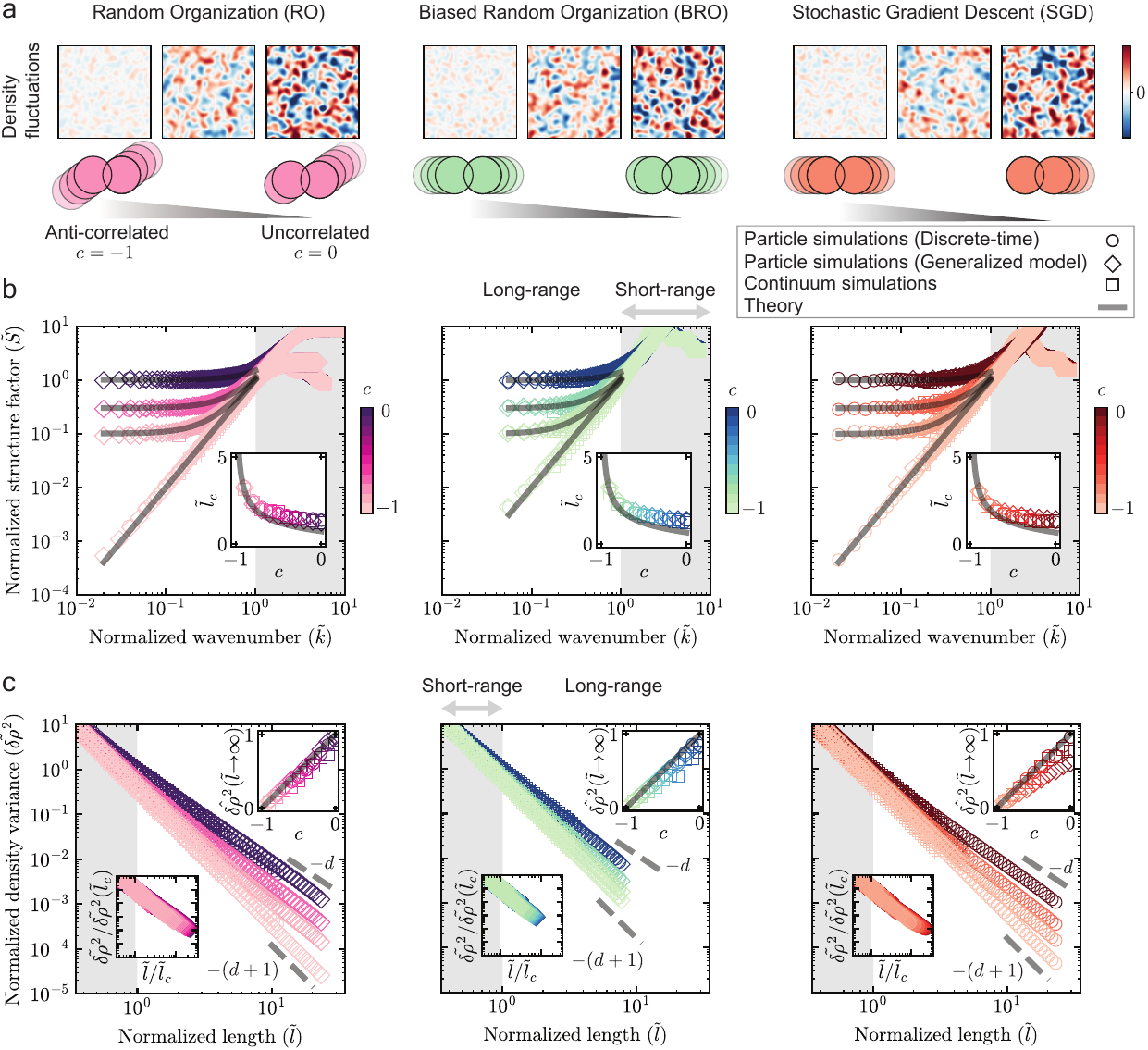}
    \caption{\small \textbf{Universal long-range structure in random-organizing systems.} (a) Coarse-grained density fluctuations $\delta \rho(c)/|\delta \rho_{\text{avg}}(c=0)|$ in random-organizing systems, where $c$ is the pairwise noise correlation, and $\delta \rho_{\text{avg}}$ denotes average density fluctuations over the whole system. The three panels denote $c=-1$, $c=-0.75$, $c=0$ (left to right) for each system. A Gaussian kernel of width $100R$ was chosen for coarse-graining all systems, where $R$ is the particle radius. As the pairwise noise correlation $c$ goes from $0$ (uncorrelated) to $-1$ (anti-correlated), the density fluctuations are suppressed for all systems. (b) Normalized radially averaged structure factor $\Tilde{S}(\Tilde{k})$ versus normalized wave number $\Tilde{k}$ for RO, BRO, and SGD (right to left). $\tilde{S} =  S(k)/S_{0}(2\pi/L)$ where $S_{0}(2\pi/L)$ is the structure factor for $c=0$ at $k = 2\pi/L$, and $L$ is the side length of the simulation box. $\tilde{k} = k/k_0$, where $k_0$ is the value at which $\tilde{S}(k_0) = 1$ for anti-correlated noise ($c=-1$) of the same system. Solid black lines show predictions of Eq.~\ref{eq:struc_fact_sim} for different values of $c$. Inset shows the normalized crossover length scale ($l_c/l_0 = \tilde{l}_c = 1/\tilde{k}_c$) versus $c$. We find the normalized crossover wavenumber ($\tilde{k}_c$) in simulations as the intersection, on a log-log plot, between a fit of slope $0$ near the $\tilde{k} \to 0$ region, and a fit of slope $2$ near the $\tilde{k} \approx 1$ region. Solid black line in the inset shows prediction of Eq.~\ref{eq:k_c}. Gray shaded regions denote short-range behavior ($\tilde{k}>1$). (c) Normalized variance of number density $\tilde{\delta \rho}^2 (\tilde{l})$ versus normalized diameter of the hypersphere ($\Tilde{l}$) used for measuring density fluctuations for RO, BRO, and SGD (right to left). $\tilde{\delta \rho}^2 (\tilde{l}) = \delta \rho^2 (l)/\delta \rho^2 (l_0)$ where $\delta \rho^2 (l_0)$ is the density variance for $c=0$ at $l=l_0$, and $\tilde{l}=l/l_0$, where $l_0 = 2\pi/k_0$. Bottom inset shows data collapse of density variances when $\tilde{l}$ is rescaled by $\tilde{l}_c$, and $\tilde{\delta \rho}^2 (\tilde{l})$ is rescaled by $\tilde{\delta \rho}^2 (\tilde{l}_c)$. Top inset shows infinite wavelength density fluctuations $\hat{\delta \rho}^2 (\tilde{l} \to \infty)$ versus $c$. $\hat{\delta \rho}^2 (c) = [\tilde{\delta \rho}^2 (c) - \tilde{\delta \rho}^2 (c=-1)]/\tilde{\delta \rho}^2 (c=0)$. Solid black line denotes $1+c$ (prediction of Eq.~\ref{eq:struc_fact_sim} in the limit $\tilde{k} \to 0$). Gray shaded regions denote short-range behavior ($\tilde{l}<1$).
}
    \label{fig:figure2}
\end{figure*}

\subsection{Universal active phase behavior}

There are three distinct sources of noise in random-organizing systems: magnitude of kicks, direction of kicks, and selection of particles. Notice that the origin of stochasticity in RO, BRO, and SGD is different; (i) for RO, the magnitude and direction of kicks are both noisy, while the selection of particles is deterministic, (ii) for BRO, the magnitude of kicks is noisy, while both the direction of kicks and selection of particles is deterministic, (iii) for SGD, the selection of particles is noisy, while the magnitude and direction of kicks are both deterministic (Figs.~\ref{fig:figure1}b-d, Eqns.~\ref{eq:discrete_ro}, \ref{eq:discrete_bro}, and \ref{eq:discrete_sgd}). We perform particle simulations for RO, BRO, and SGD in the active phase ($\phi > \phi_c$) and measure the long-range structure, quantified by the radially averaged structure factor $S(k)$ and variance in number density $\delta \rho^2 (l)$ (Methods). $l$ is the diameter of the hypershpere used for measuring density fluctuations and $k = 2\pi /l$ is the wave number. We study all properties above a threshold length scale $l_{0} = 2\pi/k_0$, below which we find system-specific short-range behavior (Fig.~\ref{fig:figure2}b,c). Hereafter, we work with normalized quantities: $\tilde{l}$, $\tilde{k}$, $\tilde{S}(\tilde{k})$, and $\tilde{\delta \rho}^2 (\tilde{l})$ (see Fig.~\ref{fig:figure2} caption for definitions).

Despite having microscopically different dynamics, all random-organizing systems display a universal long-range structure, controlled solely by the pairwise noise correlation $c$, and qualitatively independent of all other parameters---be it the kick magnitude $\epsilon$, volume fraction $\phi$, or spatial dimension $d$ in RO and BRO, or $\phi$, $d$, learning rate $\alpha$, batch fraction $b_f$, potential stiffness $p$, and energy scale $E$ in SGD. (Fig.~\ref{fig:figure2}a,b,c, Supplementary Information (SI) Figs.~\ref{fig:figure2_si}, \ref{fig:figure3_si}, \ref{fig:figure4_si}). Remarkably, the active phase behavior across all systems is independent of $d$, in contrast to their critical behavior, which is heavily dependent on $d$ \cite{hexner2015hyperuniformity, wilken2021random, zhang2024absorbing} (Figs.~\ref{fig:figure2_si}c, \ref{fig:figure3_si}c, \ref{fig:figure4_si}c). All systems self-organize to suppress density fluctuations below a normalized crossover length scale $\tilde{l}_c$; specifically, for length scales $1 < \tilde{l} < \tilde{l}_c$, the structure factor follows a power law ($\tilde{S} \sim \tilde{k}^{2}$) and $\tilde{\delta \rho} ^2 \sim \tilde{l}^{-(d+1)}$, whereas for $\tilde{l} > \tilde{l}_c$, the structure factor is constant, ($\tilde{S} \sim \text{const.}$) and $\tilde{\delta \rho} ^2 \sim \tilde{l}^{-d}$ (Figs.~\ref{fig:figure2}b,c).
Further, the crossover length scale $\tilde{l}_c$ increases monotonically as $c$ decreases: as the pairwise noise becomes more negatively correlated, density fluctuations are suppressed up to larger length scales (Figs.~\ref{fig:figure2}b inset). Consequently, the infinite wavelength density fluctuations, $\tilde{\delta \rho}^2 (\tilde{l} \to \infty) \propto \tilde{S}(\tilde{k} \to 0)$, decrease monotonically as $c$ decreases (Figs.~\ref{fig:figure2}c inset).
Finally, when the noise is anti-correlated ($c=-1$), the crossover length scale $\tilde{l_c} \to \infty$ and the system becomes strongly hyperuniform, $\tilde{S}(\tilde{k} \to 0) \sim \tilde{k}^{2}$, and $\tilde{\delta \rho} ^2 (\tilde{l} \to \infty) \sim \tilde{l}^{-(d+1)}$ (Figs.~\ref{fig:figure2}b, c). 

Our results are consistent with previous studies on RO, which focused on the specific case of uncorrelated noise ($c=0$) \cite{tjhung2015hyperuniform, tjhung2016criticality}, and on BRO, which focused on the specific case of anti-correlated noise ($c=-1$) \cite{hexner2017noise, galliano2023two}. So, why do microscopically distinct systems---RO, BRO, and SGD---exhibit universal long-range behavior?

\subsection{Generalized model}

We now develop a continuous-time model of discrete-time random-organizing systems. Using the framework of stochastic modified equations \cite{li2017stochastic, zhang2024absorbing}, we approximate the discrete-time dynamics by a continuous-time stochastic differential equation (SDE) (SI Sec. I.A). In the resulting genralized model, the dynamics of the position of particle $i$ ($\mathbf{x}_{i}$) is given by an overdamped Langevin equation,
\begin{equation}
    \frac{d\mathbf{x}_{i}(t)}{dt}
    = - \frac{1}{\gamma} \sum_{j = 1}^N \nabla_{i} V_{ji} 
    + \sum_{j = 1}^N \sqrt{\mathbf{\Lambda}_{ji}} \ \boldsymbol{\xi}_{ji}, 
    \label{eq:generic_sde}
\end{equation}
where $\gamma$ is the friction constant, $V_{ji}$, $\mathbf{\Lambda}_{ji}$ are short-range, pairwise interaction potential and diffusion matrix between particles $j$ and $i$, respectively, and $\sqrt{\mathbf{\Lambda}_{ji}}$ denotes a matrix square root. $\boldsymbol{\xi}_{ji}$ is a pairwise, Gaussian noise given by the particle $j$ to particle $i$ having mean $\left< {\xi}_{ji,\alpha} (t) \right> = 0$ and covariance matrix $\left< {\xi}_{ij,\alpha} (t) \, {\xi}_{kl,\beta} (t') \right> = \delta (t-t') \, \delta_{\alpha \beta} (\delta_{ik}\delta_{jl} + c \, \delta_{il}\delta_{jk})$, where $c \in [-1,0]$ is the Pearson correlation coefficient between ${\xi}_{ij,\alpha} (t)$ and ${\xi}_{ji,\alpha} (t)$. Eq.~\ref{eq:generic_sde}, supplemented with system-specific  $\gamma$, $V_{ij}$ and $\mathbf{\Lambda}_{ij}$, is an SDE approximating RO, BRO, and SGD (SI Sec. I.A, Table~\ref{tab:system_params}). Notice that in the generalized model, the source(s) of noise for RO and BRO remain the same, while the selection noise in SGD becomes a noise on the magnitude of the kicks (Figs.~\ref{fig:figure1}e-g, SI Sec. I.A) \cite{zhang2024absorbing}.

We perform particle simulations of the generalized model for RO, BRO, and SGD and find that the long-range structure is quantitatively the same as that for their discrete-time counterparts (Methods, Figs.~\ref{fig:figure2}b, c). Thus, the generalized model serves as an accurate continuous-time approximation for all random-organizing systems. 

\subsection{Fluctuating hydrodynamic theory}

Equipped with the generalized model, we formulate a theory for the evolution of the density field $\rho(\mathbf{x},t)$. 
Dean's method, originally introduced for Brownian particles with additive noise \cite{dean1996langevin}, and later extended to study active matter with multiplicative noise \cite{bertin2013mesoscopic, solon2015active}, is a well-known approach for directly coarse-graining microscopic dynamics. However, it has not yet been extended to systems with pairwise correlated noise between components.
Starting from Eq.~\ref{eq:generic_sde}, 
and under the assumption that the two-point density field $\rho_{2}(\mathbf{x}, \mathbf{y}) \approx \rho(\mathbf{x}) \rho(\mathbf{y})$, 
we extend Dean's method \cite{dean1996langevin} to account for multiplicative, and pairwise correlated noise, and derive the time evolution of $\rho(\mathbf{x},t)$ in arbitrary dimension $d$ to get (SI Sec. I.B.2),
\begin{equation}
    \frac{\partial \rho(\mathbf{x}, t)}{\partial t} 
    = - \underbrace{\nabla \cdot \left[ \rho(\mathbf{x})  \mathbf{v}(\mathbf{x})   \right]}_\text{drift term} + \underbrace{\nabla \nabla : \left[  \mathbf{D}(\mathbf{x}) \rho(\mathbf{x}) \right]}_\text{diffusion term} - \underbrace{\nabla \cdot \mathbf{j}_n(\mathbf{x})}_\text{noise term},  
    \label{eq:generic_drho}
\end{equation}
where $\nabla = \nabla_{\mathbf{x}}$, and $:$ denotes a double dot product. 

The velocity $\mathbf{v}(\mathbf{x})$ in Eq.~\ref{eq:generic_drho} is given by
\begin{equation}
     \mathbf{v}(\mathbf{x}) = - \frac{1}{\gamma} \bigl \langle \nabla V(\mathbf{x}, \mathbf{y}) \bigr \rangle_{\rho(\mathbf{y})},
    \label{eq:det_vel}
\end{equation}
where $\langle a \rangle_{\rho(\mathbf{y})} = \int a \rho(\mathbf{y}) d\mathbf{y}$, and $V(\mathbf{x}, \mathbf{y})$ is the ``continuous'' version of $V_{ji}$, given by replacing $\mathbf{x}_i$ and $\mathbf{x}_j$ by $\mathbf{x}$ and $\mathbf{y}$ in Eq.~\ref{eq:v_ji_family}. $\mathbf{v}(\mathbf{x})$ originates from the deterministic (first) term in Eq.~\ref{eq:generic_sde}, and can be understood as the average force at point $\mathbf{x}$ due to the local interaction potential, $- \langle \nabla V(\mathbf{x}, \mathbf{y}) \rangle_{\rho(\mathbf{y})}$, divided by the friction coefficient $\gamma$. 

The diffusion tensor $\mathbf{D}(\mathbf{x})$ in Eq.~\ref{eq:generic_drho} is given by
\begin{equation}
    \mathbf{D}(\mathbf{x}) = \frac{1}{2} \bigl \langle \mathbf{\Lambda}(\mathbf{x}, \mathbf{y}) \bigr \rangle_{\rho(\mathbf{y})}.
    \label{eq:diff_tensor}
\end{equation}
$\mathbf{D}(\mathbf{x})$ originates from the noise (second) term in Eq.~\ref{eq:generic_sde}, and can be understood as the average diffusion tensor over the local density. 

The stochastic flux $\mathbf{j}_n(\mathbf{x})$ in Eq.~\ref{eq:generic_drho} is given by
\begin{equation}
    \mathbf{j}_n(\mathbf{x})= - \sqrt{\rho(\mathbf{x})} \int \sqrt{\rho(\mathbf{y})} \sqrt{ \mathbf{\Lambda} (\mathbf{x}, \mathbf{y})} \cdot \boldsymbol{\eta}(\mathbf{x}, \mathbf{y}) \ \mathrm{d}\mathbf{y} , 
    \label{eq:fluc_flux}
\end{equation}
where $\sqrt{ \mathbf{\Lambda}}$ denotes a matrix square root, and $\mathbf{\Lambda}(\mathbf{x}, \mathbf{y})$ is the ``continuous'' version of $\mathbf{\Lambda}_{ji}$. $\boldsymbol{\eta}(\mathbf{x}, \mathbf{y}, t)$ is a vectorial, two-point, Gaussian noise field having mean $\left< {\eta}_{\alpha}(\mathbf{x}, \mathbf{y}, t) \right> = 0$, and covariance matrix $\left< {\eta}_{\alpha}(\mathbf{x}, \mathbf{y}, t) \, {\eta}_{\beta}(\mathbf{u}, \mathbf{w}, t')\right> = \delta_{\alpha \beta} \delta(t-t') \, [ \delta(\mathbf{x} - \mathbf{u}) \, \delta(\mathbf{y} - \mathbf{w}) + \, c \, \delta(\mathbf{x} - \mathbf{w}) \, \delta(\mathbf{y} - \mathbf{u}) ]$, where $c \in [-1,0]$ is the Pearson correlation coefficient between $\eta_{\alpha}(\mathbf{x}, \mathbf{y}, t)$ and $\eta_{\alpha}(\mathbf{y}, \mathbf{x}, t)$. 
$\mathbf{j}_{n}(\mathbf{x})$ originates from the noise (second) term in Eq.~\ref{eq:generic_sde}. 
Notice that the noise at any spatial location $\mathbf{x}$ can be viewed as the sum of independent kicks from $n_y \propto \rho(\mathbf{y}) \delta V$ particles at $\mathbf{y}$ given to $n_x \propto \rho(\mathbf{x}) \delta V$ particles at $\mathbf{x}$, where $\delta V$ is an infinitesimal volume. Then, since the sum of $n$ Gaussian noises yields a Gaussian noise with standard deviation $\propto\sqrt{n} \propto \sqrt{\rho}$, we have $\boldsymbol{j}_n \propto \sqrt{\rho(\mathbf{x})} \sqrt{\rho(\mathbf{y})}$. Further, similar to Eq.~\ref{eq:generic_sde}, $\sqrt{ \mathbf{\Lambda} (\mathbf{x}, \mathbf{y})}$ acts as a projection tensor, making the noise anisotropic. Finally, integrating over $\mathbf{y}$ collects contributions from different locations to the total stochastic flux at $\mathbf{x}$.

We perform finite-difference simulations of Eq.~\ref{eq:generic_drho} for RO, BRO, and SGD, and find that the long-range structure is quantitatively the same as particle simulations (Methods, Fig.~\ref{fig:figure2}b). Thus, our coarse-grained theory quantitatively captures the long-range structure for all random-organizing systems.

\begin{figure*}[htpb!]
    \centering
    \includegraphics[width=\linewidth]{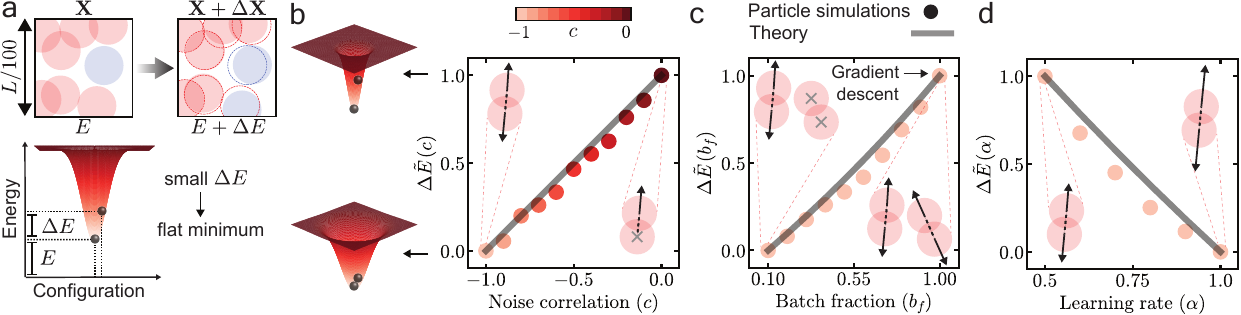}
    \caption{\small \textbf{Flatness of energy minimum in stochastic gradient descent (SGD).} (a) Zoomed-in exemplar configuration of discrete-time SGD before and after adding a small noise to all particles (Top). Schematic of a system (black ball) in an energy minimum. The system has energy $E$ at the minimum $\mathbf{X}$, and energy $E + \Delta E$ at a slightly perturbed position $\mathbf{X} + \Delta \mathbf{X}$ after the addition of a small noise $\mathbf{N}$. (b) Normalized energy change $\Delta \Tilde{E}(c)$ versus noise correlation $c$. $\Delta E(c)$ is normalized as: $\Delta \Tilde{E} (c) = [\Delta E (c) - \Delta E (c=-1)] / [\Delta E (c=0) - \Delta E (c=-1)]$. Black line shows prediction of Eq.~\ref{eq:delta_E_SI_c} (SI Sec. I.C). Schematic showing two minima with different flatness (left). (c) Normalized energy change $\Delta \Tilde{E}(b_f)$ versus batch fraction $b_f$. $\Delta E(b_f)$ is normalized as: $\Delta \Tilde{E} (b_f) = [\Delta E (b_f) - \Delta E (b_f=0.1)] / [\Delta E (b_f=1.0) - \Delta E (b_f=0.1)]$. Black line shows prediction of Eq.~\ref{eq:delta_E_SI_bf} (SI Sec. I.C). (d) Normalized energy change $\Delta \Tilde{E}(\alpha)$ versus learning rate $\alpha$. $\Delta E(\alpha)$ is normalized as: $\Delta E (\alpha) = [\Delta E (\alpha) - \Delta E (\alpha=1.0)] / [\Delta E (\alpha=0.5) - \Delta E (\alpha=1.0)]$. Black line shows prediction of Eq.~\ref{eq:delta_E_SI_alpha} (SI Sec. I.C). Data in (b), (c), and (d) denote discrete-time particle simulations.
}
    \label{fig:figure3}
\end{figure*}

To gain further analytical insights on Eq.~\ref{eq:generic_drho}, we linearize $\rho (\mathbf{x}, t)$ to first order and derive the analytical form of the static structure factor (SI Sec. I.B.2), 
\begin{equation}
    \Tilde{S}(\Tilde{k}) = (1+c) + [M(1+c) - c] \tilde{k}^2,
    \label{eq:struc_fact_sim}
\end{equation} 
where $M$ is a known system-dependent constant (Eqs.~\ref{eq:M_ro}, \ref{eq:M_bro}, and \ref{eq:M_sgd}, SI Sec. I.B.2). Eq.~\ref{eq:struc_fact_sim} directly relates $c$, the microscopic noise correlation coefficient between a pair of particles, to the long-range structure in the system, and quantitatively predicts $S(k)$ for all random-organizing systems without free parameters (Fig.~\ref{fig:figure2}b). Further, since $\tilde{\delta \rho}^2 (\tilde{l} \to \infty) \propto \tilde{S}(\tilde{k} \to 0)$ \cite{torquato2018hyperuniform}, Eq.~\ref{eq:struc_fact_sim} also predicts the behavior of infinite wavelength density fluctuations for all random-organizing systems (Fig.~\ref{fig:figure2}c inset).  

Eq.~\ref{eq:struc_fact_sim} shows a competition between two terms. The first term $(1+c)$ makes the long-range structure random, while the second term $[M(1+c) - c] \tilde{k}^2$ makes the long-range structure strongly hyperuniform. The ratio of these two competing terms gives the normalized crossover length scale,
\begin{equation}
    \tilde{l}_c = \sqrt{ M -
    \frac{c}{ 1 + c } },
    \label{eq:k_c}
\end{equation}
which quantitatively predicts the crossover length scale for all random-organizing systems (Fig.~\ref{fig:figure2}b inset). 

Two remarks are in order. First, in a variety of other systems having local center of mass conserving dynamics \cite{hexner2017noise, lei2019nonequilibrium, DeLuca2024} (equivalent to $c=-1$ in random-organizing systems) and displaying hyperuniformity in the dense phase, the coarse-grained noise term is of Laplacian form $\nabla^2 [\sqrt{\rho(\mathbf{x})} \omega (\mathbf{x})]$, where $\omega (\mathbf{x})$ is a spatially uncorrelated, one-point Gaussian noise field. In contrast, the noise term in our microscopically derived theory (Eqs.~\ref{eq:generic_drho}, \ref{eq:fluc_flux}) is fundamentally different from ``Laplacian noise'' in two ways: (i) it enters Eq.~\ref{eq:generic_drho} in the ``divergence'' form and, (ii) is spatially correlated. Second, the widely used Fokker-Planck coarse-graining approach gives a density evolution equation identical to Eq.~\ref{eq:generic_drho}, but without the stochastic flux term ($\mathbf{j}_n(\mathbf{x}) = \mathbf{0}$), which, after density linearization, trivially predicts spatially uniform steady-state density ($\rho (\mathbf{x}) = \text{constant}$ and $S(k) = 0$) independent of $c$ (SI Sec. I.B.1). This demonstrates the crucial role of noise and noise correlations \textit{even} at the coarse-grained scale in determining the long-range structure, thereby underscoring the importance of our theoretical framework. 

\subsection{Flatness of energy minima in stochastic gradient descent}

SGD is widely used for training neural networks, not only for its computational efficiency but also for its remarkable ability to steer neural networks toward flat regions of their loss landscapes \cite{bishop2023deep, keskar2016large, ballard2017energy}. This attribute is crucial for ``good'' learning algorithms since flatter minima are strongly correlated with better generalization performance on unseen data \cite{xie2020diffusion, jastrzkebski2017three, keskar2016large, huang2020understanding}. The bias towards flat minima, due to the selection noise in SGD, highlights the vital role of noise in shaping learning dynamics in neural networks. This, then, raises two questions: (i) Does SGD-driven descent of energy landscapes in particle systems similarly bias the dynamics toward flat minima, akin to neural networks? If so, (ii) Can this bias be linked to the long-range structure observed in these particle systems?

We first examine how the flatness of energy minima varies with noise correlation $c$, batch fraction $b_f$, and learning rate $\alpha$ in SGD (Eq.~\ref{eq:discrete_sgd}). We add a small noise $\mathbf{N}$ to the steady-state configurations $\mathbf{X}$ obtained in particle simulations to get a perturbed configuration $\mathbf{X} + \Delta \mathbf{X}$ and measure the change in the total energy of the system, $\Delta E = \langle  E(\mathbf{X}+\Delta \mathbf{X}) - E(\mathbf{X})\rangle_{\mathbf{N}} \propto \text{Tr}(\mathbf{H})$, where $E = \sum_i \sum_{j>i} V_{ij}$, and $\mathbf{H}$ is the Hessian matrix of the system (Eq.~\ref{eq:v_ji_family}, Fig.~\ref{fig:figure3}a) \cite{zhang2024absorbing} (Methods). $\Delta E$ is a measure of the flatness of the energy landscape: lower $\Delta E$ corresponds to a flatter minimum (Fig.~\ref{fig:figure3}a) \cite{jastrzkebski2017three, zhang2024absorbing}. We find that $\Delta E$ decreases as $c$ decreases, meaning that increase in long-range order leads to flatter minima (Figs.~\ref{fig:figure3}b, \ref{fig:figure2}b). We now fix $c=-1$ (anti-correlated noise), and find that $\Delta E$ increases with $b_f$ \cite{zhang2024absorbing} and decreases with $\alpha$, suggesting that lower batch fractions and higher learning rates lead to flatter minima---consistent with results on SGD dynamics in neural networks (Figs.~\ref{fig:figure3}c, d) \cite{jastrzkebski2017three, keskar2016large}. Notice that $c=-1$ and $b_f=1$ correspond to (noiseless) gradient descent (Fig.~\ref{fig:figure3}c). All results are qualitatively independent of particle volume fraction $\phi > \phi_c$, the potential $V_{ij}$, and the spatial dimension $d$, a useful feature since typical neural manifold dimensions are $\mathcal{O} (100)$ \cite{zhang2024absorbing} (SI Fig.~\ref{fig:figure5_si}).

Finally, we investigate the relationship between $\Delta E$ and the long-range structure formed by SGD. 
Combining the change in $S(k)$ before and after adding a small perturbation to the system \cite{gabrielli2004point} with our fluctuating hydrodynamic theory (Eq.~\ref{eq:struc_fact_sim}), we derive an expression for $\Delta E (c,b_f,\alpha)$ (SI Sec. I.C, Eqs.~\ref{eq:delta_E_SI_c}, \ref{eq:delta_E_SI_alpha}, and \ref{eq:delta_E_SI_bf})---in quantitative agreement with numerical simulations without free parameters (Figs.~\ref{fig:figure3}b,c,d). Thus, the emergent long-range structure in SGD provides a quantitative framework for understanding the flatness of energy minima, a key feature linked to generalization. This unites two seemingly disparate domains---neural networks and interacting-particle systems---by revealing that the bias of SGD towards flat minima is a universal hallmark of high-dimensional loss (or energy) landscapes, regardless of the underlying system. Beyond its relevance to machine learning algorithms, our framework may guide the design of self-organizing materials with tunable energetic and structural properties.

Random-organizing systems offer an ideal framework to probe noise-driven self-organization. Combining simulations and theory, we provide a unified, microscopic description of dynamics and emergent organization across a diverse array of random-organizing systems.
We reveal that universal long-range behavior arises \textit{away from criticality} in these microscopically distinct systems, dictated solely by interparticle noise correlations.
Finally, we demonstrate that SGD in particle systems inherently biases dynamics toward flatter energy minima, mirroring the behavior of SGD in neural networks---highlighting deep parallels between SGD dynamics in these two high-dimensional, but otherwise completely distinct, systems.

\bibliographystyle{apsrev4-2}
\bibliography{ref.bib}

\begin{thebibliography}{15}%
\makeatletter
\providecommand \@ifxundefined [1]{%
 \@ifx{#1\undefined}
}%
\providecommand \@ifnum [1]{%
 \ifnum #1\expandafter \@firstoftwo
 \else \expandafter \@secondoftwo
 \fi
}%
\providecommand \@ifx [1]{%
 \ifx #1\expandafter \@firstoftwo
 \else \expandafter \@secondoftwo
 \fi
}%
\providecommand \natexlab [1]{#1}%
\providecommand \enquote  [1]{``#1''}%
\providecommand \bibnamefont  [1]{#1}%
\providecommand \bibfnamefont [1]{#1}%
\providecommand \citenamefont [1]{#1}%
\providecommand \href@noop [0]{\@secondoftwo}%
\providecommand \href [0]{\begingroup \@sanitize@url \@href}%
\providecommand \@href[1]{\@@startlink{#1}\@@href}%
\providecommand \@@href[1]{\endgroup#1\@@endlink}%
\providecommand \@sanitize@url [0]{\catcode `\\12\catcode `\$12\catcode `\&12\catcode `\#12\catcode `\^12\catcode `\_12\catcode `\%12\relax}%
\providecommand \@@startlink[1]{}%
\providecommand \@@endlink[0]{}%
\providecommand \url  [0]{\begingroup\@sanitize@url \@url }%
\providecommand \@url [1]{\endgroup\@href {#1}{\urlprefix }}%
\providecommand \urlprefix  [0]{URL }%
\providecommand \Eprint [0]{\href }%
\providecommand \doibase [0]{https://doi.org/}%
\providecommand \selectlanguage [0]{\@gobble}%
\providecommand \bibinfo  [0]{\@secondoftwo}%
\providecommand \bibfield  [0]{\@secondoftwo}%
\providecommand \translation [1]{[#1]}%
\providecommand \BibitemOpen [0]{}%
\providecommand \bibitemStop [0]{}%
\providecommand \bibitemNoStop [0]{.\EOS\space}%
\providecommand \EOS [0]{\spacefactor3000\relax}%
\providecommand \BibitemShut  [1]{\csname bibitem#1\endcsname}%
\let\auto@bib@innerbib\@empty
\bibitem [{\citenamefont {Li}\ \emph {et~al.}(2017)\citenamefont {Li}, \citenamefont {Tai},\ and\ \citenamefont {Weinan}}]{li2017stochastic}%
  \BibitemOpen
  \bibfield  {author} {\bibinfo {author} {\bibfnamefont {Q.}~\bibnamefont {Li}}, \bibinfo {author} {\bibfnamefont {C.}~\bibnamefont {Tai}},\ and\ \bibinfo {author} {\bibfnamefont {E.}~\bibnamefont {Weinan}},\ }in\ \href@noop {} {\emph {\bibinfo {booktitle} {International Conference on Machine Learning}}}\ (\bibinfo {organization} {PMLR},\ \bibinfo {year} {2017})\ pp.\ \bibinfo {pages} {2101--2110}\BibitemShut {NoStop}%
\bibitem [{\citenamefont {Zhang}\ and\ \citenamefont {Martiniani}(2024)}]{zhang2024absorbing}%
  \BibitemOpen
  \bibfield  {author} {\bibinfo {author} {\bibfnamefont {G.}~\bibnamefont {Zhang}}\ and\ \bibinfo {author} {\bibfnamefont {S.}~\bibnamefont {Martiniani}},\ }\href@noop {} {\bibfield  {journal} {\bibinfo  {journal} {arXiv preprint arXiv:2411.11834}\ } (\bibinfo {year} {2024})}\BibitemShut {NoStop}%
\bibitem [{\citenamefont {Hansen}\ and\ \citenamefont {McDonald}(2013)}]{hansen2013theory}%
  \BibitemOpen
  \bibfield  {author} {\bibinfo {author} {\bibfnamefont {J.-P.}\ \bibnamefont {Hansen}}\ and\ \bibinfo {author} {\bibfnamefont {I.~R.}\ \bibnamefont {McDonald}},\ }\href@noop {} {\emph {\bibinfo {title} {Theory of simple liquids: with applications to soft matter}}}\ (\bibinfo  {publisher} {Academic press},\ \bibinfo {year} {2013})\BibitemShut {NoStop}%
\bibitem [{\citenamefont {Dean}(1996)}]{dean1996langevin}%
  \BibitemOpen
  \bibfield  {author} {\bibinfo {author} {\bibfnamefont {D.~S.}\ \bibnamefont {Dean}},\ }\href@noop {} {\bibfield  {journal} {\bibinfo  {journal} {Journal of Physics A: Mathematical and General}\ }\textbf {\bibinfo {volume} {29}},\ \bibinfo {pages} {L613} (\bibinfo {year} {1996})}\BibitemShut {NoStop}%
\bibitem [{\citenamefont {Bertin}\ \emph {et~al.}(2013)\citenamefont {Bertin}, \citenamefont {Chat{\'e}}, \citenamefont {Ginelli}, \citenamefont {Mishra}, \citenamefont {Peshkov},\ and\ \citenamefont {Ramaswamy}}]{bertin2013mesoscopic}%
  \BibitemOpen
  \bibfield  {author} {\bibinfo {author} {\bibfnamefont {E.}~\bibnamefont {Bertin}}, \bibinfo {author} {\bibfnamefont {H.}~\bibnamefont {Chat{\'e}}}, \bibinfo {author} {\bibfnamefont {F.}~\bibnamefont {Ginelli}}, \bibinfo {author} {\bibfnamefont {S.}~\bibnamefont {Mishra}}, \bibinfo {author} {\bibfnamefont {A.}~\bibnamefont {Peshkov}},\ and\ \bibinfo {author} {\bibfnamefont {S.}~\bibnamefont {Ramaswamy}},\ }\href@noop {} {\bibfield  {journal} {\bibinfo  {journal} {New journal of physics}\ }\textbf {\bibinfo {volume} {15}},\ \bibinfo {pages} {085032} (\bibinfo {year} {2013})}\BibitemShut {NoStop}%
\bibitem [{\citenamefont {Solon}\ \emph {et~al.}(2015)\citenamefont {Solon}, \citenamefont {Cates},\ and\ \citenamefont {Tailleur}}]{solon2015active}%
  \BibitemOpen
  \bibfield  {author} {\bibinfo {author} {\bibfnamefont {A.~P.}\ \bibnamefont {Solon}}, \bibinfo {author} {\bibfnamefont {M.~E.}\ \bibnamefont {Cates}},\ and\ \bibinfo {author} {\bibfnamefont {J.}~\bibnamefont {Tailleur}},\ }\href@noop {} {\bibfield  {journal} {\bibinfo  {journal} {The European Physical Journal Special Topics}\ }\textbf {\bibinfo {volume} {224}},\ \bibinfo {pages} {1231} (\bibinfo {year} {2015})}\BibitemShut {NoStop}%
\bibitem [{\citenamefont {Dean}\ \emph {et~al.}(2016)\citenamefont {Dean}, \citenamefont {Lu}, \citenamefont {Maggs},\ and\ \citenamefont {Podgornik}}]{dean2016nonequilibrium}%
  \BibitemOpen
  \bibfield  {author} {\bibinfo {author} {\bibfnamefont {D.~S.}\ \bibnamefont {Dean}}, \bibinfo {author} {\bibfnamefont {B.-S.}\ \bibnamefont {Lu}}, \bibinfo {author} {\bibfnamefont {A.}~\bibnamefont {Maggs}},\ and\ \bibinfo {author} {\bibfnamefont {R.}~\bibnamefont {Podgornik}},\ }\href@noop {} {\bibfield  {journal} {\bibinfo  {journal} {Physical Review Letters}\ }\textbf {\bibinfo {volume} {116}},\ \bibinfo {pages} {240602} (\bibinfo {year} {2016})}\BibitemShut {NoStop}%
\bibitem [{\citenamefont {Dean}\ and\ \citenamefont {Podgornik}(2014)}]{dean2014relaxation}%
  \BibitemOpen
  \bibfield  {author} {\bibinfo {author} {\bibfnamefont {D.~S.}\ \bibnamefont {Dean}}\ and\ \bibinfo {author} {\bibfnamefont {R.}~\bibnamefont {Podgornik}},\ }\href@noop {} {\bibfield  {journal} {\bibinfo  {journal} {Physical Review E}\ }\textbf {\bibinfo {volume} {89}},\ \bibinfo {pages} {032117} (\bibinfo {year} {2014})}\BibitemShut {NoStop}%
\bibitem [{\citenamefont {D{\'e}mery}\ \emph {et~al.}(2014)\citenamefont {D{\'e}mery}, \citenamefont {B{\'e}nichou},\ and\ \citenamefont {Jacquin}}]{demery2014generalized}%
  \BibitemOpen
  \bibfield  {author} {\bibinfo {author} {\bibfnamefont {V.}~\bibnamefont {D{\'e}mery}}, \bibinfo {author} {\bibfnamefont {O.}~\bibnamefont {B{\'e}nichou}},\ and\ \bibinfo {author} {\bibfnamefont {H.}~\bibnamefont {Jacquin}},\ }\href@noop {} {\bibfield  {journal} {\bibinfo  {journal} {New Journal of Physics}\ }\textbf {\bibinfo {volume} {16}},\ \bibinfo {pages} {053032} (\bibinfo {year} {2014})}\BibitemShut {NoStop}%
\bibitem [{\citenamefont {Dinelli}\ \emph {et~al.}(2024)\citenamefont {Dinelli}, \citenamefont {O’Byrne},\ and\ \citenamefont {Tailleur}}]{dinelli2024fluctuating}%
  \BibitemOpen
  \bibfield  {author} {\bibinfo {author} {\bibfnamefont {A.}~\bibnamefont {Dinelli}}, \bibinfo {author} {\bibfnamefont {J.}~\bibnamefont {O’Byrne}},\ and\ \bibinfo {author} {\bibfnamefont {J.}~\bibnamefont {Tailleur}},\ }\href@noop {} {\bibfield  {journal} {\bibinfo  {journal} {Journal of Physics A: Mathematical and Theoretical}\ }\textbf {\bibinfo {volume} {57}},\ \bibinfo {pages} {395002} (\bibinfo {year} {2024})}\BibitemShut {NoStop}%
\bibitem [{\citenamefont {Jastrz{\k{e}}bski}\ \emph {et~al.}(2017)\citenamefont {Jastrz{\k{e}}bski}, \citenamefont {Kenton}, \citenamefont {Arpit}, \citenamefont {Ballas}, \citenamefont {Fischer}, \citenamefont {Bengio},\ and\ \citenamefont {Storkey}}]{jastrzkebski2017three}%
  \BibitemOpen
  \bibfield  {author} {\bibinfo {author} {\bibfnamefont {S.}~\bibnamefont {Jastrz{\k{e}}bski}}, \bibinfo {author} {\bibfnamefont {Z.}~\bibnamefont {Kenton}}, \bibinfo {author} {\bibfnamefont {D.}~\bibnamefont {Arpit}}, \bibinfo {author} {\bibfnamefont {N.}~\bibnamefont {Ballas}}, \bibinfo {author} {\bibfnamefont {A.}~\bibnamefont {Fischer}}, \bibinfo {author} {\bibfnamefont {Y.}~\bibnamefont {Bengio}},\ and\ \bibinfo {author} {\bibfnamefont {A.}~\bibnamefont {Storkey}},\ }\href@noop {} {\bibfield  {journal} {\bibinfo  {journal} {arXiv preprint arXiv:1711.04623}\ } (\bibinfo {year} {2017})}\BibitemShut {NoStop}%
\bibitem [{\citenamefont {Torquato}\ \emph {et~al.}(2015)\citenamefont {Torquato}, \citenamefont {Zhang},\ and\ \citenamefont {Stillinger}}]{torquato2015ensemble}%
  \BibitemOpen
  \bibfield  {author} {\bibinfo {author} {\bibfnamefont {S.}~\bibnamefont {Torquato}}, \bibinfo {author} {\bibfnamefont {G.}~\bibnamefont {Zhang}},\ and\ \bibinfo {author} {\bibfnamefont {F.~H.}\ \bibnamefont {Stillinger}},\ }\href@noop {} {\bibfield  {journal} {\bibinfo  {journal} {Physical Review X}\ }\textbf {\bibinfo {volume} {5}},\ \bibinfo {pages} {021020} (\bibinfo {year} {2015})}\BibitemShut {NoStop}%
\bibitem [{\citenamefont {Gabrielli}(2004)}]{gabrielli2004point}%
  \BibitemOpen
  \bibfield  {author} {\bibinfo {author} {\bibfnamefont {A.}~\bibnamefont {Gabrielli}},\ }\href@noop {} {\bibfield  {journal} {\bibinfo  {journal} {Physical Review E—Statistical, Nonlinear, and Soft Matter Physics}\ }\textbf {\bibinfo {volume} {70}},\ \bibinfo {pages} {066131} (\bibinfo {year} {2004})}\BibitemShut {NoStop}%
\bibitem [{\citenamefont {Kim}\ and\ \citenamefont {Torquato}(2018)}]{kim2018effect}%
  \BibitemOpen
  \bibfield  {author} {\bibinfo {author} {\bibfnamefont {J.}~\bibnamefont {Kim}}\ and\ \bibinfo {author} {\bibfnamefont {S.}~\bibnamefont {Torquato}},\ }\href@noop {} {\bibfield  {journal} {\bibinfo  {journal} {Physical Review B}\ }\textbf {\bibinfo {volume} {97}},\ \bibinfo {pages} {054105} (\bibinfo {year} {2018})}\BibitemShut {NoStop}%
\bibitem [{\citenamefont {Casiulis}\ \emph {et~al.}(2024)\citenamefont {Casiulis}, \citenamefont {Shih},\ and\ \citenamefont {Martiniani}}]{casiulis2024gyromorphs}%
  \BibitemOpen
  \bibfield  {author} {\bibinfo {author} {\bibfnamefont {M.}~\bibnamefont {Casiulis}}, \bibinfo {author} {\bibfnamefont {A.}~\bibnamefont {Shih}},\ and\ \bibinfo {author} {\bibfnamefont {S.}~\bibnamefont {Martiniani}},\ }\href@noop {} {\bibfield  {journal} {\bibinfo  {journal} {arXiv preprint arXiv:2410.09023}\ } (\bibinfo {year} {2024})}\BibitemShut {NoStop}%
\end{thebibliography}%


\begin{thebibliography}{49}%
\makeatletter
\providecommand \@ifxundefined [1]{%
 \@ifx{#1\undefined}
}%
\providecommand \@ifnum [1]{%
 \ifnum #1\expandafter \@firstoftwo
 \else \expandafter \@secondoftwo
 \fi
}%
\providecommand \@ifx [1]{%
 \ifx #1\expandafter \@firstoftwo
 \else \expandafter \@secondoftwo
 \fi
}%
\providecommand \natexlab [1]{#1}%
\providecommand \enquote  [1]{``#1''}%
\providecommand \bibnamefont  [1]{#1}%
\providecommand \bibfnamefont [1]{#1}%
\providecommand \citenamefont [1]{#1}%
\providecommand \href@noop [0]{\@secondoftwo}%
\providecommand \href [0]{\begingroup \@sanitize@url \@href}%
\providecommand \@href[1]{\@@startlink{#1}\@@href}%
\providecommand \@@href[1]{\endgroup#1\@@endlink}%
\providecommand \@sanitize@url [0]{\catcode `\\12\catcode `\$12\catcode `\&12\catcode `\#12\catcode `\^12\catcode `\_12\catcode `\%12\relax}%
\providecommand \@@startlink[1]{}%
\providecommand \@@endlink[0]{}%
\providecommand \url  [0]{\begingroup\@sanitize@url \@url }%
\providecommand \@url [1]{\endgroup\@href {#1}{\urlprefix }}%
\providecommand \urlprefix  [0]{URL }%
\providecommand \Eprint [0]{\href }%
\providecommand \doibase [0]{https://doi.org/}%
\providecommand \selectlanguage [0]{\@gobble}%
\providecommand \bibinfo  [0]{\@secondoftwo}%
\providecommand \bibfield  [0]{\@secondoftwo}%
\providecommand \translation [1]{[#1]}%
\providecommand \BibitemOpen [0]{}%
\providecommand \bibitemStop [0]{}%
\providecommand \bibitemNoStop [0]{.\EOS\space}%
\providecommand \EOS [0]{\spacefactor3000\relax}%
\providecommand \BibitemShut  [1]{\csname bibitem#1\endcsname}%
\let\auto@bib@innerbib\@empty
\bibitem [{\citenamefont {Sagu{\'e}s}\ \emph {et~al.}(2007)\citenamefont {Sagu{\'e}s}, \citenamefont {Sancho},\ and\ \citenamefont {Garc{\'\i}a-Ojalvo}}]{sagues2007spatiotemporal}%
  \BibitemOpen
  \bibfield  {author} {\bibinfo {author} {\bibfnamefont {F.}~\bibnamefont {Sagu{\'e}s}}, \bibinfo {author} {\bibfnamefont {J.~M.}\ \bibnamefont {Sancho}},\ and\ \bibinfo {author} {\bibfnamefont {J.}~\bibnamefont {Garc{\'\i}a-Ojalvo}},\ }\href@noop {} {\bibfield  {journal} {\bibinfo  {journal} {Reviews of Modern Physics}\ }\textbf {\bibinfo {volume} {79}},\ \bibinfo {pages} {829} (\bibinfo {year} {2007})}\BibitemShut {NoStop}%
\bibitem [{\citenamefont {Corte}\ \emph {et~al.}(2008)\citenamefont {Corte}, \citenamefont {Chaikin}, \citenamefont {Gollub},\ and\ \citenamefont {Pine}}]{corte2008random}%
  \BibitemOpen
  \bibfield  {author} {\bibinfo {author} {\bibfnamefont {L.}~\bibnamefont {Corte}}, \bibinfo {author} {\bibfnamefont {P.~M.}\ \bibnamefont {Chaikin}}, \bibinfo {author} {\bibfnamefont {J.~P.}\ \bibnamefont {Gollub}},\ and\ \bibinfo {author} {\bibfnamefont {D.~J.}\ \bibnamefont {Pine}},\ }\href@noop {} {\bibfield  {journal} {\bibinfo  {journal} {Nature Physics}\ }\textbf {\bibinfo {volume} {4}},\ \bibinfo {pages} {420} (\bibinfo {year} {2008})}\BibitemShut {NoStop}%
\bibitem [{\citenamefont {Hexner}\ \emph {et~al.}(2017)\citenamefont {Hexner}, \citenamefont {Chaikin},\ and\ \citenamefont {Levine}}]{hexner2017enhanced}%
  \BibitemOpen
  \bibfield  {author} {\bibinfo {author} {\bibfnamefont {D.}~\bibnamefont {Hexner}}, \bibinfo {author} {\bibfnamefont {P.~M.}\ \bibnamefont {Chaikin}},\ and\ \bibinfo {author} {\bibfnamefont {D.}~\bibnamefont {Levine}},\ }\href@noop {} {\bibfield  {journal} {\bibinfo  {journal} {Proceedings of the National Academy of Sciences}\ }\textbf {\bibinfo {volume} {114}},\ \bibinfo {pages} {4294} (\bibinfo {year} {2017})}\BibitemShut {NoStop}%
\bibitem [{\citenamefont {Matsumoto}\ and\ \citenamefont {Tsuda}(1983)}]{matsumoto1983noise}%
  \BibitemOpen
  \bibfield  {author} {\bibinfo {author} {\bibfnamefont {K.}~\bibnamefont {Matsumoto}}\ and\ \bibinfo {author} {\bibfnamefont {I.}~\bibnamefont {Tsuda}},\ }\href@noop {} {\bibfield  {journal} {\bibinfo  {journal} {Journal of Statistical Physics}\ }\textbf {\bibinfo {volume} {31}},\ \bibinfo {pages} {87} (\bibinfo {year} {1983})}\BibitemShut {NoStop}%
\bibitem [{\citenamefont {Villain}\ \emph {et~al.}(1980)\citenamefont {Villain}, \citenamefont {Bidaux}, \citenamefont {Carton},\ and\ \citenamefont {Conte}}]{villain1980order}%
  \BibitemOpen
  \bibfield  {author} {\bibinfo {author} {\bibfnamefont {J.}~\bibnamefont {Villain}}, \bibinfo {author} {\bibfnamefont {R.}~\bibnamefont {Bidaux}}, \bibinfo {author} {\bibfnamefont {J.-P.}\ \bibnamefont {Carton}},\ and\ \bibinfo {author} {\bibfnamefont {R.}~\bibnamefont {Conte}},\ }\href@noop {} {\bibfield  {journal} {\bibinfo  {journal} {Journal de Physique}\ }\textbf {\bibinfo {volume} {41}},\ \bibinfo {pages} {1263} (\bibinfo {year} {1980})}\BibitemShut {NoStop}%
\bibitem [{\citenamefont {Jhawar}\ \emph {et~al.}(2020)\citenamefont {Jhawar}, \citenamefont {Morris}, \citenamefont {Amith-Kumar}, \citenamefont {Danny~Raj}, \citenamefont {Rogers}, \citenamefont {Rajendran},\ and\ \citenamefont {Guttal}}]{jhawar2020noise}%
  \BibitemOpen
  \bibfield  {author} {\bibinfo {author} {\bibfnamefont {J.}~\bibnamefont {Jhawar}}, \bibinfo {author} {\bibfnamefont {R.~G.}\ \bibnamefont {Morris}}, \bibinfo {author} {\bibfnamefont {U.}~\bibnamefont {Amith-Kumar}}, \bibinfo {author} {\bibfnamefont {M.}~\bibnamefont {Danny~Raj}}, \bibinfo {author} {\bibfnamefont {T.}~\bibnamefont {Rogers}}, \bibinfo {author} {\bibfnamefont {H.}~\bibnamefont {Rajendran}},\ and\ \bibinfo {author} {\bibfnamefont {V.}~\bibnamefont {Guttal}},\ }\href@noop {} {\bibfield  {journal} {\bibinfo  {journal} {Nature Physics}\ }\textbf {\bibinfo {volume} {16}},\ \bibinfo {pages} {488} (\bibinfo {year} {2020})}\BibitemShut {NoStop}%
\bibitem [{\citenamefont {Torquato}(2018)}]{torquato2018hyperuniform}%
  \BibitemOpen
  \bibfield  {author} {\bibinfo {author} {\bibfnamefont {S.}~\bibnamefont {Torquato}},\ }\href@noop {} {\bibfield  {journal} {\bibinfo  {journal} {Physics Reports}\ }\textbf {\bibinfo {volume} {745}},\ \bibinfo {pages} {1} (\bibinfo {year} {2018})}\BibitemShut {NoStop}%
\bibitem [{\citenamefont {Donev}\ \emph {et~al.}(2005)\citenamefont {Donev}, \citenamefont {Stillinger},\ and\ \citenamefont {Torquato}}]{donev2005unexpected}%
  \BibitemOpen
  \bibfield  {author} {\bibinfo {author} {\bibfnamefont {A.}~\bibnamefont {Donev}}, \bibinfo {author} {\bibfnamefont {F.~H.}\ \bibnamefont {Stillinger}},\ and\ \bibinfo {author} {\bibfnamefont {S.}~\bibnamefont {Torquato}},\ }\href@noop {} {\bibfield  {journal} {\bibinfo  {journal} {Physical review letters}\ }\textbf {\bibinfo {volume} {95}},\ \bibinfo {pages} {090604} (\bibinfo {year} {2005})}\BibitemShut {NoStop}%
\bibitem [{\citenamefont {Hexner}\ and\ \citenamefont {Levine}(2015)}]{hexner2015hyperuniformity}%
  \BibitemOpen
  \bibfield  {author} {\bibinfo {author} {\bibfnamefont {D.}~\bibnamefont {Hexner}}\ and\ \bibinfo {author} {\bibfnamefont {D.}~\bibnamefont {Levine}},\ }\href@noop {} {\bibfield  {journal} {\bibinfo  {journal} {Physical review letters}\ }\textbf {\bibinfo {volume} {114}},\ \bibinfo {pages} {110602} (\bibinfo {year} {2015})}\BibitemShut {NoStop}%
\bibitem [{\citenamefont {Wilken}\ \emph {et~al.}(2021)\citenamefont {Wilken}, \citenamefont {Guerra}, \citenamefont {Levine},\ and\ \citenamefont {Chaikin}}]{wilken2021random}%
  \BibitemOpen
  \bibfield  {author} {\bibinfo {author} {\bibfnamefont {S.}~\bibnamefont {Wilken}}, \bibinfo {author} {\bibfnamefont {R.~E.}\ \bibnamefont {Guerra}}, \bibinfo {author} {\bibfnamefont {D.}~\bibnamefont {Levine}},\ and\ \bibinfo {author} {\bibfnamefont {P.~M.}\ \bibnamefont {Chaikin}},\ }\href@noop {} {\bibfield  {journal} {\bibinfo  {journal} {Physical review letters}\ }\textbf {\bibinfo {volume} {127}},\ \bibinfo {pages} {038002} (\bibinfo {year} {2021})}\BibitemShut {NoStop}%
\bibitem [{\citenamefont {Lei}\ and\ \citenamefont {Ni}(2019)}]{lei2019hydrodynamics}%
  \BibitemOpen
  \bibfield  {author} {\bibinfo {author} {\bibfnamefont {Q.-L.}\ \bibnamefont {Lei}}\ and\ \bibinfo {author} {\bibfnamefont {R.}~\bibnamefont {Ni}},\ }\href@noop {} {\bibfield  {journal} {\bibinfo  {journal} {Proceedings of the National Academy of Sciences}\ }\textbf {\bibinfo {volume} {116}},\ \bibinfo {pages} {22983} (\bibinfo {year} {2019})}\BibitemShut {NoStop}%
\bibitem [{\citenamefont {Hexner}\ and\ \citenamefont {Levine}(2017)}]{hexner2017noise}%
  \BibitemOpen
  \bibfield  {author} {\bibinfo {author} {\bibfnamefont {D.}~\bibnamefont {Hexner}}\ and\ \bibinfo {author} {\bibfnamefont {D.}~\bibnamefont {Levine}},\ }\href@noop {} {\bibfield  {journal} {\bibinfo  {journal} {Physical review letters}\ }\textbf {\bibinfo {volume} {118}},\ \bibinfo {pages} {020601} (\bibinfo {year} {2017})}\BibitemShut {NoStop}%
\bibitem [{\citenamefont {Huang}\ \emph {et~al.}(2021)\citenamefont {Huang}, \citenamefont {Hu}, \citenamefont {Yang}, \citenamefont {Liu},\ and\ \citenamefont {Zhang}}]{huang2021circular}%
  \BibitemOpen
  \bibfield  {author} {\bibinfo {author} {\bibfnamefont {M.}~\bibnamefont {Huang}}, \bibinfo {author} {\bibfnamefont {W.}~\bibnamefont {Hu}}, \bibinfo {author} {\bibfnamefont {S.}~\bibnamefont {Yang}}, \bibinfo {author} {\bibfnamefont {Q.-X.}\ \bibnamefont {Liu}},\ and\ \bibinfo {author} {\bibfnamefont {H.}~\bibnamefont {Zhang}},\ }\href@noop {} {\bibfield  {journal} {\bibinfo  {journal} {Proceedings of the National Academy of Sciences}\ }\textbf {\bibinfo {volume} {118}},\ \bibinfo {pages} {e2100493118} (\bibinfo {year} {2021})}\BibitemShut {NoStop}%
\bibitem [{\citenamefont {Galliano}\ \emph {et~al.}(2023)\citenamefont {Galliano}, \citenamefont {Cates},\ and\ \citenamefont {Berthier}}]{galliano2023two}%
  \BibitemOpen
  \bibfield  {author} {\bibinfo {author} {\bibfnamefont {L.}~\bibnamefont {Galliano}}, \bibinfo {author} {\bibfnamefont {M.~E.}\ \bibnamefont {Cates}},\ and\ \bibinfo {author} {\bibfnamefont {L.}~\bibnamefont {Berthier}},\ }\href@noop {} {\bibfield  {journal} {\bibinfo  {journal} {Physical Review Letters}\ }\textbf {\bibinfo {volume} {131}},\ \bibinfo {pages} {047101} (\bibinfo {year} {2023})}\BibitemShut {NoStop}%
\bibitem [{\citenamefont {Tjhung}\ and\ \citenamefont {Berthier}(2015)}]{tjhung2015hyperuniform}%
  \BibitemOpen
  \bibfield  {author} {\bibinfo {author} {\bibfnamefont {E.}~\bibnamefont {Tjhung}}\ and\ \bibinfo {author} {\bibfnamefont {L.}~\bibnamefont {Berthier}},\ }\href@noop {} {\bibfield  {journal} {\bibinfo  {journal} {Physical review letters}\ }\textbf {\bibinfo {volume} {114}},\ \bibinfo {pages} {148301} (\bibinfo {year} {2015})}\BibitemShut {NoStop}%
\bibitem [{\citenamefont {Tjhung}\ and\ \citenamefont {Berthier}(2016)}]{tjhung2016criticality}%
  \BibitemOpen
  \bibfield  {author} {\bibinfo {author} {\bibfnamefont {E.}~\bibnamefont {Tjhung}}\ and\ \bibinfo {author} {\bibfnamefont {L.}~\bibnamefont {Berthier}},\ }\href@noop {} {\bibfield  {journal} {\bibinfo  {journal} {Journal of Statistical Mechanics: Theory and Experiment}\ }\textbf {\bibinfo {volume} {2016}},\ \bibinfo {pages} {033501} (\bibinfo {year} {2016})}\BibitemShut {NoStop}%
\bibitem [{\citenamefont {Milz}\ and\ \citenamefont {Schmiedeberg}(2013)}]{milz2013connecting}%
  \BibitemOpen
  \bibfield  {author} {\bibinfo {author} {\bibfnamefont {L.}~\bibnamefont {Milz}}\ and\ \bibinfo {author} {\bibfnamefont {M.}~\bibnamefont {Schmiedeberg}},\ }\href@noop {} {\bibfield  {journal} {\bibinfo  {journal} {Physical Review E}\ }\textbf {\bibinfo {volume} {88}},\ \bibinfo {pages} {062308} (\bibinfo {year} {2013})}\BibitemShut {NoStop}%
\bibitem [{\citenamefont {Menon}\ and\ \citenamefont {Ramaswamy}(2009)}]{menon2009universality}%
  \BibitemOpen
  \bibfield  {author} {\bibinfo {author} {\bibfnamefont {G.~I.}\ \bibnamefont {Menon}}\ and\ \bibinfo {author} {\bibfnamefont {S.}~\bibnamefont {Ramaswamy}},\ }\href@noop {} {\bibfield  {journal} {\bibinfo  {journal} {Physical Review E—Statistical, Nonlinear, and Soft Matter Physics}\ }\textbf {\bibinfo {volume} {79}},\ \bibinfo {pages} {061108} (\bibinfo {year} {2009})}\BibitemShut {NoStop}%
\bibitem [{\citenamefont {Wilken}\ \emph {et~al.}(2023)\citenamefont {Wilken}, \citenamefont {Guo}, \citenamefont {Levine},\ and\ \citenamefont {Chaikin}}]{wilken2023dynamical}%
  \BibitemOpen
  \bibfield  {author} {\bibinfo {author} {\bibfnamefont {S.}~\bibnamefont {Wilken}}, \bibinfo {author} {\bibfnamefont {A.~Z.}\ \bibnamefont {Guo}}, \bibinfo {author} {\bibfnamefont {D.}~\bibnamefont {Levine}},\ and\ \bibinfo {author} {\bibfnamefont {P.~M.}\ \bibnamefont {Chaikin}},\ }\href@noop {} {\bibfield  {journal} {\bibinfo  {journal} {Physical review letters}\ }\textbf {\bibinfo {volume} {131}},\ \bibinfo {pages} {238202} (\bibinfo {year} {2023})}\BibitemShut {NoStop}%
\bibitem [{\citenamefont {Lei}\ \emph {et~al.}(2023)\citenamefont {Lei}, \citenamefont {Zheng},\ and\ \citenamefont {Ni}}]{lei2023random}%
  \BibitemOpen
  \bibfield  {author} {\bibinfo {author} {\bibfnamefont {Y.}~\bibnamefont {Lei}}, \bibinfo {author} {\bibfnamefont {N.}~\bibnamefont {Zheng}},\ and\ \bibinfo {author} {\bibfnamefont {R.}~\bibnamefont {Ni}},\ }\href@noop {} {\bibfield  {journal} {\bibinfo  {journal} {The Journal of Chemical Physics}\ }\textbf {\bibinfo {volume} {159}} (\bibinfo {year} {2023})}\BibitemShut {NoStop}%
\bibitem [{\citenamefont {Zhang}\ and\ \citenamefont {Martiniani}(2024)}]{zhang2024absorbing}%
  \BibitemOpen
  \bibfield  {author} {\bibinfo {author} {\bibfnamefont {G.}~\bibnamefont {Zhang}}\ and\ \bibinfo {author} {\bibfnamefont {S.}~\bibnamefont {Martiniani}},\ }\href@noop {} {\bibfield  {journal} {\bibinfo  {journal} {arXiv preprint arXiv:2411.11834}\ } (\bibinfo {year} {2024})}\BibitemShut {NoStop}%
\bibitem [{\citenamefont {Pine}\ \emph {et~al.}(2005)\citenamefont {Pine}, \citenamefont {Gollub}, \citenamefont {Brady},\ and\ \citenamefont {Leshansky}}]{pine2005chaos}%
  \BibitemOpen
  \bibfield  {author} {\bibinfo {author} {\bibfnamefont {D.~J.}\ \bibnamefont {Pine}}, \bibinfo {author} {\bibfnamefont {J.~P.}\ \bibnamefont {Gollub}}, \bibinfo {author} {\bibfnamefont {J.~F.}\ \bibnamefont {Brady}},\ and\ \bibinfo {author} {\bibfnamefont {A.~M.}\ \bibnamefont {Leshansky}},\ }\href@noop {} {\bibfield  {journal} {\bibinfo  {journal} {Nature}\ }\textbf {\bibinfo {volume} {438}},\ \bibinfo {pages} {997} (\bibinfo {year} {2005})}\BibitemShut {NoStop}%
\bibitem [{\citenamefont {Wilken}\ \emph {et~al.}(2020)\citenamefont {Wilken}, \citenamefont {Guerra}, \citenamefont {Pine},\ and\ \citenamefont {Chaikin}}]{wilken2020hyperuniform}%
  \BibitemOpen
  \bibfield  {author} {\bibinfo {author} {\bibfnamefont {S.}~\bibnamefont {Wilken}}, \bibinfo {author} {\bibfnamefont {R.~E.}\ \bibnamefont {Guerra}}, \bibinfo {author} {\bibfnamefont {D.~J.}\ \bibnamefont {Pine}},\ and\ \bibinfo {author} {\bibfnamefont {P.~M.}\ \bibnamefont {Chaikin}},\ }\href@noop {} {\bibfield  {journal} {\bibinfo  {journal} {Physical Review Letters}\ }\textbf {\bibinfo {volume} {125}},\ \bibinfo {pages} {148001} (\bibinfo {year} {2020})}\BibitemShut {NoStop}%
\bibitem [{\citenamefont {Hinrichsen}(2000)}]{hinrichsen2000non}%
  \BibitemOpen
  \bibfield  {author} {\bibinfo {author} {\bibfnamefont {H.}~\bibnamefont {Hinrichsen}},\ }\href@noop {} {\bibfield  {journal} {\bibinfo  {journal} {Advances in physics}\ }\textbf {\bibinfo {volume} {49}},\ \bibinfo {pages} {815} (\bibinfo {year} {2000})}\BibitemShut {NoStop}%
\bibitem [{\citenamefont {Henkel}(2008)}]{henkel2008non}%
  \BibitemOpen
  \bibfield  {author} {\bibinfo {author} {\bibfnamefont {M.}~\bibnamefont {Henkel}},\ }\href@noop {} {\emph {\bibinfo {title} {Non-equilibrium phase transitions}}}\ (\bibinfo  {publisher} {Springer},\ \bibinfo {year} {2008})\BibitemShut {NoStop}%
\bibitem [{\citenamefont {Martiniani}\ \emph {et~al.}(2019)\citenamefont {Martiniani}, \citenamefont {Chaikin},\ and\ \citenamefont {Levine}}]{martiniani2019quantifying}%
  \BibitemOpen
  \bibfield  {author} {\bibinfo {author} {\bibfnamefont {S.}~\bibnamefont {Martiniani}}, \bibinfo {author} {\bibfnamefont {P.~M.}\ \bibnamefont {Chaikin}},\ and\ \bibinfo {author} {\bibfnamefont {D.}~\bibnamefont {Levine}},\ }\href@noop {} {\bibfield  {journal} {\bibinfo  {journal} {Physical Review X}\ }\textbf {\bibinfo {volume} {9}},\ \bibinfo {pages} {011031} (\bibinfo {year} {2019})}\BibitemShut {NoStop}%
\bibitem [{\citenamefont {Jastrz{\k{e}}bski}\ \emph {et~al.}(2017)\citenamefont {Jastrz{\k{e}}bski}, \citenamefont {Kenton}, \citenamefont {Arpit}, \citenamefont {Ballas}, \citenamefont {Fischer}, \citenamefont {Bengio},\ and\ \citenamefont {Storkey}}]{jastrzkebski2017three}%
  \BibitemOpen
  \bibfield  {author} {\bibinfo {author} {\bibfnamefont {S.}~\bibnamefont {Jastrz{\k{e}}bski}}, \bibinfo {author} {\bibfnamefont {Z.}~\bibnamefont {Kenton}}, \bibinfo {author} {\bibfnamefont {D.}~\bibnamefont {Arpit}}, \bibinfo {author} {\bibfnamefont {N.}~\bibnamefont {Ballas}}, \bibinfo {author} {\bibfnamefont {A.}~\bibnamefont {Fischer}}, \bibinfo {author} {\bibfnamefont {Y.}~\bibnamefont {Bengio}},\ and\ \bibinfo {author} {\bibfnamefont {A.}~\bibnamefont {Storkey}},\ }\href@noop {} {\bibfield  {journal} {\bibinfo  {journal} {arXiv preprint arXiv:1711.04623}\ } (\bibinfo {year} {2017})}\BibitemShut {NoStop}%
\bibitem [{\citenamefont {Xie}\ \emph {et~al.}(2020)\citenamefont {Xie}, \citenamefont {Sato},\ and\ \citenamefont {Sugiyama}}]{xie2020diffusion}%
  \BibitemOpen
  \bibfield  {author} {\bibinfo {author} {\bibfnamefont {Z.}~\bibnamefont {Xie}}, \bibinfo {author} {\bibfnamefont {I.}~\bibnamefont {Sato}},\ and\ \bibinfo {author} {\bibfnamefont {M.}~\bibnamefont {Sugiyama}},\ }\href@noop {} {\bibfield  {journal} {\bibinfo  {journal} {arXiv preprint arXiv:2002.03495}\ } (\bibinfo {year} {2020})}\BibitemShut {NoStop}%
\bibitem [{\citenamefont {Keskar}\ \emph {et~al.}(2016)\citenamefont {Keskar}, \citenamefont {Mudigere}, \citenamefont {Nocedal}, \citenamefont {Smelyanskiy},\ and\ \citenamefont {Tang}}]{keskar2016large}%
  \BibitemOpen
  \bibfield  {author} {\bibinfo {author} {\bibfnamefont {N.~S.}\ \bibnamefont {Keskar}}, \bibinfo {author} {\bibfnamefont {D.}~\bibnamefont {Mudigere}}, \bibinfo {author} {\bibfnamefont {J.}~\bibnamefont {Nocedal}}, \bibinfo {author} {\bibfnamefont {M.}~\bibnamefont {Smelyanskiy}},\ and\ \bibinfo {author} {\bibfnamefont {P.~T.~P.}\ \bibnamefont {Tang}},\ }\href@noop {} {\bibfield  {journal} {\bibinfo  {journal} {arXiv preprint arXiv:1609.04836}\ } (\bibinfo {year} {2016})}\BibitemShut {NoStop}%
\bibitem [{\citenamefont {Huang}\ \emph {et~al.}(2020)\citenamefont {Huang}, \citenamefont {Emam}, \citenamefont {Goldblum}, \citenamefont {Fowl}, \citenamefont {Terry}, \citenamefont {Huang},\ and\ \citenamefont {Goldstein}}]{huang2020understanding}%
  \BibitemOpen
  \bibfield  {author} {\bibinfo {author} {\bibfnamefont {W.~R.}\ \bibnamefont {Huang}}, \bibinfo {author} {\bibfnamefont {Z.}~\bibnamefont {Emam}}, \bibinfo {author} {\bibfnamefont {M.}~\bibnamefont {Goldblum}}, \bibinfo {author} {\bibfnamefont {L.}~\bibnamefont {Fowl}}, \bibinfo {author} {\bibfnamefont {J.~K.}\ \bibnamefont {Terry}}, \bibinfo {author} {\bibfnamefont {F.}~\bibnamefont {Huang}},\ and\ \bibinfo {author} {\bibfnamefont {T.}~\bibnamefont {Goldstein}},\ }in\ \href@noop {} {\emph {\bibinfo {booktitle} {Proceedings of the 37th International Conference on Machine Learning}}},\ Vol.\ \bibinfo {volume} {119}\ (\bibinfo  {publisher} {PMLR},\ \bibinfo {year} {2020})\ pp.\ \bibinfo {pages} {4499--4509}\BibitemShut {NoStop}%
\bibitem [{\citenamefont {Averbeck}\ \emph {et~al.}(2006)\citenamefont {Averbeck}, \citenamefont {Latham},\ and\ \citenamefont {Pouget}}]{averbeck2006neural}%
  \BibitemOpen
  \bibfield  {author} {\bibinfo {author} {\bibfnamefont {B.~B.}\ \bibnamefont {Averbeck}}, \bibinfo {author} {\bibfnamefont {P.~E.}\ \bibnamefont {Latham}},\ and\ \bibinfo {author} {\bibfnamefont {A.}~\bibnamefont {Pouget}},\ }\href@noop {} {\bibfield  {journal} {\bibinfo  {journal} {Nature reviews neuroscience}\ }\textbf {\bibinfo {volume} {7}},\ \bibinfo {pages} {358} (\bibinfo {year} {2006})}\BibitemShut {NoStop}%
\bibitem [{\citenamefont {Pal}\ \emph {et~al.}(2022)\citenamefont {Pal}, \citenamefont {Deb},\ and\ \citenamefont {Dutta}}]{pal2022tipping}%
  \BibitemOpen
  \bibfield  {author} {\bibinfo {author} {\bibfnamefont {K.}~\bibnamefont {Pal}}, \bibinfo {author} {\bibfnamefont {S.}~\bibnamefont {Deb}},\ and\ \bibinfo {author} {\bibfnamefont {P.~S.}\ \bibnamefont {Dutta}},\ }\href@noop {} {\bibfield  {journal} {\bibinfo  {journal} {Physical Review E}\ }\textbf {\bibinfo {volume} {106}},\ \bibinfo {pages} {054412} (\bibinfo {year} {2022})}\BibitemShut {NoStop}%
\bibitem [{\citenamefont {Tsimring}(2014)}]{tsimring2014noise}%
  \BibitemOpen
  \bibfield  {author} {\bibinfo {author} {\bibfnamefont {L.~S.}\ \bibnamefont {Tsimring}},\ }\href@noop {} {\bibfield  {journal} {\bibinfo  {journal} {Reports on Progress in Physics}\ }\textbf {\bibinfo {volume} {77}},\ \bibinfo {pages} {026601} (\bibinfo {year} {2014})}\BibitemShut {NoStop}%
\bibitem [{\citenamefont {Torquato}\ \emph {et~al.}(2000)\citenamefont {Torquato}, \citenamefont {Truskett},\ and\ \citenamefont {Debenedetti}}]{torquato2000random}%
  \BibitemOpen
  \bibfield  {author} {\bibinfo {author} {\bibfnamefont {S.}~\bibnamefont {Torquato}}, \bibinfo {author} {\bibfnamefont {T.~M.}\ \bibnamefont {Truskett}},\ and\ \bibinfo {author} {\bibfnamefont {P.~G.}\ \bibnamefont {Debenedetti}},\ }\href@noop {} {\bibfield  {journal} {\bibinfo  {journal} {Physical review letters}\ }\textbf {\bibinfo {volume} {84}},\ \bibinfo {pages} {2064} (\bibinfo {year} {2000})}\BibitemShut {NoStop}%
\bibitem [{\citenamefont {Kamien}\ and\ \citenamefont {Liu}(2007)}]{kamien2007random}%
  \BibitemOpen
  \bibfield  {author} {\bibinfo {author} {\bibfnamefont {R.~D.}\ \bibnamefont {Kamien}}\ and\ \bibinfo {author} {\bibfnamefont {A.~J.}\ \bibnamefont {Liu}},\ }\href@noop {} {\bibfield  {journal} {\bibinfo  {journal} {Physical review letters}\ }\textbf {\bibinfo {volume} {99}},\ \bibinfo {pages} {155501} (\bibinfo {year} {2007})}\BibitemShut {NoStop}%
\bibitem [{\citenamefont {Parisi}\ and\ \citenamefont {Zamponi}(2005)}]{parisi2005ideal}%
  \BibitemOpen
  \bibfield  {author} {\bibinfo {author} {\bibfnamefont {G.}~\bibnamefont {Parisi}}\ and\ \bibinfo {author} {\bibfnamefont {F.}~\bibnamefont {Zamponi}},\ }\href@noop {} {\bibfield  {journal} {\bibinfo  {journal} {The Journal of chemical physics}\ }\textbf {\bibinfo {volume} {123}} (\bibinfo {year} {2005})}\BibitemShut {NoStop}%
\bibitem [{\citenamefont {Martiniani}\ \emph {et~al.}(2017)\citenamefont {Martiniani}, \citenamefont {Schrenk}, \citenamefont {Ramola}, \citenamefont {Chakraborty},\ and\ \citenamefont {Frenkel}}]{martiniani2017numerical}%
  \BibitemOpen
  \bibfield  {author} {\bibinfo {author} {\bibfnamefont {S.}~\bibnamefont {Martiniani}}, \bibinfo {author} {\bibfnamefont {K.~J.}\ \bibnamefont {Schrenk}}, \bibinfo {author} {\bibfnamefont {K.}~\bibnamefont {Ramola}}, \bibinfo {author} {\bibfnamefont {B.}~\bibnamefont {Chakraborty}},\ and\ \bibinfo {author} {\bibfnamefont {D.}~\bibnamefont {Frenkel}},\ }\href@noop {} {\bibfield  {journal} {\bibinfo  {journal} {Nature physics}\ }\textbf {\bibinfo {volume} {13}},\ \bibinfo {pages} {848} (\bibinfo {year} {2017})}\BibitemShut {NoStop}%
\bibitem [{\citenamefont {Anzivino}\ \emph {et~al.}(2023)\citenamefont {Anzivino}, \citenamefont {Casiulis}, \citenamefont {Zhang}, \citenamefont {Moussa}, \citenamefont {Martiniani},\ and\ \citenamefont {Zaccone}}]{anzivino2023estimating}%
  \BibitemOpen
  \bibfield  {author} {\bibinfo {author} {\bibfnamefont {C.}~\bibnamefont {Anzivino}}, \bibinfo {author} {\bibfnamefont {M.}~\bibnamefont {Casiulis}}, \bibinfo {author} {\bibfnamefont {T.}~\bibnamefont {Zhang}}, \bibinfo {author} {\bibfnamefont {A.~S.}\ \bibnamefont {Moussa}}, \bibinfo {author} {\bibfnamefont {S.}~\bibnamefont {Martiniani}},\ and\ \bibinfo {author} {\bibfnamefont {A.}~\bibnamefont {Zaccone}},\ }\href@noop {} {\bibfield  {journal} {\bibinfo  {journal} {The Journal of Chemical Physics}\ }\textbf {\bibinfo {volume} {158}} (\bibinfo {year} {2023})}\BibitemShut {NoStop}%
\bibitem [{\citenamefont {Bishop}\ and\ \citenamefont {Bishop}(2023)}]{bishop2023deep}%
  \BibitemOpen
  \bibfield  {author} {\bibinfo {author} {\bibfnamefont {C.~M.}\ \bibnamefont {Bishop}}\ and\ \bibinfo {author} {\bibfnamefont {H.}~\bibnamefont {Bishop}},\ }\href@noop {} {\emph {\bibinfo {title} {Deep learning: Foundations and concepts}}}\ (\bibinfo  {publisher} {Springer Nature},\ \bibinfo {year} {2023})\BibitemShut {NoStop}%
\bibitem [{\citenamefont {Li}\ \emph {et~al.}(2017)\citenamefont {Li}, \citenamefont {Tai},\ and\ \citenamefont {Weinan}}]{li2017stochastic}%
  \BibitemOpen
  \bibfield  {author} {\bibinfo {author} {\bibfnamefont {Q.}~\bibnamefont {Li}}, \bibinfo {author} {\bibfnamefont {C.}~\bibnamefont {Tai}},\ and\ \bibinfo {author} {\bibfnamefont {E.}~\bibnamefont {Weinan}},\ }in\ \href@noop {} {\emph {\bibinfo {booktitle} {International Conference on Machine Learning}}}\ (\bibinfo {organization} {PMLR},\ \bibinfo {year} {2017})\ pp.\ \bibinfo {pages} {2101--2110}\BibitemShut {NoStop}%
\bibitem [{\citenamefont {Dean}(1996)}]{dean1996langevin}%
  \BibitemOpen
  \bibfield  {author} {\bibinfo {author} {\bibfnamefont {D.~S.}\ \bibnamefont {Dean}},\ }\href@noop {} {\bibfield  {journal} {\bibinfo  {journal} {Journal of Physics A: Mathematical and General}\ }\textbf {\bibinfo {volume} {29}},\ \bibinfo {pages} {L613} (\bibinfo {year} {1996})}\BibitemShut {NoStop}%
\bibitem [{\citenamefont {Bertin}\ \emph {et~al.}(2013)\citenamefont {Bertin}, \citenamefont {Chat{\'e}}, \citenamefont {Ginelli}, \citenamefont {Mishra}, \citenamefont {Peshkov},\ and\ \citenamefont {Ramaswamy}}]{bertin2013mesoscopic}%
  \BibitemOpen
  \bibfield  {author} {\bibinfo {author} {\bibfnamefont {E.}~\bibnamefont {Bertin}}, \bibinfo {author} {\bibfnamefont {H.}~\bibnamefont {Chat{\'e}}}, \bibinfo {author} {\bibfnamefont {F.}~\bibnamefont {Ginelli}}, \bibinfo {author} {\bibfnamefont {S.}~\bibnamefont {Mishra}}, \bibinfo {author} {\bibfnamefont {A.}~\bibnamefont {Peshkov}},\ and\ \bibinfo {author} {\bibfnamefont {S.}~\bibnamefont {Ramaswamy}},\ }\href@noop {} {\bibfield  {journal} {\bibinfo  {journal} {New journal of physics}\ }\textbf {\bibinfo {volume} {15}},\ \bibinfo {pages} {085032} (\bibinfo {year} {2013})}\BibitemShut {NoStop}%
\bibitem [{\citenamefont {Solon}\ \emph {et~al.}(2015)\citenamefont {Solon}, \citenamefont {Cates},\ and\ \citenamefont {Tailleur}}]{solon2015active}%
  \BibitemOpen
  \bibfield  {author} {\bibinfo {author} {\bibfnamefont {A.~P.}\ \bibnamefont {Solon}}, \bibinfo {author} {\bibfnamefont {M.~E.}\ \bibnamefont {Cates}},\ and\ \bibinfo {author} {\bibfnamefont {J.}~\bibnamefont {Tailleur}},\ }\href@noop {} {\bibfield  {journal} {\bibinfo  {journal} {The European Physical Journal Special Topics}\ }\textbf {\bibinfo {volume} {224}},\ \bibinfo {pages} {1231} (\bibinfo {year} {2015})}\BibitemShut {NoStop}%
\bibitem [{\citenamefont {Lei}\ \emph {et~al.}(2019)\citenamefont {Lei}, \citenamefont {Ciamarra},\ and\ \citenamefont {Ni}}]{lei2019nonequilibrium}%
  \BibitemOpen
  \bibfield  {author} {\bibinfo {author} {\bibfnamefont {Q.-L.}\ \bibnamefont {Lei}}, \bibinfo {author} {\bibfnamefont {M.~P.}\ \bibnamefont {Ciamarra}},\ and\ \bibinfo {author} {\bibfnamefont {R.}~\bibnamefont {Ni}},\ }\href@noop {} {\bibfield  {journal} {\bibinfo  {journal} {Science advances}\ }\textbf {\bibinfo {volume} {5}},\ \bibinfo {pages} {eaau7423} (\bibinfo {year} {2019})}\BibitemShut {NoStop}%
\bibitem [{\citenamefont {De~Luca}\ \emph {et~al.}(2024)\citenamefont {De~Luca}, \citenamefont {Ma}, \citenamefont {Nardini},\ and\ \citenamefont {Cates}}]{DeLuca2024}%
  \BibitemOpen
  \bibfield  {author} {\bibinfo {author} {\bibfnamefont {F.}~\bibnamefont {De~Luca}}, \bibinfo {author} {\bibfnamefont {X.}~\bibnamefont {Ma}}, \bibinfo {author} {\bibfnamefont {C.}~\bibnamefont {Nardini}},\ and\ \bibinfo {author} {\bibfnamefont {M.~E.}\ \bibnamefont {Cates}},\ }\href@noop {} {\bibfield  {journal} {\bibinfo  {journal} {Journal of Physics: Condensed Matter}\ }\textbf {\bibinfo {volume} {36}},\ \bibinfo {pages} {405101} (\bibinfo {year} {2024})}\BibitemShut {NoStop}%
\bibitem [{\citenamefont {Ballard}\ \emph {et~al.}(2017)\citenamefont {Ballard}, \citenamefont {Das}, \citenamefont {Martiniani}, \citenamefont {Mehta}, \citenamefont {Sagun}, \citenamefont {Stevenson},\ and\ \citenamefont {Wales}}]{ballard2017energy}%
  \BibitemOpen
  \bibfield  {author} {\bibinfo {author} {\bibfnamefont {A.~J.}\ \bibnamefont {Ballard}}, \bibinfo {author} {\bibfnamefont {R.}~\bibnamefont {Das}}, \bibinfo {author} {\bibfnamefont {S.}~\bibnamefont {Martiniani}}, \bibinfo {author} {\bibfnamefont {D.}~\bibnamefont {Mehta}}, \bibinfo {author} {\bibfnamefont {L.}~\bibnamefont {Sagun}}, \bibinfo {author} {\bibfnamefont {J.~D.}\ \bibnamefont {Stevenson}},\ and\ \bibinfo {author} {\bibfnamefont {D.~J.}\ \bibnamefont {Wales}},\ }\href@noop {} {\bibfield  {journal} {\bibinfo  {journal} {Physical Chemistry Chemical Physics}\ }\textbf {\bibinfo {volume} {19}},\ \bibinfo {pages} {12585} (\bibinfo {year} {2017})}\BibitemShut {NoStop}%
\bibitem [{\citenamefont {Gabrielli}(2004)}]{gabrielli2004point}%
  \BibitemOpen
  \bibfield  {author} {\bibinfo {author} {\bibfnamefont {A.}~\bibnamefont {Gabrielli}},\ }\href@noop {} {\bibfield  {journal} {\bibinfo  {journal} {Physical Review E—Statistical, Nonlinear, and Soft Matter Physics}\ }\textbf {\bibinfo {volume} {70}},\ \bibinfo {pages} {066131} (\bibinfo {year} {2004})}\BibitemShut {NoStop}%
\bibitem [{\citenamefont {Barnett}\ \emph {et~al.}(2019)\citenamefont {Barnett}, \citenamefont {Magland},\ and\ \citenamefont {af~Klinteberg}}]{barnett2019parallel}%
  \BibitemOpen
  \bibfield  {author} {\bibinfo {author} {\bibfnamefont {A.~H.}\ \bibnamefont {Barnett}}, \bibinfo {author} {\bibfnamefont {J.}~\bibnamefont {Magland}},\ and\ \bibinfo {author} {\bibfnamefont {L.}~\bibnamefont {af~Klinteberg}},\ }\href@noop {} {\bibfield  {journal} {\bibinfo  {journal} {SIAM Journal on Scientific Computing}\ }\textbf {\bibinfo {volume} {41}},\ \bibinfo {pages} {C479} (\bibinfo {year} {2019})}\BibitemShut {NoStop}%
\bibitem [{\citenamefont {Barnett}(2021)}]{barnett2021aliasing}%
  \BibitemOpen
  \bibfield  {author} {\bibinfo {author} {\bibfnamefont {A.~H.}\ \bibnamefont {Barnett}},\ }\href@noop {} {\bibfield  {journal} {\bibinfo  {journal} {Applied and Computational Harmonic Analysis}\ }\textbf {\bibinfo {volume} {51}},\ \bibinfo {pages} {1} (\bibinfo {year} {2021})}\BibitemShut {NoStop}%
\end{thebibliography}%

\section{Acknowledgements}

We thank Mathias Casiulis, David Heeger, Flaviano Morone, and Aaron Shih for fruitful discussions. This work was supported by the National Science Foundation grant IIS-2226387, National Institute of Health under award number R01MH137669, Simons Center for Computational Physical Chemistry, and in part by the NYU IT High Performance Computing resources, services, and staff expertise.

\section{Author contributions}

S.A., G.Z. and S.M. conceived the project. S.A. and G.Z. performed the research. S.A., G.Z. and S.M. analyzed data and wrote the manuscript. S.M. supervised the research.

\section{Methods}

\subsection{Particle simulations}

Our system consisted of $N$ hyperspherical particles of radius $R$ in a $d$-dimensional hypercubic box of side length $L$ with periodic boundary conditions. The unit of length, time, and energy were chosen as $2R$, $\delta t$, and $E$, respectively. $\delta t$ is the simulation time-step for continuous-time simulations and $E=1$ is the characteristic energy scale of the potential given by Eq.~\ref{eq:v_ji_family}. $N = 318309$ and $R = 1$ were kept fixed and the particle volume fraction $\phi = N V_s / V_c$ was varied by changing $L$. $V_s$ and $V_c$ denote volumes of a $d$-dimensional hypersphere of radius $R$, and hypercube of side length $L$, respectively. Particles were randomly distributed in the simulation box at $t=0$. All simulations were run until the system reached a steady-state and all measurements were performed at steady-state. All results were averaged over $100$ steady-state configurations.

\textbf{Discrete-time simulations.} For discrete-time simulations of RO, BRO, and SGD, the dynamics of particles were evolved according to Eqs.~\ref{eq:discrete_ro}, \ref{eq:discrete_bro}, and \ref{eq:discrete_sgd}, respectively. As evident from Eqs.~\ref{eq:discrete_ro}, \ref{eq:discrete_bro}, and \ref{eq:discrete_sgd}, positions of isolated particles (particles with no overlapping neighbors) do not evolve at any given time-step. 

Dynamics for RO are controlled by four parameters, the kick magnitude $\epsilon$, the particle volume fraction $\phi$, the spatial dimension $d$, and the correlation of pairwise noise $c$. For the results reported in the main text, parameters were set as $\epsilon = 1$, $\phi = 1.0$, $d = 2$, and $c \in [-1, 0]$. The critical volume fraction $\phi_c \approx 0.375$ for this set of parameters. 

Dynamics for BRO are controlled by four parameters, the kick magnitude $\epsilon$, the particle volume fraction $\phi$, the spatial dimension $d$, and the correlation of pairwise noise $c$. For the results reported in the main text, parameters were set as $\epsilon = 1$, $\phi = 1.0$, $d = 2$, and $c \in [-1, 0]$. The critical volume fraction $\phi_c \approx 0.395$ for this set of parameters. 

Dynamics for SGD are controlled by six parameters, the learning rate $\alpha$, the particle volume fraction $\phi$, the batch fraction $b_f$, the ``stiffness'' of the potential $p$, the spatial dimension $d$, and the correlation of pairwise noise $c$. For the results reported in the main text, parameters were set as $\alpha = 0.5$, $\phi = 1.0$, $b_f = 0.5$, $p = 1$, $d = 2$, and $c \in [-1, 0]$. The critical volume fraction $\phi_c \approx 0.615$ for this set of parameters. 

For the measurement of flatness of energy minima in SGD, a configuration $\mathbf{X} \equiv \{ \mathbf{x}_1, \mathbf{x}_2,...,\mathbf{x}_N \}$ was perturbed by adding an independent Gaussian noise $\mathbf{N}_i(\mathbf{0}, \sigma^2 \mathbf{I})$ to the position of each particle to get $\mathbf{X} + \Delta \mathbf{X} \equiv \{ \mathbf{x}_1+\mathbf{N}_1, \mathbf{x}_2+\mathbf{N}_2,...,\mathbf{x}_N+\mathbf{N}_N \}$. All results were averaged over $5000$ independent noise realizations. For the results reported in Fig.~\ref{fig:figure3}b, parameters were set as $\alpha = 0.5$, $b_f = 0.5$, $\phi = 1.0$, $p = 1$, $d = 2$, $\sigma = 0.02$, and $c \in [-1, 0]$. For the results reported in Fig.~\ref{fig:figure3}c, parameters were set as $\alpha = 0.5$, $c=-1$, $\phi = 1.0$, $p = 1$, $d = 2$, $\sigma = 0.01$, and $b_f \in [0.1, 1.0]$. For the results reported in Fig.~\ref{fig:figure3}d, parameters were set as $b_f = 0.5$, $c=-1$, $\phi = 1.0$, $p = 1$, $d = 2$, $\sigma = 0.01$, and $\alpha \in [0.5, 1.0]$.

\textbf{Continuous-time simulations.} For the continuous-time simulations of the generalized model of RO, BRO, and SGD, the dynamics of particles were evolved according to Eq.~\ref{eq:generic_sde}, supplemented with appropriate values of the friction constant $\gamma$, and the short-range interaction potential $V_{ij}$ and matrix $\mathbf{\Lambda}_{ij}$. The Euler–Maruyama method was used to solve Eq.~\ref{eq:generic_sde}, with a time-step $\delta t = 1.0$. 

For the generalized model of RO, $V_{ji} = 0$, and $\Lambda_{ji, \alpha \beta} = \mathbf{1}_{(0, 2R)}(r_{ij}) (\epsilon^2/3d\tau) \delta_{\alpha \beta}$, where $\mathbf{1}_{(0, 2R)}(r_{ij})$ is an indicator function such that $\mathbf{1}_{(0, 2R)}(r_{ij}) = 1 \ \forall \ r_{ij} \in (0, 2R)$ and $\mathbf{1}_{(0, 2R)}(r_{ij}) = 0$ otherwise. $\tau$ is the time scale quantifying the time elapsed in a discrete step. For the results reported in the main text, parameters were set as $\epsilon = 1$, $\tau = 1.0$, $\phi = 1.0$, $d = 2$, and $c \in [-1, 0]$. 

For the generalized model of BRO, $\gamma = E \tau / \epsilon R$, $\Lambda_{ji, \alpha \beta} = (\epsilon^2 R^2 / 3 \tau E^2) \, \partial_{\alpha} V_{ji} \partial_{\beta} V_{ji}$, and $V_{ji}$ is a short-range, pairwise, linear, repulsive potential (Eq.~\ref{eq:v_ji_family} with $p=1$). For the results reported in the main text, parameters were set as $\epsilon = 1$, $\tau = 1.0$, $\phi = 1.0$, $d = 2$, and $c \in [-1, 0]$. 

For the generalized model of SGD, $\gamma = \tau/\alpha b_f$, $\Lambda_{ji, \alpha \beta} = (\alpha^2 b_f (1 - b_f)/\tau) \, \partial_{\alpha} V_{ji} \partial_{\beta} V_{ji}$, and $V_{ji}$ is given by Eq.~\ref{eq:v_ji_family}. For the results reported in the main text, parameters were set as $\alpha = 1$, $b_f = 0.5$, $\tau = 1$, $p = 1$, $\phi = 1.0$, $d = 2$, and $c \in [-1, 0]$.

\subsection{Continuum simulations}

Our system consisted of density $\rho(\mathbf{x},t)$ evolving in a $d$-dimensional hypercubic box of side length $L$ with periodic boundary conditions. The unit of length, time, and energy were chosen as $R$, $\delta t$, and $E$, respectively. $\delta t$ is the simulation time-step, $E=1$ is the characteristic energy scale of the continuous version of the potential given by Eq.~\ref{eq:v_ji_family}, and $2R$ is the cutoff length of the potential. Space was discretized with a square grid of grid spacing $\delta l = L/512$. $L = 200$ and $R = 1$ were kept fixed and $\int \rho (\mathbf{x}, t) d \mathbf{x} = N$ was fixed at all times (density conservation). Density at all grid points was sampled from a Gaussian distribution with mean $N/V_c$, and standard deviation $0.01N/V_c$ at $t=0$. All other parameters for RO, BRO, and SGD were chosen to be the same as that of continuous-time particle simulations. All simulations were run until the system reached a steady-state and all measurements were performed at steady-state. 

The finite-difference method combined with forward Euler time stepping was used to solve Eq.~\ref{eq:generic_drho}. Spatial derivatives for the stochastic flux and the diffusion term in Eq.~\ref{eq:generic_drho} were approximated by the second-order central difference scheme. We split the velocity term (Eqs.~\ref{eq:generic_drho}, \ref{eq:det_vel}) into two parts using chain rule,
\begin{align}
    \frac{1}{\gamma} \nabla \cdot \left[ \rho(\mathbf{x}) \nabla \left< V(\mathbf{x},\mathbf{y}) \right>_{\rho(\mathbf{y})} \right] &= \frac{1}{\gamma} \nabla \rho(\mathbf{x}) \cdot \nabla \left< V(\mathbf{x},\mathbf{y}) \right>_{\rho(\mathbf{y})} \nonumber \\
    &+ \frac{1}{\gamma} \rho(\mathbf{x}) \ \nabla^2 \left< V(\mathbf{x},\mathbf{y}) \right>_{\rho(\mathbf{y}) }.
    \label{det_vel_chain}
\end{align}
Spatial derivatives in the first and second term on the right hand side of Eq.~\ref{det_vel_chain} were approximated using the first-order forward difference, and the second-order central difference scheme, respectively. The finite-difference schemes were chosen to ensure strict density conservation and prevent checkerboard artifacts.

\subsection{Structure factor and density fluctuations}

The structure factor for a particle system is defined as $S(\mathbf{k}) = |\hat{\rho}(\mathbf{k})|^2/N$, where $\rho(\mathbf{x}) = \sum_{i=1}^{N} \delta (\mathbf{x} - \mathbf{x}_i)$, and $\hat{f}$ is the spatial Fourier transform of any arbitrary function $f(\mathbf{x})$, given by $\hat{f}(\mathbf{k}) = \int f(\mathbf{x}) e^{-i \mathbf{k} \cdot \mathbf{x}} d \mathbf{x}$. $S(\mathbf{k})$ was calculated using the nonuniform fast Fourier transform \cite{barnett2019parallel, barnett2021aliasing}. The radial structure factor $S(k)$ was then calculated by radially averaging $S(\mathbf{k})$.

The structure factor for a continuous density field $\rho(\mathbf{x})$ is defined as $S(\mathbf{k}) = |\hat{\delta \rho}(\mathbf{k})|^2/\bar{\rho}$, where $\delta \rho(\mathbf{x}) = \rho(\mathbf{x}) - \bar{\rho}$, and $\bar{\rho} = \int \rho(\mathbf{x}) d\mathbf{x} / \int  d\mathbf{x}$.

Number density variance $\delta \rho^2(l)$ in a hyperspherical window of diameter $l$ was measured by exploiting the exact relationship between $S(k)$ and $\delta \rho^2(l)$ \cite{torquato2018hyperuniform}, 
\begin{equation}
    \delta \rho^2(l) = \frac{\bar{\rho} d 2^d \Gamma(1 + \frac{d}{2})}{( \sqrt{\pi} l)^d} \int_0^{\infty} \frac{1}{k} S(k) \left[ J_{d/2} \left( \frac{kl}{2}\right) \right]^2 \, dk,
    \label{sk_den_fluc_rel}
\end{equation}
where $\bar{\rho} = N/V_c$, $\Gamma$ is the Gamma function, and $J$ is the Bessel function of the first kind. The integral in Eq.~\ref{sk_den_fluc_rel} was evaluated numerically using Simpson's rule, based on the measured $S(k)$.

\end{document}


\preprint{APS/123-QED}

\title{\textbf{\LARGE{Supplementary Information}} \\ Emergent universal long-range structure in random-organizing systems}

\author{Satyam Anand}
\thanks{Equal contribution}
\email{sa7483@nyu.edu}
\affiliation{Courant Institute of Mathematical Sciences, New York University, New York, NY 10003, USA}
\affiliation{\mbox{Center for Soft Matter Research, Department of Physics, New York University, New York, NY 10003, USA}}

\author{Guanming Zhang}
\thanks{Equal contribution}
\email{gz2241@nyu.edu}
\affiliation{\mbox{Center for Soft Matter Research, Department of Physics, New York University, New York, NY 10003, USA}}
\affiliation{\mbox{Simons Center for Computational Physical Chemistry, Department of Chemistry, New York University, New York, NY 10003, USA}}

\author{Stefano Martiniani}
\email{sm7683@nyu.edu}
\affiliation{Courant Institute of Mathematical Sciences, New York University, New York, NY 10003, USA}
\affiliation{\mbox{Center for Soft Matter Research, Department of Physics, New York University, New York, NY 10003, USA}}
\affiliation{\mbox{Simons Center for Computational Physical Chemistry, Department of Chemistry, New York University, New York, NY 10003, USA}}
\affiliation{\mbox{Center for Neural Science, New York University, New York, NY 10003, USA}}

\maketitle    

\tableofcontents

\section{Theory}

\subsection{Continuous-time approximation of discrete-time dynamics}

Random-organizing systems are discrete-time systems (Eqs.~\ref{eq:discrete_ro}, \ref{eq:discrete_bro}, and \ref{eq:discrete_sgd} in main text). Following the framework of stochastic modified equations \cite{li2017stochastic, zhang2024absorbing}, we first approximate these systems by a continuous-time dynamics. Consider a generic discrete-time evolution equation,
\begin{align}
    \mathbf{x}^{m+1} &= \mathbf{x}^{m} + \boldsymbol{\vartheta} \nonumber \\
    &= \mathbf{x}^{m} + \left< \boldsymbol{\vartheta} \right> + \left( \boldsymbol{\vartheta} - \left< \boldsymbol{\vartheta} \right> \right),
    \label{eq:dis_gen_1}
\end{align}
where $\mathbf{x}^{m}$ is the value of variable $\mathbf{x}$ at time-step $m$, $\boldsymbol{\vartheta}$ is an arbitrary noise with mean $\left< \vartheta_{\alpha} \right>$ and covariance matrix $\left< (\vartheta_{\alpha} - \left< \vartheta_{\alpha} \right>) (\vartheta_{\beta} - \left< \vartheta_{\beta} \right>) \right>$, and $\left< . \right>$ denotes the expectation value. We approximate $\left( \boldsymbol{\vartheta} - \left< \boldsymbol{\vartheta} \right> \right)$ in Eq.~\ref{eq:dis_gen_1} by a Gaussian noise $\sqrt{\left< (\boldsymbol{\vartheta} - \left< \boldsymbol{\vartheta} \right>) (\boldsymbol{\vartheta} - \left< \boldsymbol{\vartheta} \right>)^\intercal \right>} \boldsymbol{\varrho}$, where $\boldsymbol{\varrho}$ is a Gaussian noise with $\left< \varrho_{\alpha} \right> = 0$ and $\left< \varrho_{\alpha} \varrho_{\beta} \right> = \delta_{\alpha \beta}$ to get,
\begin{align}
    \mathbf{x}^{m+1} &= \mathbf{x}^{m} + \left< \boldsymbol{\vartheta} \right> + \sqrt{\left< (\boldsymbol{\vartheta} - \left< \boldsymbol{\vartheta} \right>) (\boldsymbol{\vartheta} - \left< \boldsymbol{\vartheta} \right>)^\intercal \right>} \boldsymbol{\varrho},
    \label{eq:dis_gen_2}
\end{align}
where $\sqrt{.}$ is the matrix square root. We assume the characteristic time of one discrete time-step to be $\tau$, giving time elapsed after $m$ discrete time-steps as $t=m\tau$. We can now rewrite Eq.~\ref{eq:dis_gen_2} as,
\begin{align}
    \mathbf{x}^{(m+1)\tau} &= \mathbf{x}^{m\tau} + \tau \left( \frac{\left< \boldsymbol{\vartheta} \right>}{\tau} \right) + \sqrt{\tau} \left( \frac{\sqrt{\left< (\boldsymbol{\vartheta} - \left< \boldsymbol{\vartheta} \right>) (\boldsymbol{\vartheta} - \left< \boldsymbol{\vartheta} \right>)^\intercal \right>}}{\sqrt{\tau}} \right) \boldsymbol{\varrho}.
    \label{eq:dis_gen_3}
\end{align}
Eq.~\ref{eq:dis_gen_3} is nothing but the Euler-Maruyama discretization of a continuous-time stochastic differential equation (SDE) given by,
\begin{align}
    \frac{d\mathbf{x}(t)}{dt}
    &= \underbrace{\frac{\left< \boldsymbol{\vartheta} \right>}{\tau}}_{\text{deterministic term}} + \underbrace{\frac{\sqrt{\left< (\boldsymbol{\vartheta} - \left< \boldsymbol{\vartheta} \right>) (\boldsymbol{\vartheta} - \left< \boldsymbol{\vartheta} \right>)^\intercal \right>}}{\sqrt{\tau}} \boldsymbol{\varphi} (t)}_{\text{noise term}},
    \label{eq:cont_gen}
\end{align}
where $\boldsymbol{\varphi} (t)$ is a Gaussian noise with mean $\left< \varphi_{\alpha} (t) \right> = 0$, and covariance matrix $\left< \varphi_{\alpha} (t) \varphi_{\beta} (t) \right> = \delta_{\alpha \beta} \delta(t-t')$. Thus, Eq.~\ref{eq:cont_gen} is a continuous-time approximation of the discrete-time Eq.~\ref{eq:dis_gen_1}. We now apply this generic procedure to RO, BRO, and SGD.

In RO, the dynamics of the position of particle $i$ at time-step $m+1$ ($\mathbf{x}_{i}^{m+1}$) is given by,
\begin{equation}
    \mathbf{x}_{i}^{m+1} = \mathbf{x}_{i}^{m} + \epsilon \sum_{j \in \Gamma_i^m} u_{ji}^m \boldsymbol{\zeta}^m_{ji}, 
    \label{eq:discrete_ro_SI}
\end{equation}
where $\epsilon$ controls the magnitude of the pairwise kick given by particle $j$ to $i$, $u_{ji}^m$ is a random number sampled from a standard uniform distribution ($U[0,1]$) at time-step $m$, $\boldsymbol{\zeta}^m_{ji}$ is a random unit vector sampled uniformly on the surface of a $d$-dimensional unit hypersphere at time-step $m$, and $\Gamma_i^m = \{j \mid |\mathbf{x}_j^m - \mathbf{x}_i^m| < 2R, j \neq i \}$ is the set containing all particles that overlap with particle $i$ at time-step $m$ (Fig.~\ref{fig:figure1}b). The covariance matrix of the complete noise vector $\boldsymbol{\omega}^m_{ij} = \epsilon u_{ij}^m \boldsymbol{\zeta}^m_{ij}$ is $\text{cov}[ \omega_{ij, \alpha}^m \omega_{kl, \beta}^n] = (\epsilon^2/3d) \, \delta^{mn} \delta_{\alpha \beta}  (\delta_{ik}\delta_{jl} + c \, \delta_{il}\delta_{jk})$, where $c$ is the Pearson correlation coefficient between $\omega_{ij, \alpha}^m$ and $\omega_{ji, \alpha}^m$. Comparing Eq.~\ref{eq:discrete_ro_SI} with Eq.~\ref{eq:dis_gen_1} to get $\boldsymbol{\vartheta} = \epsilon \sum_{j \in \Gamma_i^m} u_{ji}^m \boldsymbol{\zeta}^m_{ji}$, we can write the SDE for RO as,
\begin{equation}
    \frac{d\mathbf{x}_{i}(t)}{dt}
    = \underbrace{ \sqrt{\frac{\epsilon^2}{3d\tau}} \sum_{j = 1}^N \sqrt{ \mathbf{1}_{(0, 2R)}(r_{ij}) \mathbf{I}} \ \boldsymbol{\xi}_{ji} (t)}_{\text{noise term}},
    \label{eq:sde_ro}
\end{equation}
where $\mathbf{1}_{(0, 2R)}(r_{ij})$ is an indicator function such that $\mathbf{1}_{(0, 2R)}(r_{ij}) = 1 \ \forall \ r_{ij} \in (0, 2R)$ and $\mathbf{1}_{(0, 2R)}(r_{ij}) = 0$ otherwise, $r_{ij} = |\mathbf{x}_j^m - \mathbf{x}_i^m|$, $\mathbf{I}$ is the identity matrix, $\boldsymbol{\xi}_{ji}$ is a pairwise, Gaussian noise given by the particle $j$ to particle $i$ having mean $\left< {\xi}_{ji,\alpha} (t) \right> = 0$ and covariance matrix $\left< {\xi}_{ij,\alpha} (t) \, {\xi}_{kl,\beta} (t') \right> = \delta (t-t') \, \delta_{\alpha \beta} (\delta_{ik}\delta_{jl} + c \, \delta_{il}\delta_{jk})$, where $c$ is the Pearson correlation coefficient between ${\xi}_{ij,\alpha} (t)$ and ${\xi}_{ji,\alpha} (t)$. 

In BRO, the dynamics of the position of particle $i$ at time-step $m+1$ ($\mathbf{x}_{i}^{m+1}$) is given by,
\begin{equation}
    \mathbf{x}_{i}^{m+1} = \mathbf{x}_{i}^{m} + \epsilon \sum_{j \in \Gamma_i^m} u_{ji}^m \mathbf{\hat{x}}^m_{ji}, 
    \label{eq:discrete_bro_SI}
\end{equation}
where $\epsilon$ controls the magnitude of the pairwise kick given by particle $j$ to $i$, $u_{ji}^m$ is a random number sampled from a standard uniform distribution ($U[0,1]$) at time-step $m$, and $\mathbf{\hat{x}}^m_{ji} = -(\mathbf{x}^m_{j} - \mathbf{x}^m_{i})/|\mathbf{x}^m_{j} - \mathbf{x}^m_{i}|$ is the deterministic unit vector pointing from the center of particle $j$ to $i$ at time-step $m$. The covariance of the complete noise vector $\boldsymbol{\omega}^m_{ij} = \epsilon u_{ij}^m \mathbf{\hat{x}}^m_{ij}$ is $\text{cov}[ \omega_{ij, \alpha}^m \omega_{kl, \beta}^n] = (\epsilon^2/12) \, \delta^{mn} \hat{x}^m_{ji,\alpha} \hat{x}^n_{ji,\beta} (\delta_{ik}\delta_{jl} + c \, \delta_{il}\delta_{jk})$, where $c$ is the Pearson correlation coefficient between $\omega_{ij, \alpha}^m$ and $\omega_{ji, \alpha}^m$. Comparing Eq.~\ref{eq:discrete_bro_SI} with Eq.~\ref{eq:dis_gen_1} to get $\boldsymbol{\vartheta} = \epsilon \sum_{j \in \Gamma_i^m} u_{ji}^m \mathbf{\hat{x}}^m_{ji}$, we can write the SDE for BRO as,
\begin{equation}
    \frac{d\mathbf{x}_{i}(t)}{dt}
    = \underbrace{ - \frac{\epsilon R}{E \tau} \sum_{j = 1}^N \nabla_{i} V_{ji}}_{\text{deterministic term}} + \underbrace{ \sqrt{\frac{\epsilon^2 R^2}{3\tau E^2}} \sum_{j = 1}^N \sqrt{\nabla_{i} V_{ji} \nabla_{i} V_{ji}^\intercal} \ \boldsymbol{\xi}_{ji} (t)}_\text{noise term}, 
    \label{eq:sde_bro}
\end{equation}
where $\nabla_i = \nabla_{\mathbf{x}_i}$, and $V_{ji}$ is a pairwise, linear, repulsive potential given by 
\begin{equation}
    V_{ij}(r) =     
    \begin{cases}
      E\left(1 - \frac{r_{ij}}{2R}\right), & \text{if}\ 0 < r_{ij} < 2R, \\
      0, & \text{otherwise},
    \end{cases}  
    \label{eq:v_ji_bro}
\end{equation}
where $E$ is the characteristic energy scale.

In SGD, we consider particles interacting via a pairwise potential. The dynamics of the position of particle $i$ at time-step $m+1$ ($\mathbf{x}_{i}^{m+1}$) is given by,
\begin{equation}
        \mathbf{x}_{i}^{m+1} = \mathbf{x}_{i}^{m} - \alpha \sum_{j \in \Gamma_i^m} \theta_{ji}^m \nabla_{i} V_{ji}^{m}
\label{eq:discrete_sgd_SI}
\end{equation}
where $\alpha$ is the learning rate, $V_{ji}^m = V(|\mathbf{x}^m_{j} - \mathbf{x}^m_{i}|)$ is the interaction potential, and $\theta_{ji}^m$ is a random number sampled from a Bernoulli distribution having parameter $b_f$ (batch fraction) at time-step $m$. The covariance of the complete noise vector $\boldsymbol{\omega}^m_{ij} = - \alpha \theta_{ij}^m \nabla_{i} V_{ij}^{m}$ is $\text{cov}[ \omega_{ij, \alpha}^m \omega_{kl, \beta}^n] = \alpha^2 b_f(1-b_f) \, \delta^{mn} \partial_{i,\alpha} V_{ji}^{m} \, \partial_{i,\beta} V_{ji}^{n} (\delta_{ik}\delta_{jl} + c \, \delta_{il}\delta_{jk})$, where $c$ is the Pearson correlation coefficient between $\omega_{ij, \alpha}^m$ and $\omega_{ji, \alpha}^m$. While $V_{ji}$ can be any pairwise potential in SGD, either short- or long-range, here, we consider a class of short-range, repulsive potentials given by
\begin{equation}
    V_{ij}(r) =     
    \begin{cases}
      \frac{E}{p}\left(1 - \frac{r_{ij}}{2R}\right)^p, & \text{if}\ 0 < r_{ij} < 2R, \\
      0, & \text{otherwise},
    \end{cases}  
    \label{eq:v_ji_family_SI}
\end{equation}
where $E$ is the characteristic energy scale, and $p$ controls the stiffness of the potential. Comparing Eq.~\ref{eq:discrete_sgd_SI} with Eq.~\ref{eq:dis_gen_1} to get $\boldsymbol{\vartheta} = - \theta_{ij}^m \alpha \nabla_{i} V_{ji}^{m}$, we can write the SDE for SGD as,
\begin{equation}
    \frac{d\mathbf{x}_{i}(t)}{dt}
    = \underbrace{ - \frac{\alpha b_f}{\tau} \sum_{j = 1}^N \nabla_{i} V_{ji}}_{\text{deterministic term}} + \underbrace{\sqrt{\frac{\alpha^2 b_f(1-b_f)}{\tau}} \sum_{j = 1}^N \sqrt{\nabla_{i} V_{ji} \nabla_{i} V_{ji}^\intercal} \ \boldsymbol{\xi}_{ji} (t)}_{\text{noise term}}. 
    \label{eq:sde_sgd}
\end{equation}

The SDE for RO, BRO, and SGD (Eqs.~\ref{eq:sde_ro}, \ref{eq:sde_bro}, and \ref{eq:sde_sgd}) can be generically written as an overdamped Langevin equation of the form
\begin{equation}
    \frac{d\mathbf{x}_{i}(t)}{dt}
    = \underbrace{ - \frac{1}{\gamma} \sum_{j = 1}^N \nabla_{i} V_{ji}}_{\text{deterministic term}} 
    + \underbrace{ \sum_{j = 1}^N \sqrt{\mathbf{\Lambda}_{ji}} \ \boldsymbol{\xi}_{ji}}_{\text{noise term}}, 
    \label{eq:generic_sde_SI}
\end{equation}
where $\gamma$ is the friction constant, $V_{ji}$, $\mathbf{\Lambda}_{ji}$ are short-range, pairwise interaction potential and diffusion matrix between particles $j$ and $i$, respectively. $\boldsymbol{\xi}_{ji}$ is a pairwise, Gaussian noise given by particle $j$ to particle $i$ with mean $\left< {\xi}_{ji,\alpha} (t) \right> = 0$ and covariance matrix $\left< {\xi}_{ij,\alpha} (t) \, {\xi}_{kl,\beta} (t') \right> = \delta (t-t') \, \delta_{\alpha \beta} (\delta_{ik}\delta_{jl} + c \, \delta_{il}\delta_{jk})$, where $c$ is the Pearson correlation coefficient between ${\xi}_{ij,\alpha} (t)$ and ${\xi}_{ji,\alpha} (t)$.
$\gamma$, $V_{ji}$, and $\mathbf{\Lambda}_{ji}$ take different values and functional forms for different systems (Table~\ref{tab:system_params}, Eqs.~\ref{eq:sde_ro}, \ref{eq:sde_bro}, and \ref{eq:sde_sgd}). 
$V_{ii} =\Lambda_{ii, \alpha \beta} =0$ for BRO, RO and SGD due to the absence of self-interaction in all three systems. Further, for all three systems, both $V_{ij}$ and $\mathbf{\Lambda}_{ij}$ are invariant under particle exchange, i.e., $V_{ij} = V_{ji}$ and $\mathbf{\Lambda}_{ij} = \mathbf{\Lambda}_{ji}$, and $\mathbf{\Lambda}_{ij}$ is a symmetric matrix (${\Lambda}_{ij,{\alpha \beta}} = {\Lambda}_{ij,{\beta \alpha}}$). 
\begin{table}[h!]
\centering
\renewcommand{\arraystretch}{3.0} 
\begin{tabular}{|l|c|c|c|}
\hline
\textbf{} & Friction coefficient ($\gamma$) & {Interaction potential ($V_{ji}$)} & {Diffusion matrix ($\boldsymbol{\mathbf{\Lambda}}_{ji}$)} \\ \hline
RO & --- & $0$ & $\displaystyle \frac{\epsilon^2}{3d\tau} \mathbf{1}_{(0, 2R)}(r_{ij}) \mathbf{I}$ \\ \hline
BRO & $\displaystyle \frac{E \tau}{\epsilon R}$ & $    V_{ij}(r) =     
    \begin{cases}
      E\left(1 - \frac{r_{ij}}{2R}\right), & \text{if}\ 0 < r_{ij} < 2R, \\
      0, & \text{otherwise}
    \end{cases}$ & $\displaystyle \frac{\epsilon^2 R^2}{3\tau E^2} \nabla_{i} V_{ji} \nabla_{i} V_{ji}^{\intercal}$ \\ \hline
SGD & $\displaystyle \frac{\tau}{\alpha b_f}$ &
    $V_{ij}(r) =     
    \begin{cases}
      \frac{E}{p}\left(1 - \frac{r_{ij}}{2R}\right)^p, & \text{if}\ 0 < r_{ij} < 2R, \\
      0, & \text{otherwise}
    \end{cases} $ & $\displaystyle \frac{\alpha^2 b_f(1-b_f)}{\tau} \nabla_{i} V_{ji} \nabla_{i} V_{ji}^{\intercal}$ \\ \hline
\end{tabular}
\caption{Parameters for the continuous-time generalized model of random-organizing systems.}
\label{tab:system_params}
\end{table}

Notice that for RO, the deterministic (first) term in Eq.~\ref{eq:generic_sde_SI} vanishes, and the noise (second) term is isotropic---consistent with the fact that pairwise kicks are randomly directed in RO (Eqs.~\ref{eq:discrete_ro_SI}, and \ref{eq:sde_ro}). The noise term in BRO, however, is anisotropic---directed along the line joining the center of particles (Eq.~\ref{eq:sde_bro}). The noise term in SGD is also anisotropic and directed along the line joining the center of particles---in fact, SGD with a linear potential ($p=1$ in Eq.~\ref{eq:v_ji_family_SI}) maps exactly to BRO (except for prefactors) (Eqs.~\ref{eq:sde_bro}, and \ref{eq:sde_sgd}) \cite{zhang2024absorbing}. Finally, it is worth noting that the noise in discrete-time SGD is on the selection of particles, while in the SDE, it appears as an interaction noise (Eqs.~\ref{eq:discrete_sgd_SI}, and \ref{eq:sde_sgd}).

\subsection{Fluctuating hydrodynamics}

Starting from the continuous-time description of random-organizing systems (Eq.~\ref{eq:generic_sde_SI}), we now coarse-grain the microscopic dynamics to get a continuum description.

\subsubsection{Fokker-Planck method}

Defining the $N$-particle density as $\rho_N (\{\mathbf{x}_N\},t)$, where $\{ \mathbf{x}_N \} \equiv \{ \mathbf{x}_1,\mathbf{x}_2,..., \mathbf{x}_N\}$, the dynamics of $\rho_N (\{\mathbf{x}_N\},t)$ is given by the Fokker-Planck equation, 
\begin{equation}
    \frac{\partial \rho_N (\{\mathbf{x}_N\},t)}{ \partial t} \Delta t = - \sum_i \partial_{{x}_i, \alpha}(\avg{\Delta x_{i,\alpha}}\rho_N ) + \frac{1}{2} \sum_{i} \sum_j \partial_{{x}_i, \alpha} \partial_{{x}_j, \beta} (\avg{\Delta x_{i,\alpha} \Delta x_{j,\beta}} \rho_N ),
    \label{eq:fp_1}
\end{equation}
where $\Delta x_{i,\alpha} = x_{i,\alpha}(t + \Delta t ) - x_{i,\alpha}(t) $. Using Ito calculus, we evaluate $\avg{\Delta x_{i,\alpha}}$ and $\avg{\Delta x_{i,\alpha} \Delta x_{j,\beta}}$ to get,
\begin{align}
    \avg{\Delta x_{i,\alpha}} &= -\frac{1}{\gamma} \partial_{{x}_i, \alpha} \left(\sum_j
    V_{ji} \right) \Delta t + \mathcal{O}(\Delta t^{3/2}), \nonumber \\
    \avg{\Delta x_{i,\alpha} \Delta x_{j,\beta}} &= \delta_{ij} \sum_k  {\Lambda}_{ki,\alpha \beta}(t) \Delta t + c {\Lambda}_{ij,\alpha \beta}(t) \Delta t + \mathcal{O}(\Delta t^{3/2}).
    \label{eq:moments}
\end{align}
Finally, after neglecting terms of order higher than $\mathcal{O}(\Delta t)$, we get,
\begin{equation}
    \frac{\partial \rho_N (\{\mathbf{x}_N\},t)}{ \partial t} = \frac{1}{\gamma} \sum_i \partial_{{x}_i, \alpha} \left[ \partial_{{x}_i, \alpha} 
    \left(\sum_j
    V_{ji} \right) \rho_N \right] + \frac{1}{2} \sum_{i} \sum_j \partial_{{x}_i, \alpha} \partial_{{x}_j, \beta} \left[ \left( \delta_{ij} \sum_k  {\Lambda}_{ki,\alpha \beta}(t)  \rho_N + c {\Lambda}_{ij,\alpha \beta} \right) \rho_N \right]. 
    \label{eq:fp_2}
\end{equation}
Notice that the $n$-particle density $\rho_n(\mathbf{x}_1,\mathbf{x}_2,...,\mathbf{x}_n)$ is obtained by integrating the $N$-particle density $\rho_N$ over the remaining $N-n$ particle coordinates: $\rho_n(\mathbf{x}_1,\mathbf{x}_2,...,\mathbf{x}_n) = \frac{N!}{(N-n)!}\int \rho_N(\mathbf{x}_1,\mathbf{x}_2,...,\mathbf{x}_N) \ d\mathbf{x}_{n+1}d\mathbf{x}_{n+2} ... d\mathbf{x}_{N}$. Integrating Eq.~\ref{eq:fp_2} over particle coordinates yields a hierarchy of coupled equations analogous to the BBGKY hierarchy in kinetic theory, where the equation for $\rho_n$ depends on $\rho_{n+1}$ \cite{hansen2013theory}. The first equation in this hierarchy, governing the one-particle density $\rho_1(\mathbf{x}_1)$, is given by,
\begin{equation}
    \frac{\partial \rho_1(\mathbf{x}_1, t)}{\partial t} = \frac{1}{\gamma} \partial_{{x}_1, \alpha} \int \partial_{{x}_1, \alpha} (V_{21}) \rho_2(\mathbf{x}_1,\mathbf{x}_2) d\mathbf{x}_2 + \frac{1}{2} \partial_{{x}_1, \alpha} \partial_{{x}_1, \beta} \int \Lambda_{12, \alpha \beta} \ \rho_2(\mathbf{x}_1,\mathbf{x}_2) d\mathbf{x}_2,
    \label{eq:fp_3}
\end{equation}
where we have used the fact that the boundary terms vanish, and $\Lambda_{ii}=0$ (no self-interactions). We now use the mean-field closure approximation $\rho_2(\mathbf{x}_1,\mathbf{x}_2) = \rho_1(\mathbf{x}_1)\rho_1(\mathbf{x}_2)$ on Eq.~\ref{eq:fp_3} to get,
\begin{align}
    \frac{\partial \rho_1(\mathbf{x}_1, t)}{\partial t} &= \frac{1}{\gamma} \partial_{{x}_1, \alpha} \left( \rho_1(\mathbf{x}_1) \int \partial_{{x}_1, \alpha} (V_{21}) \rho_1(\mathbf{x}_2) d\mathbf{x}_2 \right) + \frac{1}{2} \partial_{{x}_1, \alpha} \partial_{{x}_1, \beta} \left( \rho_1(\mathbf{x}_1) \int \Lambda_{12, \alpha \beta} \ \rho_1(\mathbf{x}_2) d\mathbf{x}_2 \right). 
    \label{eq:fp_4}
\end{align}
Since all particles are identical and particle indices $1$ and $2$ are arbitrary, we replace $\mathbf{x}_1 \to \mathbf{x}$ and $\mathbf{x}_2 \to \mathbf{y}$ to get,
\begin{align}
    \frac{\partial \rho(\mathbf{x}, t)}{\partial t} &= \frac{1}{\gamma} \partial_{{x}, \alpha} \left( \rho(\mathbf{x}) \int \rho(\mathbf{y}) \ \partial_{{x}, \alpha} V(\mathbf{x}, \mathbf{y}) d\mathbf{y} \right) + \frac{1}{2} \partial_{{x}, \alpha} \partial_{{x}, \beta} \left( \rho(\mathbf{x}) \int \Lambda (\mathbf{x}, \mathbf{y})_{\alpha \beta} \ \rho(\mathbf{y}) d\mathbf{y} \right). 
    \label{eq:fp_5}
\end{align}
Eq.~\ref{eq:fp_5} is the Fokker-Planck equation describing the coarse-grained density evolution for all random-organizing systems.

\textbf{Linearization.} To gain further analytical insights, we now linearize the density to first order around a spatiotemporally constant mean density $\bar{\rho}$ as $\rho (\mathbf{x}, t) = \bar{\rho} + \delta \rho (\mathbf{x}, t)$. Plugging this into Eq.~\ref{eq:fp_5}, and retaining terms up to the first order, we get the evolution of density fluctuation $\delta \rho (\mathbf{x}, t)$ as,
\begin{align}
    \frac{\partial \delta \rho(\mathbf{x}, t)}{\partial t} 
    &= \frac{\bar{\rho}}{\gamma} \ \partial_{\alpha} \left( \int \delta \rho(\mathbf{y}, t) \partial_{\alpha} V(\mathbf{x} - \mathbf{y}) \, d\mathbf{y} \right) + 
    \frac{ \bar{\rho}}{2} \partial_{\alpha} \partial_{\beta} \left( \int \delta \rho(\mathbf{y}, t) {\Lambda(\mathbf{x}-\mathbf{y})}_{\alpha \beta} \, d\mathbf{y} + \delta \rho(\mathbf{x}, t) \int {\Lambda(\mathbf{x}-\mathbf{y})}_{\alpha \beta} \, d\mathbf{y} \right).
    \label{eq:d_deltarho_dt_fp}
\end{align}
Defining the Fourier transform in space and time for any function $f(\mathbf{x}, t)$ as $\hat{f}(\mathbf{k}, \omega) = \int \int f(\mathbf{x}, t) e^{-i \mathbf{k} \cdot \mathbf{x}} e^{-i \omega t} d \mathbf{x} dt$, and taking Fourier transform of Eq.~\ref{eq:d_deltarho_dt_fp}, we get,
\begin{align}
    i \omega \hat{\delta \rho} (\mathbf{k}, \omega)
    &= - \frac{\bar{\rho}}{\gamma} |\mathbf{k}|^2 \hat{\delta \rho}(\mathbf{k}, \omega) \hat{V}(\mathbf{k}) - 
    \frac{ \bar{\rho}}{2} \hat{\delta \rho}(\mathbf{k}, \omega) k_{\alpha} k_{\beta} \left( \hat{\Lambda}(\mathbf{k})_{\alpha \beta} + A_{\alpha \beta} \right),
    \label{eq:deltarho_ft_fp_1}
\end{align}
where $A_{\alpha \beta} = \int {\Lambda(\mathbf{r})}_{\alpha \beta} \, d\mathbf{r}$. Rearranging Eq.~\ref{eq:deltarho_ft_fp_1}, we get the trivial solution,
\begin{equation}
    \hat{\delta \rho} (\mathbf{k}, \omega)
    = 0.
    \label{eq:deltarho_ft_fp_2}    
\end{equation}
Using Eq.~\ref{eq:deltarho_ft_fp_2}, we now write the dynamic structure factor $S(\mathbf{k}, \omega)$ as,
\begin{align}
    S(\mathbf{k}, \omega)
    &= \frac{1}{2 \pi T N} \left< \hat{\delta \rho} (\mathbf{k}, \omega) \hat{\delta \rho}^* (\mathbf{k}, \omega) \right> = 0
\label{eq:sk_dyn_fp}  
\end{align}
where $\int dt = T$ is the total observation time, and $N = \int \rho (\mathbf{x}, t) \, d \mathbf{x}$ is the total number of particles. We now write the static structure factor $S(\mathbf{k})$ as,
\begin{align}
    S(\mathbf{k}) 
    &= \int_{-\infty}^{\infty} S(\mathbf{k}, \omega) \, d\omega 
    = 0.
\label{eq:sk_static_fp}  
\end{align}
It is evident from Eq.~\ref{eq:deltarho_ft_fp_2} that at steady-state, $\rho(\mathbf{x},t) = \bar{\rho}$, i.e., a spatiotemporally constant density. This leads to the trivial result $S(\mathbf{k}) = 0$, which is inconsistent with particle simulations (Fig.~\ref{fig:figure2}). Thus, the widely used Fokker-Planck method is not sufficient to describe the density evolution and long-range structure for random-organizing systems. This motivates us to use a method which can retain an important feature of our system---the pairwise noise correlation between particles---even at the coarse-grained level.

\subsubsection{Dean's method} 
We now use Dean's method to reach a coarse-grained description. Dean's method was first introduced to systematically coarse-grain an equilibrium system of interacting Brownian particles, starting from an overdamped Langevin equation with an additive white noise term \cite{dean1996langevin}. This method has since been adapted to study active matter systems having multiplicative noise terms \cite{bertin2013mesoscopic, solon2015active}. Here, we extend this method to account for multiplicative noise which is also pairwise correlated between particles. 

We define the density function for a single particle $\rho_{i} (\mathbf{x}, t)$ as
\begin{equation}
    \rho_{i} (\mathbf{x}, t)
    = \delta(\mathbf{x}_i (t) - \mathbf{x} (t)).
    \label{eq:density_1_part}
\end{equation}
The global density $\rho (\mathbf{x}, t)$ can then be written as,
\begin{equation}
    \rho (\mathbf{x}, t)
    = \sum_{i=1}^{N} \delta(\mathbf{x}_i (t) - \mathbf{x} (t)).
    \label{eq:density_global}
\end{equation}
Consider an arbitrary function $f(\mathbf{x}_i (t))$ defined on the coordinate space of the system. 
Using the definition of $\rho_{i} (\mathbf{x}, t)$, we can write,
\begin{equation}
    f(\mathbf{x}_i (t))
    = \int_{\mathbb{R}^d} f(\mathbf{x}) \rho_{i} (\mathbf{x}, t) d\mathbf{x}.
    \label{eq:arbt_func_f}
\end{equation}
Expanding $f(\mathbf{x}_i (t))$ using Eq.~\ref{eq:generic_sde_SI} and Ito's lemma, we get,
\begin{align}
    \frac{df(\mathbf{x}_i (t))}{dt} 
    &= - \frac{1}{\gamma} \sum_{j} \partial_{{x}_i, \alpha} V(\mathbf{x}_i, \mathbf{x}_j) \, \partial_{{x}_i, \alpha} f(\mathbf{x}_i (t)) +
    \sum_{j} \sqrt{\Lambda(\mathbf{x}_i, \mathbf{x}_j)}_{\alpha \beta} \ \xi_{ji, \beta}(t)  \partial_{{x}_i, \alpha} f(\mathbf{x}_i (t)) \nonumber \\
    &\quad + \frac{1}{2} \partial_{{x}_i, \alpha} \partial_{{x}_i, \gamma} f(\mathbf{x}_i (t)) \sum_{j} \sqrt {\Lambda(\mathbf{x}_i, \mathbf{x}_j)}_{\alpha \beta} \sqrt{\Lambda(\mathbf{x}_i, \mathbf{x}_j)}_{\gamma \beta}.
    \label{eq:df_dt_1}
\end{align}
Using Eq.~\ref{eq:arbt_func_f}, the fact that $\sqrt {\Lambda(\mathbf{x}_i, \mathbf{x}_j)}_{\alpha \beta} \sqrt{\Lambda(\mathbf{x}_i, \mathbf{x}_j)}_{\gamma \beta} = {\Lambda(\mathbf{x}_i, \mathbf{x}_j)}_{\alpha \gamma}$, $\partial_{{x}_i, \alpha} f(\mathbf{x}_{i}) = \int \rho_{i}(\mathbf{x}) \partial_{{x}, \alpha} f(\mathbf{x}) d\mathbf{x}$, and that $\partial_{{x}_i, \alpha} \partial_{{x}_i, \gamma} f(\mathbf{x}_{i}) = \int \rho_{i}(\mathbf{x}) \partial_{{x}, \alpha} \partial_{{x}, \gamma} f(\mathbf{x}) d\mathbf{x}$, we get,
\begin{align}
    \frac{df(\mathbf{x}_i (t))}{dt} 
    &= \int d\mathbf{x} \, \rho_{i}(\mathbf{x}, t) \Bigg[ - \frac{1}{\gamma} \sum_{j} \partial_{{x}, \alpha} V(\mathbf{x}, \mathbf{x}_j) \, \partial_{{x}, \alpha} f(\mathbf{x}_i (t)) + \sum_{j} \sqrt{\Lambda(\mathbf{x}, \mathbf{x}_j)}_{\alpha \beta} \ \xi_{ji,\beta}(t) \, \partial_{{x}, \alpha} f(\mathbf{x} (t)) \nonumber \\ 
    &\quad + \frac{1}{2} \partial_{{x}, \alpha} \partial_{{x}, \gamma} f(\mathbf{x} (t)) \sum_{j} {\Lambda(\mathbf{x}, \mathbf{x}_j)}_{\alpha \gamma} \Bigg].
    \label{eq:df_dt_2}
\end{align}
Using integration by parts on Eq.~\ref{eq:df_dt_2} and assuming $f(|\mathbf{x}|\to \infty)=0$, we get,
\begin{align}
    \frac{df(\mathbf{x}_i (t))}{dt} 
    &= \int d\mathbf{x} \, f(\mathbf{x}) \Bigg[ \frac{1}{\gamma} \partial_{\alpha} \left( \rho_{i}(\mathbf{x}) \sum_{j} \partial_{\alpha} V(\mathbf{x}, \mathbf{x}_j) \right) - \partial_{\alpha} \left( \rho_{i}(\mathbf{x}) \sum_{j} \sqrt{ \Lambda(\mathbf{x}, \mathbf{x}_j)}_{\alpha \beta} \ \xi_{ji, \beta}(t) \right) \nonumber \\
    &\quad + \frac{1}{2} \partial_{\alpha} \partial_{\gamma} \left( \rho_{i}(\mathbf{x}) \sum_{j} {\Lambda(\mathbf{x}, \mathbf{x}_j)}_{\alpha \gamma} \right) \Bigg],
    \label{eq:df_dt_3}
\end{align}
where $\partial_{\alpha} = \partial_{{x}, \alpha}$ hereafter. But, from Eq.~\ref{eq:arbt_func_f}, we also have,
\begin{equation}
    \frac{df(\mathbf{x}^i (t))}{dt} 
    = \int d\mathbf{x} \frac{\partial \rho^{i}(\mathbf{x}, t)}{\partial t} f(\mathbf{x}).
    \label{eq:df_dt_4}
\end{equation}
Comparing Eqs.~\ref{eq:df_dt_3} and \ref{eq:df_dt_4}, we get,
\begin{equation}
    \frac{\partial \rho_{i}(\mathbf{x}, t)}{\partial t} 
    = \frac{1}{\gamma} \partial_{\alpha} \left( \rho_{i}(\mathbf{x}) \sum_{j} \partial_{\alpha} V(\mathbf{x}, \mathbf{x}_j) \right) - \partial_{\alpha} \left( \rho_{i}(\mathbf{x}) \sum_{j} \sqrt{ \Lambda(\mathbf{x}, \mathbf{x}_j)}_{\alpha \beta} \ \xi_{ji,\beta}(t) \right) + 
    \frac{1}{2} \partial_{\alpha} \partial_{\gamma} \left( \rho_{i}(\mathbf{x}) \sum_{j} \sqrt {\Lambda(\mathbf{x}, \mathbf{x}_j)}_{\alpha \gamma} \right).
    \label{eq:drho_i_dt}
\end{equation}
Summing Eq.~\ref{eq:drho_i_dt} over $i = 1$ to $N$, we get,
\begin{equation}
    \frac{\partial \rho(\mathbf{x}, t)}{\partial t} 
    = \sum_{i} \frac{1}{\gamma} \partial_{\alpha} \left( \rho_{i}(\mathbf{x}) \sum_{j} \partial_{\alpha} V(\mathbf{x}, \mathbf{x}_j) \right) 
     - \sum_{i} \partial_{\alpha} \left( \rho_{i}(\mathbf{x}) \sum_{j} \sqrt{ \Lambda(\mathbf{x}, \mathbf{x}_j)}_{\alpha \beta} \ \xi_{ji, \beta}(t) \right) + 
    \sum_{i} \frac{1}{2} \partial_{\alpha} \partial_{\gamma} \left( \rho_{i}(\mathbf{x}) \sum_{j} {\Lambda(\mathbf{x}, \mathbf{x}_j)}_{\alpha \gamma} \right).
    \label{eq:drho_dt_1}
\end{equation}
Consider the first term on the right hand side (RHS) of Eq.~\ref{eq:drho_dt_1},
\begin{align}
    \sum_{i} \frac{1}{\gamma} \partial_{\alpha} \left( \rho_{i}(\mathbf{x}) \sum_{j} \partial_{\alpha} V(\mathbf{x}, \mathbf{x}_j) \right) 
    &= \frac{1}{\gamma} \partial_{\alpha} \left( \rho(\mathbf{x}) \sum_{j} \partial_{\alpha} V(\mathbf{x}, \mathbf{x}_j) \right) \nonumber \\
    &= \frac{1}{\gamma} \partial_{\alpha} \left( \rho(\mathbf{x}) \int \sum_{j}  \partial_{\alpha} V(\mathbf{x}, \mathbf{y}) 
    \delta(\mathbf{y} - \mathbf{x}_j) d\mathbf{y} \right) \nonumber \\
    &= \frac{1}{\gamma} \partial_{\alpha} \left( \rho(\mathbf{x}) \int \sum_{j} \partial_{\alpha} V(\mathbf{x}, \mathbf{y}) 
    \rho_{j}(\mathbf{y}) d\mathbf{y} \right) \nonumber \\
    &= \frac{1}{\gamma} \partial_{\alpha} \left( \rho(\mathbf{x}) \int \rho(\mathbf{y})  \partial_{\alpha} V(\mathbf{x}, \mathbf{y}) d\mathbf{y} \right),
    \label{eq:term_1}
\end{align}
where we have used Eq.~\ref{eq:density_1_part}, and the property of the delta function that $f(\mathbf{x}_{i}) \delta(\mathbf{x} - \mathbf{x}_i) = f(\mathbf{x}) \delta(\mathbf{x} - \mathbf{x}_i)$. Consider the third term on the RHS of Eq.~\ref{eq:drho_dt_1},
\begin{align}
    \sum_{i} \frac{1}{2} \partial_{\alpha} \partial_{\gamma} \left( \rho_{i}(\mathbf{x}) \sum_{j} {\Lambda(\mathbf{x}, \mathbf{x}_j)}_{\alpha \gamma} \right) &= 
    \frac{1}{2} \partial_{\alpha} \partial_{\gamma} \left( \rho(\mathbf{x}) \sum_{j} {\Lambda(\mathbf{x}, \mathbf{x}_j)}_{\alpha \gamma} \right) \nonumber \\
    &= \frac{1}{2} \partial_{\alpha} \partial_{\gamma} \left( \rho(\mathbf{x}) \int \sum_{j}  {\Lambda(\mathbf{x}, \mathbf{y})}_{\alpha \gamma} 
    \delta(\mathbf{y} - \mathbf{x}_j) d\mathbf{y} \right) \nonumber \\
    &= \frac{1}{2} \partial_{\alpha} \partial_{\gamma} \left( \rho(\mathbf{x}) \int \sum_{j}  {\Lambda(\mathbf{x}, \mathbf{y})}_{\alpha \gamma} 
    \rho_{j}(\mathbf{y}) d\mathbf{y} \right) \nonumber \\
    &= \frac{1}{2} \partial_{\alpha} \partial_{\gamma} \left( \rho(\mathbf{x}) \int \rho(\mathbf{y})  {\Lambda(\mathbf{x}, \mathbf{y})}_{\alpha \beta} d\mathbf{y} \right).
    \label{eq:term_3}
\end{align}
We next consider the second term on the RHS of Eq.~\ref{eq:drho_dt_1},
\begin{equation}
    \theta(\mathbf{x}, t) = - \sum_{i} \partial_{\alpha} \left( \rho_{i}(\mathbf{x}) \sum_{j} \sqrt{ \Lambda(\mathbf{x}, \mathbf{x}_j)}_{\alpha \beta} \ \xi_{ji,\beta}(t) \right) =
    - \sum_{i} \sum_{j} \partial_{\alpha} \left( \rho_{i}(\mathbf{x}) \int \rho_j(\mathbf{y}) \sqrt {\Lambda(\mathbf{x}, \mathbf{y})}_{\alpha \beta} d\mathbf{y} \ \xi_{ji, \beta}(t) \right).
    \label{eq:term_2}
\end{equation}
Writing the correlation of $\theta(\mathbf{x}, t)$, we get,
\begin{align}
   \avg{\theta(\mathbf{x}, t) \theta(\mathbf{y}, t')} &= \sum_{i} \sum_{j} \sum_{m} \sum_{n} \partial_{x,\alpha} \partial_{y, \mu} \left[  \rho_{i}(\mathbf{x}) \, \rho_{m}(\mathbf{y}) \int \rho_j(\mathbf{u}) \sqrt {\Lambda(\mathbf{x}, \mathbf{u})}_{\alpha \beta} d\mathbf{u} \int \rho_n(\mathbf{w}) \sqrt {\Lambda(\mathbf{y}, \mathbf{w})}_{\mu \nu} d\mathbf{w} \ \avg{\xi_{ji, \beta} (t) \xi_{nm, \nu} (t')} \right] \nonumber \\
   &= \delta(t-t') \sum_{i} \sum_{j} \partial_{x, \alpha} \partial_{y, \mu} \, \left[ \rho_{i}(\mathbf{x}) \, \rho_{i}(\mathbf{y}) \int \rho_j(\mathbf{u}) \sqrt {\Lambda(\mathbf{x}, \mathbf{u})}_{\alpha \beta} d\mathbf{u} \int \rho_j(\mathbf{w}) \sqrt {\Lambda(\mathbf{y}, \mathbf{w})}_{\mu \beta} d\mathbf{w} \right] \nonumber \\
   &\quad + c \ \delta(t-t') \sum_{i} \sum_{j} \partial_{x, \alpha} \partial_{y, \mu} \, \left[ \rho_{i}(\mathbf{x}) \, \rho_{j}(\mathbf{y}) \int \rho_j(\mathbf{u}) \sqrt {\Lambda(\mathbf{x}, \mathbf{u})}_{\alpha \beta} d\mathbf{u} \int \rho_i(\mathbf{w}) \sqrt {\Lambda(\mathbf{y}, \mathbf{w})}_{\mu \beta} d\mathbf{w} \right] \nonumber \\ 
   &= \delta(t-t') \partial_{x, \alpha} \partial_{y, \mu} \, \left[ \delta(\mathbf{x} - \mathbf{y}) \rho(\mathbf{x}) \int \int \rho(\mathbf{u}) \sqrt {\Lambda(\mathbf{x}, \mathbf{u})}_{\alpha \beta} \sqrt {\Lambda(\mathbf{y}, \mathbf{w})}_{\mu \beta} \, \delta(\mathbf{u} - \mathbf{w})  d\mathbf{u} d\mathbf{w} \right] \nonumber \\
   &\quad + c \ \delta(t-t') \partial_{x, \alpha} \partial_{y, \mu} \, \left[ \rho(\mathbf{x}) \, \rho(\mathbf{y}) \int \int \sqrt {\Lambda(\mathbf{x}, \mathbf{u})}_{\alpha \beta} \sqrt {\Lambda(\mathbf{y}, \mathbf{w})}_{\mu \beta} \, \delta(\mathbf{x} - \mathbf{w}) \delta(\mathbf{y} - \mathbf{u}) d\mathbf{u} d\mathbf{w} \right] \nonumber \\ 
   &= \delta(t-t') \partial_{x, \alpha} \partial_{y, \mu} \, \left[ \delta(\mathbf{x} - \mathbf{y}) \rho(\mathbf{x}) \int \rho(\mathbf{u}) {\Lambda(\mathbf{x}, \mathbf{u})}_{\alpha \mu} d\mathbf{u} + c \, \rho(\mathbf{x}) \rho(\mathbf{y}) {\Lambda(\mathbf{x}, \mathbf{y})}_{\alpha \mu} \right], 
    \label{eq:term_2_cov}
\end{align}
where we have used the property of the delta function that $\delta(\mathbf{x} - \mathbf{x}_i) \delta(\mathbf{y} - \mathbf{x}_i) = \delta(\mathbf{x} - \mathbf{x}_i) \delta(\mathbf{x} - \mathbf{y})$. We aim to define a Gaussian noise field having a correlation function identical to Eq.~\ref{eq:term_2_cov} and is thus, statistically identical to $\theta(\mathbf{x}, t)$. Consider a global noise $\chi(\mathbf{x}, t)$ given as,
\begin{equation}
    \chi(\mathbf{x}, t) = \partial_{\alpha} \left( \sqrt{\rho(\mathbf{x})} \int \sqrt{\rho(\mathbf{u})} \sqrt{\Lambda(\mathbf{x}, \mathbf{u})}_{\alpha \mu} \, \eta_{\mu}(\mathbf{x}, \mathbf{u}, t) \, d\mathbf{u} \right),
    \label{eq:global_noise}
\end{equation}
where $\boldsymbol{\eta}(\mathbf{x}, \mathbf{u}, t)$ is a zero mean two-point Gaussian noise with correlation $\left< \eta_{\alpha}(\mathbf{x}, \mathbf{u}, t) \,\eta_{\beta}(\mathbf{y}, \mathbf{w}, t')\right> = \delta_{\alpha \beta} \delta(t-t') \delta(\mathbf{x} - \mathbf{y}) \delta(\mathbf{u} - \mathbf{w}) + c \delta_{\alpha \beta} \delta(t-t') \delta(\mathbf{x} - \mathbf{w}) \delta(\mathbf{u} - \mathbf{y})$. The correlation function of $\chi(\mathbf{x}, t)$ is then given by,
\begin{align}
   \avg{\chi(\mathbf{x}, t) \chi(\mathbf{y}, t')}  
    &= \partial_{x, \alpha} \partial_{y, \mu} \, \left[ \sqrt{\rho(\mathbf{x})\rho(\mathbf{y})} \int \int \sqrt{\rho(\mathbf{u})\rho(\mathbf{w})} \sqrt {\Lambda(\mathbf{x}, \mathbf{u})}_{\alpha \beta} \sqrt {\Lambda(\mathbf{y}, \mathbf{w})}_{\mu \nu} \,  d\mathbf{u} d\mathbf{w} \left< \eta_{\beta}(\mathbf{x}, \mathbf{u}, t) \,\eta_{\nu}(\mathbf{y}, \mathbf{w}, t')\right> \right] \nonumber \\
   &= \delta(t-t') \partial_{x, \alpha} \partial_{y, \mu} \, \left[ \delta(\mathbf{x} - \mathbf{y}) \sqrt{\rho(\mathbf{x})\rho(\mathbf{y})} \int \int \sqrt{\rho(\mathbf{u})\rho(\mathbf{w})} \sqrt {\Lambda(\mathbf{x}, \mathbf{u})}_{\alpha \beta} \sqrt {\Lambda(\mathbf{y}, \mathbf{w})}_{\mu \beta} \, \delta(\mathbf{u} - \mathbf{w})  d\mathbf{u} d\mathbf{w} \right] \nonumber \\
   &\quad + c \delta(t-t') \partial_{x, \alpha} \partial_{y, \mu} \, \left[ \rho(\mathbf{x}) \, \rho(\mathbf{y}) \int \int \sqrt {\Lambda(\mathbf{x}, \mathbf{u})}_{\alpha \beta} \sqrt {\Lambda(\mathbf{y}, \mathbf{w})}_{\mu \beta} \, \delta(\mathbf{x} - \mathbf{w}) \delta(\mathbf{y} - \mathbf{u}) d\mathbf{u} d\mathbf{w} \right] \nonumber \\
   &= \delta(t-t') \partial_{x, \alpha} \partial_{y, \mu} \, \left[ \delta(\mathbf{x} - \mathbf{y}) \rho(\mathbf{x}) \int \rho(\mathbf{u}) {\Lambda(\mathbf{x}, \mathbf{u})}_{\alpha \mu} d\mathbf{u} + c \, \rho(\mathbf{x}) \rho(\mathbf{y}) {\Lambda(\mathbf{x}, \mathbf{y})}_{\alpha \mu} \right]. 
   \label{eq:global_noise_cov}
\end{align}
Eq.~\ref{eq:global_noise_cov} is precisely the same as Eq.~\ref{eq:term_2_cov}, showing that the noise terms $\theta (\mathbf{x}, t)$ and $\chi (\mathbf{x}, t)$ are statistically identical. Combining Eqs.~\ref{eq:term_1}, \ref{eq:term_3}, and \ref{eq:global_noise}, we get the equation for the evolution of global density as,
\begin{align}
    \frac{\partial \rho(\mathbf{x}, t)}{\partial t} 
    &= \frac{1}{\gamma} \ \partial_{\alpha} \left( \rho(\mathbf{x}) \int \rho(\mathbf{y}) \partial_{\alpha} V(\mathbf{x}, \mathbf{y}) d\mathbf{y} \right) + \frac{1}{2} \partial_{\alpha} \partial_{\gamma} \left( \rho(\mathbf{x}) \int \rho(\mathbf{y}) {\Lambda(\mathbf{x}, \mathbf{y})}_{\alpha \beta} d\mathbf{y} \right) \nonumber \\
    &\quad + \partial_{\alpha} \left( \sqrt{\rho(\mathbf{x})} \int \sqrt{\rho(\mathbf{y})} \sqrt{\Lambda(\mathbf{x}, \mathbf{y})}_{\alpha \mu} \, \eta_{\mu}(\mathbf{x}, \mathbf{y}, t) \, d\mathbf{y} \right) \nonumber \\    
    &= - \underbrace{ \nabla \cdot \left[ -\rho(\mathbf{x})  \frac{\left< \nabla V(\mathbf{x}, \mathbf{y}) \right>_{\rho(\mathbf{y})}}{\gamma}  \right] }_\text{drift term} + \underbrace{ \nabla \nabla : \left( \frac{1}{2} \left< {\boldsymbol{\Lambda}(\mathbf{x}, \mathbf{y})} \right>_{\rho(\mathbf{y})}\rho(\mathbf{x})  \right) }_\text{diffusion term}
    - \underbrace{\nabla \cdot \left[ - \sqrt{\rho(\mathbf{x})} \int \sqrt{\rho(\mathbf{y})} \sqrt{\boldsymbol{\Lambda}(\mathbf{x}, \mathbf{y})} \cdot \boldsymbol{\eta}(\mathbf{x}, \mathbf{y}, t) d \mathbf{y} \right] }_\text{noise term},    
    \label{eq:drho_dt_2}
\end{align} 
where $:$ is the double dot product, and $\langle a \rangle_{\rho(\mathbf{y})} = \int a \rho(\mathbf{y}) d\mathbf{y}$. Eq.~\ref{eq:drho_dt_2} is Eq.~\ref{eq:generic_drho} of the main text. Note that the density evolution equation derived using Dean's method (Eq.~\ref{eq:drho_dt_2}) becomes identical to the one derived previously using the Fokker-Planck method (Eq.~\ref{eq:fp_5}, Sec. I.B.I), if the noise term is neglected. 

\textbf{Linearization.} To make further analytical progress, we linearize the density to first order around a spatiotemporally constant mean density $\bar{\rho}$ as $\rho (\mathbf{x}, t) = \bar{\rho} + \delta \rho (\mathbf{x}, t)$. We plug this into Eq.~\ref{eq:drho_dt_2} and retain terms up to the first order for the deterministic (first two on RHS) terms of Eq.~\ref{eq:drho_dt_2}. Further, we only retain the zeroth-order terms in the expansion of the noise (third on RHS) term in Eq.~\ref{eq:drho_dt_2}---since all the other terms in the expansion of the noise become $\mathcal{O} (\delta \rho^2)$ when taking correlations \cite{dean2016nonequilibrium, dean2014relaxation, demery2014generalized, dinelli2024fluctuating}. Notice that while the zeroth order contribution vanishes for the for the deterministic (first two on RHS) terms, it is non-zero for the noise (third on RHS) term (Eq.~\ref{eq:drho_dt_2}). We can then write the evolution of density fluctuation $\delta \rho (\mathbf{x}, t)$ as,
\begin{align}
    \frac{\partial \delta \rho(\mathbf{x}, t)}{\partial t} 
    &= \frac{\bar{\rho}}{\gamma} \ \partial_{\alpha} \left( \int \delta \rho(\mathbf{y}, t) \partial_{\alpha} V(\mathbf{x} - \mathbf{y}) \, d\mathbf{y} \right) + 
    \frac{ \bar{\rho}}{2} \partial_{\alpha} \partial_{\beta} \left( \int \delta \rho(\mathbf{y}, t) {\Lambda(\mathbf{x}-\mathbf{y})}_{\alpha \beta} \, d\mathbf{y} + \delta \rho(\mathbf{x}, t) \int {\Lambda(\mathbf{x}-\mathbf{y})}_{\alpha \beta} \, d\mathbf{y} \right) \nonumber \\
    &\quad + \bar{\rho} \, \partial_{\alpha} \left( \int \sqrt{\Lambda(\mathbf{x}-\mathbf{y})}_{\alpha \mu} \, \eta_{\mu}(\mathbf{x}, \mathbf{y}, t) \, d\mathbf{y} \right). 
    \label{eq:d_deltarho_dt}
\end{align}
We now define the Fourier transform in space and time for any function $f(\mathbf{x}, t)$ as $\hat{f}(\mathbf{k}, \omega) = \int \int f(\mathbf{x}, t) e^{-i \mathbf{k} \cdot \mathbf{x}} e^{-i \omega t} d \mathbf{x} dt$. Taking Fourier transform of Eq.~\ref{eq:d_deltarho_dt}, we get,
\begin{align}
    i \omega \hat{\delta \rho} (\mathbf{k}, \omega)
    &= - \frac{\bar{\rho}}{\gamma} |\mathbf{k}|^2 \hat{\delta \rho}(\mathbf{k}, \omega) \hat{V}(\mathbf{k}) - 
    \frac{ \bar{\rho}}{2} \hat{\delta \rho}(\mathbf{k}, \omega) k_{\alpha} k_{\beta} \left( \hat{\Lambda}(\mathbf{k})_{\alpha \beta} + A_{\alpha \beta} \right) \nonumber \\
    &\quad + \bar{\rho} i k_{\alpha} \, \int e^{-i \mathbf{k} \cdot \mathbf{x}} d\mathbf{x} \int e^{-i \omega t} dt \int \sqrt{\Lambda(\mathbf{x}-\mathbf{y})}_{\alpha \mu} \, \eta_{\mu}(\mathbf{x}, \mathbf{y}, t) \, d\mathbf{y},
    \label{eq:deltarho_ft_1}
\end{align}
where $A_{\alpha \beta} = \int {\Lambda(\mathbf{r})}_{\alpha \beta} \, d\mathbf{r}$. Rearranging Eq.~\ref{eq:deltarho_ft_1}, we get,
\begin{equation}
    \hat{\delta \rho} (\mathbf{k}, \omega)
    = \frac{\bar{\rho} i k_{\alpha} \, \int e^{-i \mathbf{k} \cdot \mathbf{x}} d\mathbf{x} \int e^{-i \omega t} dt \int \sqrt{\Lambda(\mathbf{x}-\mathbf{y})}_{\alpha \mu} \, \eta_{\mu}(\mathbf{x}, \mathbf{y}, t) \, d\mathbf{y}} {i\omega + \bar{\rho} \left[ \frac{1}{\gamma} |\mathbf{k}|^2 \hat{V}(\mathbf{k}) + \frac{1}{2} k_{\alpha} k_{\beta} \left( \hat{\Lambda}(\mathbf{k})_{\alpha \beta} + A_{\alpha \beta} \right) \right]}.
    \label{eq:deltarho_ft_2}    
\end{equation}

\textbf{Structure factor.} Defining $f^*$ as the complex conjugate of any function $f$ and using Eq.~\ref{eq:deltarho_ft_2}, we can write the dynamic structure factor $S(\mathbf{k}, \omega)$ as,
\begin{align}
    S(\mathbf{k}, \omega)
    &= \frac{1}{2 \pi T N} \left< \hat{\delta \rho} (\mathbf{k}, \omega) \hat{\delta \rho}^* (\mathbf{k}, \omega) \right> \nonumber \\
    &= \frac{ \bar{\rho}^2 k_{\alpha} k_{\beta} \left< \int e^{-i \mathbf{k} \cdot \mathbf{x}} d\mathbf{x} \int e^{-i \omega t} dt \int \sqrt{\Lambda(\mathbf{x}-\mathbf{y})}_{\alpha \mu} \, \eta_{\mu}(\mathbf{x}, \mathbf{y}, t) \, d\mathbf{y} \int e^{i \mathbf{k} \cdot \mathbf{x'}} d\mathbf{x'} \int e^{i \omega t'} dt' \int \sqrt{\Lambda(\mathbf{x'}-\mathbf{y'})}_{\beta \delta} \, \eta_{\delta}(\mathbf{x'}, \mathbf{y'}, t') \, d\mathbf{y'} \right>}{2 \pi T N \left( \omega^2 + \bar{\rho}^2 \left[ \frac{1}{\gamma} |\mathbf{k}|^2 \hat{V}(\mathbf{k}) + \frac{1}{2} k_{\alpha} k_{\beta} \left( \hat{\Lambda}(\mathbf{k})_{\alpha \beta} + A_{\alpha \beta} \right) \right]^2 \right)} \nonumber \\
    &= \frac{ \bar{\rho}^2 k_{\alpha} k_{\beta} \int \int \int \int \int \int e^{-i \mathbf{k} \cdot (\mathbf{x}-\mathbf{x'})} e^{-i \omega (t-t')} \sqrt{\Lambda(\mathbf{x}-\mathbf{y})}_{\alpha \mu} \sqrt{\Lambda(\mathbf{x'}-\mathbf{y'})}_{\beta \delta} \left< \eta_{\mu}(\mathbf{x}, \mathbf{y}, t) \eta_{\delta}(\mathbf{x'}, \mathbf{y'}, t') \right> d\mathbf{x} d\mathbf{y} d\mathbf{x'} d\mathbf{y'} dt dt' }{2 \pi T N \left( \omega^2 + \bar{\rho}^2 \left[ \frac{1}{\gamma} |\mathbf{k}|^2 \hat{V}(\mathbf{k}) + \frac{1}{2} k_{\alpha} k_{\beta} \left( \hat{\Lambda}(\mathbf{k})_{\alpha \beta} + A_{\alpha \beta} \right) \right]^2 \right) } \nonumber \\
    &= \frac{ \bar{\rho}^2 k_{\alpha} k_{\beta} \int \int \int \int \int \int e^{-i \mathbf{k} \cdot (\mathbf{x}-\mathbf{x'})} e^{-i \omega (t-t')} \sqrt{\Lambda(\mathbf{x}-\mathbf{y})}_{\alpha \mu} \sqrt{\Lambda(\mathbf{x'}-\mathbf{y'})}_{\beta \delta} \delta_{\mu \delta} \delta(t-t') \delta(\mathbf{x} - \mathbf{x'}) \delta(\mathbf{y} - \mathbf{y'}) d\mathbf{x} d\mathbf{y} d\mathbf{x'} d\mathbf{y'} dt dt' }{2 \pi T N \left( \omega^2 + \bar{\rho}^2 \left[ \frac{1}{\gamma} |\mathbf{k}|^2 \hat{V}(\mathbf{k}) + \frac{1}{2} k_{\alpha} k_{\beta} \left( \hat{\Lambda}(\mathbf{k})_{\alpha \beta} + A_{\alpha \beta} \right) \right]^2 \right) } \nonumber \\
    &\quad + \frac{ \bar{\rho}^2 k_{\alpha} k_{\beta} c \int \int \int \int \int \int e^{-i \mathbf{k} \cdot (\mathbf{x}-\mathbf{x'})} e^{-i \omega (t-t')} \sqrt{\Lambda(\mathbf{x}-\mathbf{y})}_{\alpha \mu} \sqrt{\Lambda(\mathbf{x'}-\mathbf{y'})}_{\beta \delta} \delta_{\mu \delta} \delta(t-t') \delta(\mathbf{x} - \mathbf{y'}) \delta(\mathbf{x'} - \mathbf{y})  d\mathbf{x} d\mathbf{y} d\mathbf{x'} d\mathbf{y'} dt dt' }{2 \pi T N \left( \omega^2 + \bar{\rho}^2 \left[ \frac{1}{\gamma} |\mathbf{k}|^2 \hat{V}(\mathbf{k}) + \frac{1}{2} k_{\alpha} k_{\beta} \left( \hat{\Lambda}(\mathbf{k})_{\alpha \beta} + A_{\alpha \beta} \right) \right]^2 \right) } \nonumber \\
    &= \frac{ \bar{\rho} k_{\alpha} k_{\beta} \left( A_{\alpha \beta} + c \hat{\Lambda}(\mathbf{k})_{\alpha \beta} \right) } {2 \pi \left( \omega^2 + \bar{\rho}^2 \left[ \frac{1}{\gamma} |\mathbf{k}|^2 \hat{V}(\mathbf{k}) + \frac{1}{2} k_{\alpha} k_{\beta} \left( \hat{\Lambda}(\mathbf{k})_{\alpha \beta} + A_{\alpha \beta} \right) \right]^2 \right) },
\label{eq:sk_dyn}  
\end{align}
where $\int dt = T$ is the total observation time, $N = \int \rho (\mathbf{x}, t) \, d \mathbf{x}$ is the total number of particles, and we have used the fact that $\bar{\rho} = N/\int d \mathbf{x}$. We now write the static structure factor $S(\mathbf{k})$ as,
\begin{align}
    S(\mathbf{k}) 
    &= \int S(\mathbf{k}, \omega) \, d\omega 
    = \frac{ k_{\alpha} k_{\beta} \left( A_{\alpha \beta} + c \hat{\Lambda}(\mathbf{k})_{\alpha \beta} \right) } {2 \left[ \frac{1}{\gamma} |\mathbf{k}|^2 \hat{V}(\mathbf{k}) + \frac{1}{2} k_{\alpha} k_{\beta} \left( \hat{\Lambda}(\mathbf{k})_{\alpha \beta} + A_{\alpha \beta} \right) \right] }.
\label{eq:sk_static_1}  
\end{align}
Eq.~\ref{eq:sk_static_1} shows that $S(\mathbf{k})$ depends on the specific functional forms of $A_{\alpha \beta} = \int {\Lambda(\mathbf{r})}_{\alpha \beta} \, d\mathbf{r}$, $\hat{\Lambda}(\mathbf{k})_{\alpha \beta}$, and $\hat{V}(\mathbf{k})$. Since we are only interested in the long-range ($|\mathbf{k}| \to 0$) behavior of $S(\mathbf{k})$, we can approximate $\hat{V}(\mathbf{k})$ as,
\begin{align}
    \hat{V}(\mathbf{k})
    &= \int V(\mathbf{r}) e^{-i\mathbf{k} \cdot \mathbf{r}} d\mathbf{r}
    \approx \int V(\mathbf{r}) \left( 1 - ik_{\alpha}r_{\alpha} - \frac{k_{\alpha}r_{\alpha}k_{\beta}r_{\beta}}{2} \right) d\mathbf{r}
    = \int V(r) d\mathbf{r} - \frac{k_{\alpha}k_{\beta}}{2} \int V(r) r^2 n_{\alpha} n_{\beta} d\mathbf{r} \nonumber \\
    &= \underbrace{S_{d} \int V(r) r^{d-1} dr}_{V_1} - k^2 \underbrace{\frac{S_{d}}{2d} \int V(r) r^{d+1} dr}_{V_2}
    = V_1 - k^2 V_2,
\label{eq:V_ft}  
\end{align}
where $k = |\mathbf{k}|$, $r = |\mathbf{r}|$, $\mathbf{n} = \mathbf{r}/|\mathbf{r}|$ is the unit vector, $V_1$ and $V_2$ are constants, $S_d$ is the surface area of a unit $d$-dimensional hypersphere. We have also used the fact that $V(\mathbf{r})$ only depends on $r$ and is an even function of $r$---true for RO, BRO, and SGD (Eqs.~\ref{eq:v_ji_bro}, and \ref{eq:v_ji_family_SI}).

Notice that $\mathbf{\Lambda}(\mathbf{r})$ for RO, BRO, and SGD can be generically written as $\Lambda_{\alpha \beta}(\mathbf{r}) = f_1 (r) \delta_{\alpha \beta} + f_2(r) n_{\alpha} n_{\beta}$, where $\mathbf{n} = \mathbf{r}/|\mathbf{r}|$ is the unit vector (Eqs.~\ref{eq:sde_ro}, \ref{eq:sde_bro}, and \ref{eq:sde_sgd}). Specifically, (i) for RO, $f_1 (r)= \mathbf{1}_{(0, 2R)}(r_{ij}) \ \epsilon^2/3d \tau$, and $f_2 (r) = 0$, (ii) for BRO, $f_1 (r)=0$, and $f_2 (r) = \mathbf{1}_{(0, 2R)}(r_{ij}) \ \epsilon^2/12 \tau$, and (iii) for SGD, $f_1 (r)=0$, and $f_2 (r) = \mathbf{1}_{(0, 2R)}(r_{ij}) \ [E^2\alpha^2b_f(1-b_f)/4R^2\tau] (1-r/2R)^{2p-2}$. Then,
\begin{align}
    A_{\alpha \beta} 
    &= \int {\Lambda(\mathbf{r})}_{\alpha \beta} d\mathbf{r}
    = \int \int \left[ f_1 (r) \delta_{\alpha \beta} + f_2(r) n_{\alpha} n_{\beta} \right] r^{d-1} dS dr
    = \delta_{\alpha \beta} \underbrace{S_{d} \int \left( f_1 (r) + \frac{f_2 (r)}{d} \right) r^{d-1} dr}_{A_1}
    = A_1 \delta_{\alpha \beta},
\label{eq:A}  
\end{align}
where $A_1$ is a constant. Since we are only interested in the long-range ($|\mathbf{k}| \to 0$) behavior of $S(\mathbf{k})$, we approximate $\hat{\Lambda}(\mathbf{k})_{\alpha \beta}$ as,
\begin{align}
    \hat{\Lambda}(\mathbf{k})_{\alpha \beta}
    &= \int \Lambda(\mathbf{r})_{\alpha \beta} e^{-i\mathbf{k} \cdot \mathbf{r}} d\mathbf{r}
    \approx \int \Lambda(\mathbf{r})_{\alpha \beta} \left( 1 - ik_{\gamma}r_{\gamma} - \frac{k_{\gamma}r_{\gamma}k_{\delta}r_{\delta}}{2} \right) d\mathbf{r}
    = A_{\alpha \beta} - \frac{k_{\gamma} k_{\delta}}{2} \int \Lambda(\mathbf{r})_{\alpha \beta} r^2 n_{\gamma} n_{\delta} d\mathbf{r} \nonumber \\
    &= A_1 \delta_{\alpha \beta} - \frac{k_{\gamma} k_{\delta}}{2} \int \int \left[ f_1 (r) \delta_{\alpha \beta} + f_2(r) n_{\alpha} n_{\beta} \right] r^{d+1} n_{\gamma} n_{\delta} dS dr \nonumber \\
    &= A_1 \delta_{\alpha \beta} - k^2 \delta_{\alpha \beta} \underbrace{\frac{S_d}{2d} \int f_1 (r) r^{d+1} dr}_{\lambda_1} - \left( k^2 \delta_{\alpha \beta} + 2 k_{\alpha} k_{\beta} \right) \underbrace{\frac{S_d}{2d(d+2)} \int f_2(r) r^{d+1} dr}_{\lambda_2} \nonumber \\
    &= A_1 \delta_{\alpha \beta} - \lambda_1 k^2 \delta_{\alpha \beta} - \lambda_2 \left( k^2 \delta_{\alpha \beta} + 2 k_{\alpha} k_{\beta} \right),
\label{eq:lamda_ft}  
\end{align}
where $\lambda_1$ and $\lambda_2$ are constants, and we have used the fact that $\Lambda(\mathbf{r})_{\alpha \beta}$ only depends on $r$ and is an even function of $r$---true for RO, BRO, and SGD (Eqs.~\ref{eq:sde_ro}, \ref{eq:sde_bro}, and \ref{eq:sde_sgd}). Finally, we combine Eqs.~\ref{eq:V_ft}, \ref{eq:A}, and \ref{eq:lamda_ft} with Eq.~\ref{eq:sk_static_1} to get,
\begin{align}
    S(\mathbf{k}) 
    &= \frac{ k_{\alpha} k_{\beta} \left( A_1 \delta_{\alpha \beta} + c \left[ A_1 \delta_{\alpha \beta} - \lambda_1 k^2 \delta_{\alpha \beta} - \lambda_2 \left( k^2 \delta_{\alpha \beta} + 2 k_{\alpha} k_{\beta} \right) \right] \right) } {2 \left[ \frac{1}{\gamma} k^2 (V_1 - k^2 V_2) + \frac{1}{2} k_{\alpha} k_{\beta} \left( A_1 \delta_{\alpha \beta} - \lambda_1 k^2 \delta_{\alpha \beta} - \lambda_2 \left( k^2 \delta_{\alpha \beta} + 2 k_{\alpha} k_{\beta} \right) + A_1 \delta_{\alpha \beta} \right) \right] } \nonumber \\
    &= \frac{ \left[ (1+c)A_1 - c(\lambda_1 + 3 \lambda_2)k^2 \right]} {2 \left[ \frac{1}{\gamma} \left( V_1 - V_2 k^2 \right) + \frac{1}{2} \left( 2A_1 - (\lambda_1 + 3\lambda_2)k^2 \right) \right]} \nonumber \\
    &= \frac{1}{2} \left[ (1+c)A_1 - c(\lambda_1 + 3 \lambda_2)k^2 \right] \left[ \left(\frac{V_1}{\gamma} + A_1 \right) - \left( \frac{V_2}{\gamma} + \frac{(\lambda_1 + 3\lambda_2)}{2} \right) k^2 \right]^{-1} \nonumber \\
    &= \frac{1}{2 \left(\frac{V_1}{\gamma A_1} + 1 \right)} \left[ (1+c) - \frac{c(\lambda_1 + 3 \lambda_2)}{A_1} k^2 \right] \left[ 1 + \left( \frac{ \frac{V_2}{\gamma} + \frac{(\lambda_1 + 3\lambda_2)}{2}}{\frac{V_1}{\gamma} + A_1 } \right) k^2 + \mathcal{O}(k^4) \right] \nonumber \\
    &= \frac{1}{2 \left(\frac{V_1}{\gamma A_1} + 1 \right)} \left[ (1+c) + \left( \frac{ \frac{V_2}{\gamma} + \frac{(\lambda_1 + 3\lambda_2)}{2}}{\frac{V_1}{\gamma} + A_1 } \right) \left( 1 + \left( 1 - \frac{\frac{V_1}{\gamma A_1} + 1 }{\frac{V_2}{\gamma (\lambda_1 + 3 \lambda_2)}+ \frac{1}{2}}  \right) c \right) k^2 + \mathcal{O}(k^4) \right].
\label{eq:sk_static_2}  
\end{align}
\begin{figure}[htbp!]
    \centering
    \includegraphics[width=\linewidth]{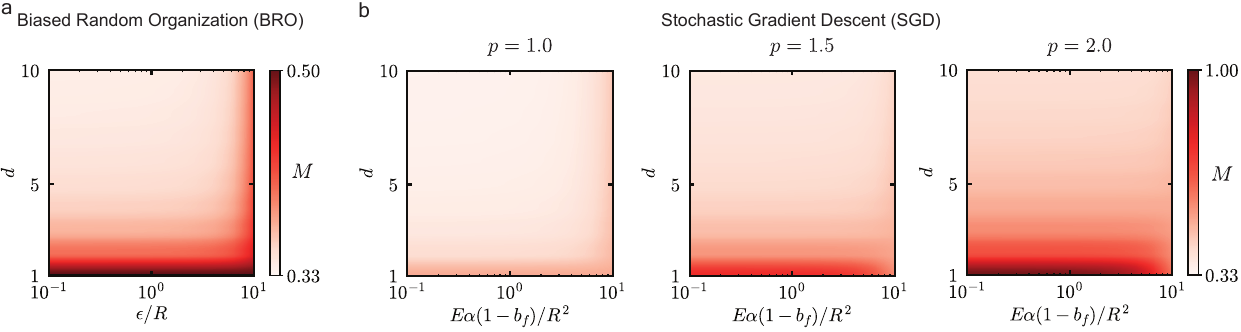}
    \caption{\small (a) $M$ (Eq.~\ref{eq:M_bro}) for Biased Random Organization (BRO) as a function of spatial dimension $d$ and $\epsilon/R$, where $R$ is the particle radius and $\epsilon$ is the kick magnitude. (b) $M$ (Eq.~\ref{eq:M_sgd}) for Stochastic Gradient Descent (SGD) as a function of spatial dimension $d$ and $E \alpha (1-b_f) / R^2$ for varying stiffness $p$ of potential (Eq.~\ref{eq:v_ji_family_SI}). $R$ is the particle radius, $E$ is the characteristic energy scale, $\alpha$ is the learning rate, and $b_f$ is the batch fraction. 
}
    \label{fig:figure1_si}
\end{figure}
We normalize $S$ by the value of $S$ for uncorrelated noise ($c=0$) when $k \to 0$ to get $\tilde{S}(k) = S(k)/S_0$, where 
\begin{equation}
    S_0 = \frac{1}{2 \left(\frac{V_1}{\gamma A_1} + 1 \right)}.
\label{eq:s0}  
\end{equation}
We further normalize $k$ by the value of $k$ for which $\tilde{S} (c=-1) = \lim_{k\to0} \tilde{S} (c=0) = 1.0$ to get $\tilde{k} = k/k_0$, where 
\begin{equation}
    k_0 = \sqrt{ \frac{A_1}{\lambda_1 + 3 \lambda_2}}.
\label{eq:k0}  
\end{equation}
We then get,
\begin{equation}
    \tilde{S}(\tilde{k}) 
    = (1+c) + \left[ \underbrace{\frac{\frac{V_2}{\gamma (\lambda_1 + 3 \lambda_2)}+ \frac{1}{2} }{\frac{V_1}{\gamma A_1} + 1 }}_{M} \left( 1+c \right) - c \right] \tilde{k}^2 + \mathcal{O}(\tilde{k}^4) = (1+c) + [M(1+c) - c] \tilde{k}^2 + \mathcal{O}(\tilde{k}^4),
\label{eq:sk_static_3}  
\end{equation}
where $M$ is a system-specific constant. Neglecting the $\mathcal{O}(\tilde{k}^4)$ term, Eq.~\ref{eq:sk_static_3} yields Eq.~\ref{eq:struc_fact_sim} of the main text. 

We now calculate $M$ for RO, BRO, and SGD. For RO, $V_1^{\text{RO}} = 0$ and $V_2^{\text{RO}} = 0$. Using the definition of $M$, we then get,
\begin{equation}
    M^{\text{RO}} = \frac{1}{2}.
\label{eq:M_ro}  
\end{equation}
For BRO,
\begin{align}
    \gamma^{\text{BRO}} &= \frac{E \tau}{\epsilon R}, \nonumber \\
    V_1^{\text{BRO}} &= \frac{E S_d (2R)^d}{d(d+1)}, \nonumber  \\
    V_2^{\text{BRO}} &= \frac{E S_d (2R)^{d+2}}{2d(d+2)(d+3)}, \nonumber \\
    A_1^{\text{BRO}} &= \frac{S_d \epsilon^2 (2R)^d}{12 \tau d^2}, \nonumber
 \\
     \lambda_1^{\text{BRO}} &= 0, \nonumber \\
     \lambda_2^{\text{BRO}} &= \frac{S_d (2R)^{d+2}, \epsilon^2}{24 d (d+2)^2 \tau}.
    \label{eq:consts_bro}  
\end{align}
Using the definition of $M$, we then get,
\begin{equation}
    M^{\text{BRO}} = \frac{(d+1) [8 (d+2) + (d+3) \frac{\epsilon}{R}]}{2 (d+3) [12 d + (d+1) \frac{\epsilon}{R}]}.
\label{eq:M_bro}  
\end{equation}
Fig.~\ref{fig:figure1_si} shows that $M^{\text{BRO}} \in (\frac{1}{3}, \frac{1}{2})$ for arbitrary spatial dimension $d$ and $\epsilon/R$, where $R$ is the particle radius and $\epsilon$ is the kick magnitude. This is also evident by taking the limits $\lim_{\epsilon/R \to 0}$ and $\lim_{\epsilon/R \to \infty}$ in Eq.~\ref{eq:M_bro}. Similarly, for SGD,
\begin{align}
    \gamma^{\text{SGD}} &= \frac{\tau}{\alpha b_f}, \nonumber \\
    V_1^{\text{SGD}} &= \frac{E S_d (2R)^d \Gamma(p+1) \Gamma(d)}{p\Gamma(p+d+1)}, \nonumber \\
    V_2^{\text{SGD}} &= \frac{E S_d (2R)^{d+2} \Gamma(p+1) \Gamma(d+2)}{2 d p \ \Gamma(p+d+3)}, \nonumber \\
    A_1^{\text{SGD}} &= \frac{S_d E^2 \alpha^2 b_f (1-b_f) (2R)^d \Gamma(2p-1) \Gamma(d)}{4 R^2 \tau d \ \Gamma(2p+d-1)}, \nonumber
 \\
     \lambda_1^{\text{SGD}} &= 0, \nonumber \\
     \lambda_2^{\text{SGD}} &= \frac{S_d E^2 \alpha^2 b_f (1-b_f) (2R)^{d+2} \Gamma(2p-1) \Gamma(d+2)}{8 d (d+2) R^2 \tau \ \Gamma(2p+d+1)},
    \label{eq:consts_sgd}  
\end{align}
where $\Gamma$ is the Gamma function. Using the definition of $M$, we then get,
\begin{equation}
    M^{\text{SGD}} = \frac{\frac{E \alpha (1-b_f)}{R^2} 3 p  \Gamma(2p-1) \Gamma(p+d+3) + 8 (d+2) \Gamma(p+1) \Gamma(2p+d+1)}{6 (p+d+2)(p+d+1) \left[\frac{E \alpha (1-b_f)}{R^2} p  \Gamma(2p-1) \Gamma(p+d+1) + 4 d \Gamma(p+1) \Gamma(2p+d-1) \right]}.
\label{eq:M_sgd}  
\end{equation}
Fig.~\ref{fig:figure1_si} shows that for a linear potential ($p=1$ in Eq.~\ref{eq:v_ji_family_SI}), $M^{\text{SGD}} \in (\frac{1}{3}, \frac{1}{2})$ for arbitrary spatial dimension $d$ and $E \alpha (1-b_f) / R^2$, where $R$ is the particle radius, $E$ is the characteristic energy scale, $\alpha$ is the learning rate, and $b_f$ is the batch fraction. In general, for any non-linear potential ($p > 1$ in Eq.~\ref{eq:v_ji_family_SI}), $M^{\text{SGD}} \in (\frac{1}{3}, 4)$ for arbitrary spatial dimension $d$ and $E \alpha (1-b_f) / R^2$ (Fig.~\ref{fig:figure1_si}). This is also evident by taking the limits $\lim_{E \alpha (1-b_f) / R^2 \to 0}$ and $\lim_{E \alpha (1-b_f) / R^2 \to \infty}$ in Eq.~\ref{eq:M_sgd}. 

\subsection{Relationship between the structure and flatness of energy minima in SGD}

Given a particle configuration ($\mathbf{X} \equiv \{ \mathbf{x}_1, \mathbf{x}_2,...,\mathbf{x}_N \}$) at an energy minimum having energy $E({\mathbf{X}}) = \sum_i \sum_{j>i} V_{ij}$, we aim to relate the change in energy $\Delta E({\mathbf{X}}) = \langle E(\mathbf{X} + \Delta \mathbf{X}) - E({\mathbf{X}}) \rangle $ under a small Gaussian perturbation $\Delta \mathbf{X}$ to the structure factor $S(\mathbf{k})$ of the unperturbed configuration $\mathbf{X}$. $\Delta E({\mathbf{X}})$ is a measure of the flatness of the energy minimum since, given $V_{ij}$ is second-order smooth, $\Delta E({\mathbf{X}}) \propto \text{Tr}(\mathbf{H}({\mathbf{X}}))$, where $\mathbf{H({\mathbf{X}}})$ is the Hessian matrix \cite{jastrzkebski2017three, zhang2024absorbing}.

Since our system is in steady state, we are interested in the average energy over an ensemble of configurations $\left< . \right>_{\mathbf{X}}$. We can then write \cite{hansen2013theory},
\begin{equation}
    E = \left< E({\mathbf{X}}) \right>_{\mathbf{X}} = \frac{\rho N}{2} \int V(\mathbf{r}) g(\mathbf{r}) d \mathbf{r},
\label{eq:energy_1}  
\end{equation}
where $g(\mathbf{r})$ is the pair correlation function, $V(\mathbf{r}) = V(|\mathbf{r}_j - \mathbf{r}_i|)$ is the pairwise potential, $\rho$ is the number density, and $N$ is the total number of particles. Using $g(\mathbf{r}) - 1 = (1/\rho (2 \pi)^d) \int [S(\mathbf{k}) - 1] e^{i \mathbf{k} \cdot \mathbf{r}} d \mathbf{k}$, we get \cite{torquato2015ensemble},
\begin{align}
    E &= \frac{\rho N}{2} \int V(\mathbf{r}) d \mathbf{r} + \frac{\rho N}{2} \int V(\mathbf{r}) \left( g(\mathbf{r}) - 1 \right)  d \mathbf{r} \nonumber \\
    &= \frac{\rho N}{2} \int V(\mathbf{r}) d \mathbf{r} + \frac{N}{2 (2 \pi)^d} \int \hat{V}(\mathbf{k}) \left( S(\mathbf{k}) - 1 \right)  d \mathbf{k} \nonumber \\
    &= \frac{\rho N}{2} \int V(\mathbf{r}) d \mathbf{r} - \frac{N}{2} V(\mathbf{r}=\mathbf{0}) + \frac{N}{2 (2 \pi)^d} \int \hat{V}(\mathbf{k}) S(\mathbf{k}) d \mathbf{k},
\label{eq:energy_2}  
\end{align}
where we have used Parseval's theorem, the fact that $V(\mathbf{r})$ is an even function of $\mathbf{r}$, and that for any function $f(\mathbf{r})$, $1/(2 \pi)^d \int \hat{f}(\mathbf{k}) d \mathbf{k} = f(\mathbf{r} = \mathbf{0})$.

We now perturb the configuration $\mathbf{X} \equiv \{ \mathbf{x}_1, \mathbf{x}_2,...,\mathbf{x}_N \}$ by adding an independent Gaussian noise $\mathbf{N}_i(\mathbf{0}, \sigma^2 \mathbf{I})$ to the position of each particle to get $\mathbf{X} + \Delta \mathbf{X} \equiv \{ \mathbf{x}_1+\mathbf{N}_1, \mathbf{x}_2+\mathbf{N}_2,...,\mathbf{x}_N+\mathbf{N}_N \}$. The new structure factor $S'(\mathbf{k})$ of the system after the addition of independent noise to all particles is given by \cite{gabrielli2004point, kim2018effect, casiulis2024gyromorphs},
\begin{align}
    S'(\mathbf{k}) &= 1 + |F(\mathbf{k})|^2 \left( S(\mathbf{k}) - 1 \right) \nonumber \\
    &= 1 + e^{-\sigma^2 |\mathbf{k}|^2} \left( S(\mathbf{k}) - 1 \right) \nonumber \\
    &= 1 + (1-\sigma^2k^2 + \mathcal{O}(k^4)) \left( S(\mathbf{k}) - 1 \right) \nonumber \\
    &= \sigma^2k^2 + \left( 1 - \sigma^2k^2 \right) S(\mathbf{k}),
\label{eq:s_noise}  
\end{align}
where $F(\mathbf{k}) = e^{-(\sigma^2/2) |\mathbf{k}|^2}$ is the characteristic function of distribution of noise added to particle positions (here, Gaussian noise), and we have dropped terms of order higher than $\mathcal{O}(k^2)$. Using Eqs.~\ref{eq:energy_2} and \ref{eq:s_noise}, we can now write $\Delta E$ as,
\begin{align}
   \Delta E &= \left< E({\mathbf{X}+\Delta \mathbf{X}}) - E({\mathbf{X}}) \right>_{\mathbf{X}, \mathbf{N}} \nonumber \\
   &= \frac{N}{2 (2 \pi)^d} \int \hat{V}(\mathbf{k}) \left[ S'(\mathbf{k}) - S(\mathbf{k}) \right] d \mathbf{k},  \nonumber \\
   &= \frac{N \sigma^2}{2 (2 \pi)^d} \int \hat{V}(\mathbf{k}) k^2 \left[ 1 - S(\mathbf{k}) \right] d \mathbf{k}.
\label{eq:delta_E_SI_1}  
\end{align}
Given a system quantified by $S(\mathbf{k})$, Eq.~\ref{eq:delta_E_SI_1} relates $S(\mathbf{k})$ to the energy change $\Delta E$ when the system is perturbed by a small amount. 

We now aim to relate $\Delta E$ to the Pearson correlation coefficient $c$ between the pairwise noise. To proceed, we combine $S(\mathbf{k})$ derived from the linearized fluctuating hydrodynamic theory (Eq.~\ref{eq:sk_static_3}) with Eq.~\ref{eq:delta_E_SI_1} to get,
\begin{align}
   \Delta E (c) &= \frac{N \sigma^2}{2 (2 \pi)^d} \int \hat{V}(\mathbf{k}) k^2 \left( 1 - S_0 - cS_0 - \frac{MS_0}{k_0^2}k^2 - \frac{cMS_0}{k_0^2}k^2 + \frac{cS_0}{k_0^2}k^2 \right) d \mathbf{k} \nonumber \\
   &= \underbrace{ \left( \frac{N \sigma^2}{2 (2 \pi)^d} \int \hat{V}(\mathbf{k}) k^2 \left( 1 - S_0 - \frac{MS_0}{k_0^2}k^2 \right) d \mathbf{k} \right)}_{\Delta E_1} + c \underbrace{\left( \frac{N \sigma^2 S_0}{2 (2 \pi)^d} \int \hat{V}(\mathbf{k}) k^2 \left[ \left( 1-M \right) \frac{k^2}{k_0^2} - 1 \right] d \mathbf{k} \right)}_{\Delta E_2} \nonumber \\
   &= \Delta E_1 + c \ \Delta E_2,
\label{eq:delta_E_SI_2}  
\end{align}
where $\Delta E_1$ and $\Delta E_2$ are system-dependent constants independent of $c$. To remove the dependence on $\Delta E_1$ and $\Delta E_2$, we normalize $\Delta E$ as,
\begin{equation}
    \Delta \tilde{E} (c) = \frac{\Delta E(c) - \Delta E(c=-1)}{\Delta E(c=0) - \Delta E(c=-1)} = 1+c.
\label{eq:delta_E_SI_c}  
\end{equation}

We next aim to relate $\Delta E$ to the batch fraction $b_f$ and learning rate $\alpha$ for anti-correlated noise ($c=-1$). Note that $c=-1$ and $b_f=1$ correspond to (noiseless) gradient descent. We combine $S(\mathbf{k})$ derived from the linearized fluctuating hydrodynamic theory (Eq.~\ref{eq:sk_static_3}) with Eq.~\ref{eq:delta_E_SI_1} to get,
\begin{align}
   \Delta E (\alpha, b_f) &= \frac{N \sigma^2}{2 (2 \pi)^d} \int \hat{V}(\mathbf{k}) k^2 \left( 1 - S_0 - cS_0 - \frac{MS_0}{k_0^2}k^2 - \frac{cMS_0}{k_0^2}k^2 + \frac{cS_0}{k_0^2}k^2 \right) d \mathbf{k} \nonumber \\
   &= \left( \frac{N \sigma^2}{2 (2 \pi)^d} \int \hat{V}(\mathbf{k}) k^2 d \mathbf{k} \right) - S_0 \left( \frac{N \sigma^2}{2 (2 \pi)^d} \int \hat{V}(\mathbf{k}) k^2 \left[ 1 + c + \left[ M + (M - 1)c \right] \frac{k^2}{k_0^2} \right] d \mathbf{k} \right) \nonumber \\
   &= \underbrace{ \left( \frac{N \sigma^2}{2 (2 \pi)^d} \int \hat{V}(\mathbf{k}) k^2 d \mathbf{k} \right)}_{\Delta E_3} - S_0 \underbrace{ \left( \frac{N \sigma^2}{2 (2 \pi)^d} \int \hat{V}(\mathbf{k}) \frac{k^4}{k_0^2} d \mathbf{k} \right)}_{\Delta E_4} \nonumber \\
   &= \Delta E_3 - \underbrace{ \frac{1}{2 \left( 1 + \frac{4 R^2 d \, \Gamma(p+1) \Gamma(2p+d-1)}{E \alpha (1-b_f) p \Gamma(2p-1) \Gamma(p+d+1)} \right)} }_{g(\alpha, b_f)} \Delta E_4,
\label{eq:delta_E_SI_3}  
\end{align} 
where we substitute $c=-1$ on the third step, use Eqs.~\ref{eq:s0} and the fact that $k_0$ is independent of $\alpha$ and $b_f$ (Eqs.~\ref{eq:k0}, and \ref{eq:consts_sgd}). $\Delta E_3$ and $\Delta E_4$ are constants independent of $\alpha$ and $b_f$, and $g(\alpha, b_f)$ depends on $\alpha$ and $b_f$. It is evident by taking $\lim_{E \alpha (1-b_f)/R^2 \to 0}$ and $\lim_{E \alpha (1-b_f)/R^2 \to \infty}$ in Eq.~\ref{eq:delta_E_SI_3} that $g(\alpha, b_f) \in (0, \frac{1}{2})$ for any arbitrary $p$ and $d$. It is also evident from Eq.~\ref{eq:delta_E_SI_3} that $\Delta E (\alpha, b_f)$ increases monotonically with $b_f$, and decreases monotonically with $\alpha$. 
For a fixed $b_f$, to remove the dependence on $\Delta E_3$ and $\Delta E_4$, we normalize $\Delta E (\alpha)$ as,
\begin{align}
   \Delta \tilde{E} (\alpha)
   = \frac{\Delta E(\alpha) - \Delta E(\alpha_{\text{max}})}{\Delta E(\alpha_{\text{min}}) - \Delta E(\alpha_{\text{max}})} = \frac{g(\alpha) - g(\alpha_{\text{max}})}{g(\alpha_{\text{min}}) - g(\alpha_{\text{max}})}.
\label{eq:delta_E_SI_alpha}  
\end{align} 
Similarly, for a fixed $\alpha$, we normalize $\Delta E (b_f)$ as,
\begin{align}
   \Delta \tilde{E} (b_f)
   = \frac{\Delta E(b_f) - \Delta E({b_f}_{\text{min}})}{\Delta E({b_f}_{\text{max}}) - \Delta E({b_f}_{\text{min}})} = \frac{g(b_f) - g({b_f}_{\text{min}})}{g({b_f}_{\text{max}}) - g({b_f}_{\text{min}})}.
\label{eq:delta_E_SI_bf}  
\end{align}

\bibliographystyle{apsrev4-2}
\bibliography{ref.bib}

\clearpage

\begin{figure}[htbp!]
    \centering
    \includegraphics[width=\linewidth]{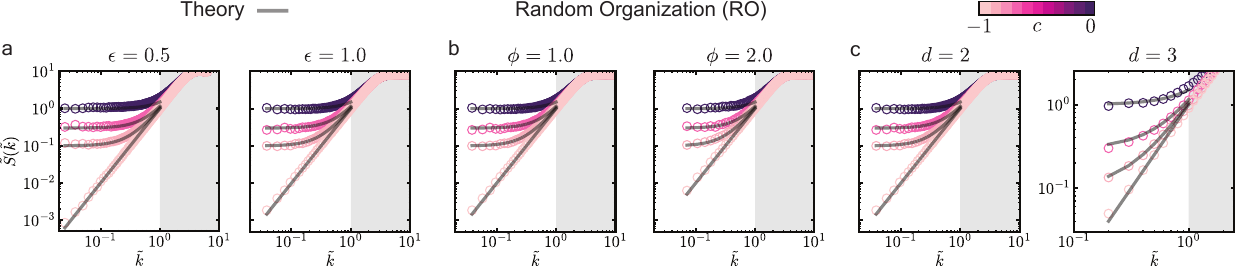}
    \caption{\small Dependence of the structure factor on system parameters for Random Organization (RO). Normalized radially averaged structure factor $\Tilde{S}(\Tilde{k})$ versus normalized radial wave number $\Tilde{k}$ for different kick magnitude $\epsilon$ (a), packing fraction $\phi$ (b), and spatial dimension $d$ (c) in discrete-time particle simulations. Baseline parameters are set to $N = 318309$, $\epsilon = 1.0$, $\phi = 1.0$, $d = 2$, and each panel varies one parameter. For $d=3$, we choose $N = 2546472$ to better explore large length scales. $\tilde{S} =  S(k)/S_{0}(2\pi/L)$ where $S_{0}(2\pi/L)$ is the structure factor for $c=0$ at $k = 2\pi/L$, and $L$ is the side length of the simulation box. $\tilde{k} = k/k_0$ where $k_0$ is the value at which $\tilde{S}(k_0) = 1$ for anti-correlated noise ($c=-1$) of the same system. Solid black lines show predictions of Eq.~\ref{eq:sk_static_3} for different values of $c$, where $M$ is substituted from Eq.~\ref{eq:M_ro}. Gray shaded regions denote short-range behavior ($\tilde{k}>1$).
}
    \label{fig:figure2_si}
\end{figure}

\clearpage

\begin{figure}[htbp!]
    \centering
    \includegraphics[width=\linewidth]{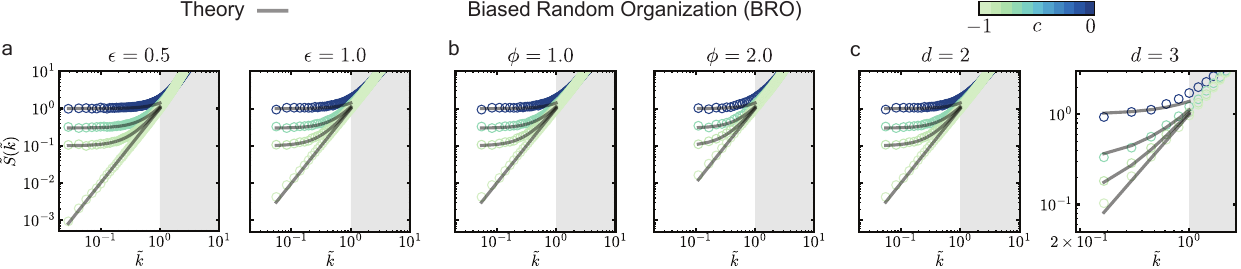}
    \caption{\small Dependence of the structure factor on system parameters for Biased Random Organization (BRO). Normalized radially averaged structure factor $\Tilde{S}(\Tilde{k})$ versus normalized radial wave number $\Tilde{k}$ for different kick magnitude $\epsilon$ (a), packing fraction $\phi$ (b), and spatial dimension $d$ (c) in discrete-time particle simulations. Baseline parameters are set to $N = 318309$, $\epsilon = 1.0$, $\phi = 1.0$, $d = 2$, and each panel varies one parameter. For $d=3$, we choose $N = 2546472$ to better explore large length scales. $\tilde{S} =  S(k)/S_{0}(2\pi/L)$ where $S_{0}(2\pi/L)$ is the structure factor for $c=0$ at $k = 2\pi/L$, and $L$ is the side length of the simulation box. $\tilde{k} = k/k_0$ where $k_0$ is the value at which $\tilde{S}(k_0) = 1$ for anti-correlated noise ($c=-1$) of the same system. Solid black lines show predictions of Eq.~\ref{eq:sk_static_3} for different values of $c$, where $M$ is substituted from Eq.~\ref{eq:M_bro}. Gray shaded regions denote short-range behavior ($\tilde{k}>1$).  
}
    \label{fig:figure3_si}
\end{figure}

\clearpage

\begin{figure}[htbp!]
    \centering
    \includegraphics[width=\linewidth]{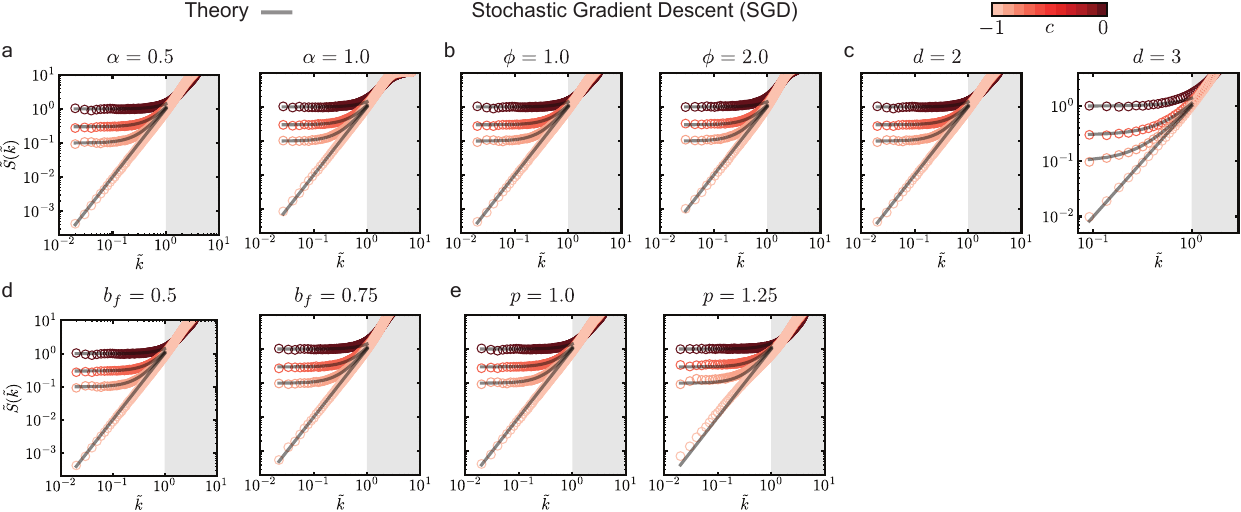}
    \caption{\small Dependence of the structure factor on system parameters for Stochastic Gradient Descent (SGD). Normalized radially averaged structure factor $\Tilde{S}(\Tilde{k})$ versus normalized radial wave number $\Tilde{k}$ for different learning rate $\alpha$ (a), packing fraction $\phi$ (b), spatial dimension $d$ (c), batch fraction $b_f$ (d), and stiffness of the potential $p$ (Eq.~\ref{eq:v_ji_family_SI}) (e) in discrete-time particle simulations. Baseline parameters are set to $N = 318309$, $\alpha = 0.5$, $\phi = 1.0$, $d = 2$, $b_f = 0.5$, $p=1.0$, and each panel varies one parameter. For $d=3$, we choose $N = 2546472$ to better explore large length scales. $\tilde{S} =  S(k)/S_{0}(2\pi/L)$ where $S_{0}(2\pi/L)$ is the structure factor for $c=0$ at $k = 2\pi/L$, and $L$ is the side length of the simulation box. $\tilde{k} = k/k_0$ where $k_0$ is the value at which $\tilde{S}(k_0) = 1$ for anti-correlated noise ($c=-1$) of the same system. Solid black lines show predictions of Eq.~\ref{eq:sk_static_3} for different values of $c$, where $M$ is substituted from Eq.~\ref{eq:M_sgd}. Gray shaded regions denote short-range behavior ($\tilde{k}>1$).
}
    \label{fig:figure4_si}
\end{figure}

\clearpage

\begin{figure}[htbp!]
    \centering
    \includegraphics[width=\linewidth]{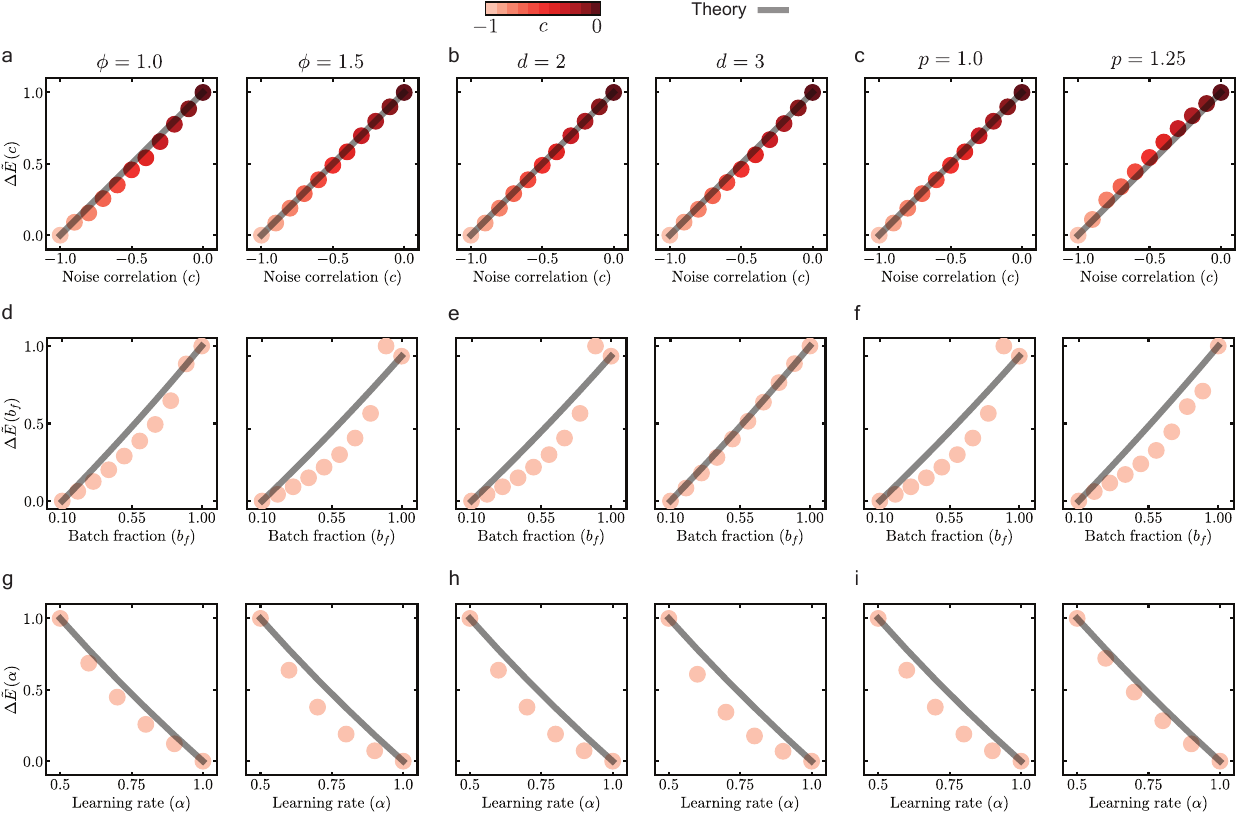}
    \caption{\small Dependence of flatness of energy minimum  on system parameters for Stochastic Gradient Descent (SGD). Normalized energy change $\Delta \Tilde{E}(c)$ versus noise correlation $c$ for different packing fraction $\phi$ (a), spatial dimension $d$ (b), and stiffness of the potential $p$ (Eq.~\ref{eq:v_ji_family_SI}) (c). $\Delta E(c)$ is normalized as: $\Delta \Tilde{E} (c) = [\Delta E (c) - \Delta E (c=-1)] / [\Delta E (c=0) - \Delta E (c=-1)]$. Black lines show prediction of Eq.~\ref{eq:delta_E_SI_c} (SI Sec. I.C). Normalized energy change $\Delta \Tilde{E}(b_f)$ versus batch fraction $b_f$ for different packing fraction $\phi$ (d), spatial dimension $d$ (e), and stiffness of the potential $p$ (Eq.~\ref{eq:v_ji_family_SI}) (f). $\Delta E(b_f)$ is normalized as: $\Delta \Tilde{E} (b_f) = [\Delta E (b_f) - \Delta E (b_f=0.1)] / [\Delta E (b_f=1.0) - \Delta E (b_f=0.1)]$. Black lines show prediction of Eq.~\ref{eq:delta_E_SI_bf} (SI Sec. I.C). Normalized energy change $\Delta \Tilde{E}(\alpha)$ versus learning rate $\alpha$ for different packing fraction $\phi$ (a), spatial dimension $d$ (b), and stiffness of the potential $p$ (Eq.~\ref{eq:v_ji_family_SI}) (c). $\Delta E(\alpha)$ is normalized as: $\Delta E (\alpha) = [\Delta E (\alpha) - \Delta E (\alpha=1.0)] / [\Delta E (\alpha=0.5) - \Delta E (\alpha=1.0)]$. Black lines show prediction of Eq.~\ref{eq:delta_E_SI_alpha} (SI Sec. I.C). The baseline parameters are set to $N=100000$, $\alpha = 0.25$, $\phi = 1.5$, $d = 2$, $b_f = 0.5$, $p = 1.0$, and $c = -1$, with each panel varying a single parameter. For $d = 3$, however, we use $\phi = 0.75$ instead, as higher packing fractions such as $\phi = 1.5$ result in very small values of $\Delta E$, making it difficult to measure accurately in simulations. All symbols denote discrete-time particle simulations.
}
    \label{fig:figure5_si}
\end{figure}